\documentclass[11pt]{article}

\PassOptionsToPackage{hyphens}{url} 
\usepackage[final]{acl}
\usepackage{times}  
\usepackage{helvet}  
\usepackage{courier}  
\usepackage[hyphens]{url}  
\usepackage{graphicx} 
\usepackage{natbib}  
\usepackage{caption} 
\usepackage{algorithm}
\usepackage{algorithmic}
\usepackage{enumitem}
\usepackage{makecell}
\usepackage{multirow}
\usepackage{booktabs,array}

\usepackage{amsmath}

\usepackage[skins,breakable]{tcolorbox}
 \tcbset{colback=gray!3,colframe=black!15,boxrule=0.5pt,arc=2pt}
\usepackage{newfloat}
\usepackage{listings}
\DeclareCaptionStyle{ruled}{labelfont=normalfont,labelsep=colon,strut=off} 
\floatstyle{ruled}
\newfloat{listing}{tb}{lst}{}
\floatname{listing}{Listing}
\usepackage{times}
\usepackage{latexsym}

\usepackage[T1]{fontenc}

\usepackage[utf8]{inputenc}
\usepackage{tabularx,booktabs,seqsplit} \newcommand{\ttwrap}[1]{\begingroup\ttfamily\small\seqsplit{#1}\endgroup}

\usepackage{microtype}

\usepackage{inconsolata}

\usepackage{graphicx}

\usepackage[dvipsnames,table]{xcolor} 
\usepackage{listings}
\usepackage{tcolorbox}
\tcbuselibrary{listings,skins,breakable}

\newcommand{\DiffAdd}[1]{\textcolor{green!70!black}{\texttt{#1}}}
\newcommand{\DiffDel}[1]{\textcolor{red!70!black}{\texttt{#1}}}

\lstdefinelanguage{Diff}{
  morecomment=[f][\color{green!70!black}]{+},
  morecomment=[f][\color{red!70!black}]{-},
  morecomment=[f][\color{gray!70!black}]{@@}, 
}

\lstdefinestyle{diffstyle}{
  language=Diff,
  basicstyle=\ttfamily\footnotesize,
  columns=fullflexible,
  keepspaces=true,
  showstringspaces=false,
  breaklines=true,
  breakatwhitespace=false,
  postbreak=\mbox{\textcolor{gray}{$\hookrightarrow$}\space},
  escapeinside={(*@}{@*)},
}

\usepackage{pgfplots}\pgfplotsset{compat=1.18}

\usetikzlibrary{patterns}
\title{ATLAS: Adaptive Trading with LLM AgentS \\Through Dynamic Prompt Optimization and Multi-Agent Coordination}

  \author{
    Charidimos Papadakis, Angeliki Dimitriou, Giorgos Filandrianos, \\
    \textbf{Maria Lymperaiou, Konstantinos Thomas, Giorgos Stamou}\\ 
    School of Electrical and Computer Engineering, AILS Laboratory\\
    National Technical University of Athens \\
    \texttt{\href{mailto:harrypapadakis02@gmail.com}{harrypapadakis02@gmail.com}}, \\ 
    \texttt{\{\href{mailto:angelikidim@ails.ece.ntua.gr}
    {angelikidim}, \href{mailto:geofila@ails.ece.ntua.gr}{geofila}, \href{mailto:marialymp@ails.ece.ntua.gr}{marialymp}, \href{mailto:kthomas@ails.ece.ntua.gr}{kthomas}}\}@ails.ece.ntua.gr,
    \\
    \texttt{\href{mailto:gstam@cs.ntua.gr}{gstam@cs.ntua.gr}}\\
}

\begin{document}
\maketitle

\begin{abstract}

Large language models (LLMs) offer promising capabilities for financial decision-making, yet their deployment in sequential trading settings faces two key challenges: synthesizing heterogeneous information sources and adapting agent behavior under delayed and noisy reward signals. 
We address these challenges by introducing \emph{ATLAS} (\emph{Adaptive Trading with LLM AgentS}), a unified agentic framework for systematic integration of market data, financial news, and corporate fundamentals, and \emph{Adaptive-OPRO}, a novel prompt optimization method that dynamically updates agent instructions using real-time stochastic feedback. 
We evaluate our approach across regime-specific equity trading scenarios and multiple LLM families. Results demonstrate that Adaptive-OPRO consistently outperforms existing methods, particularly in highly volatile regimes. Moreover, our analysis reveals that increased information availability does not necessarily translate to improved performance, highlighting the importance of careful modality integration in noisy market environments.

\end{abstract}


\section{Introduction}
Financial markets represent one of humanity's most complex decision-making environments, requiring synthesis of vast information, from technical indicators and fundamental analysis to breaking news and market sentiment. LLMs introduce new possibilities for financial decision-making through their ability to process diverse data sources and reason over complex scenarios.

From the model's perspective, financial trading serves as an ideal testbed: it combines unambiguous metrics, sequential complexity, multimodal reasoning requirements, and inherent stochasticity. Unlike synthetic benchmarks, markets provide extensive historical data without simulation bias and reward genuine understanding over pattern memorization. LLMs can therefore be tasked to make decisions under uncertainty, revealing capabilities in complex reasoning \cite{he-etal-2025-breaking}, market understanding \cite{li-etal-2025-investorbench}, and high-risk decision-making \cite{hung-etal-2023-walking}.

Despite this potential, stock market decision-making introduces inherent challenges beyond stochasticity. Decisions require synthesizing heterogeneous signals such as price dynamics, market conditions, and firm-specific developments into coherent actions. Moreover, in high-stakes financial environments where capital is continuously at risk, static decision policies are insufficient; decision patterns must be revised by incorporating market feedback as it unfolds, enabling continuous behavioral adaptation.

Consequently, turning LLM capabilities into reliable trading systems raises two key queries: (i) how diverse signals are synthesized into coherent guidance, and (ii) how models adapt their behavior through continuous market interaction. While recent work explores these issues partly, their systematic study in realistic trading settings is limited.

In this work, (i) we propose ATLAS, a multi-agent framework that provides a foundational structure for experimentation in LLM-based stock market decision-making; (ii) we introduce Adaptive-OPRO, a prompt optimization mechanism for sequential settings that supports behavioral adaptation through ongoing market interaction and achieves state-of-the-art performance across multiple market regimes and LLM families. Through extensive regime-aware evaluations, we show that additional input modalities are not uniformly beneficial and depend critically on market conditions.

\section{Related Work}

\paragraph{LLM Agents in Financial Markets}
Recent work explores several LLM-based trading agents, from sentiment-driven \cite{kirtac-germano-2024-enhanced} to coordinated, multi-component systems \cite{zhou2025multi, yang2025agentnet, liu2023dynamic}. Examples include CryptoTrade, which integrates on/off-chain signals with reflection \cite{li-etal-2024-cryptotrade}, and TradingAgents, which coordinates specialized roles via structured debate and synthesis \cite{xiao2025tradingagentsmultiagentsllmfinancial}. Memory-centric designs such as FinMem emphasize persistent, task-specific recall \cite{yu2023finmemperformanceenhancedllmtrading}, while FINCON introduces conceptual verbal reinforcement to shape multi-agent collaboration \cite{yu2024finconsynthesizedllmmultiagent}. 
Others incorporate learning signals \cite{xiong2025flagtraderfusionllmagentgradientbased} or mixture-of-experts routing \cite{ding2025tradexpertrevolutionizingtradingmixture}, and focus on document-centric analysis, e.g., filings and earnings calls \cite{fatouros2025marketsenseai20enhancingstock}.
However, key limitations persist: prompts are mainly hand-crafted even when feedback is delayed and noisy, and many setups collapse execution into directional scores. ATLAS pairs a prompt-tuning component with order-level evaluation (type, size, timing, price) in a simulator built for such interfaces \cite{papadakis2025stocksim}, using multi-run reporting to account for stochastic variability \cite{song-etal-2025-good, atil2025nondeterminismdeterministicllmsettings}.

\paragraph{Prompt Optimization} enhances LLM performance beyond manual tuning. Optimization by PROmpting (OPRO) treats the model as a meta-optimizer over instruction text, achieving gains on single-turn tasks with immediate feedback \cite{yang2024largelanguagemodelsoptimizers}. Extensions explore evolutionary search and reinforcement-style updates \cite{guo2025evopromptconnectingllmsevolutionary, do2024largelanguagemodelsprompting, austin2024gradsumleveraginggradientsummarization}. These settings typically assume fast, unambiguous scoring and independent instances. In contrast, trading provides deferred, noisy reward signals and sequentially coupled decisions. Adaptive-OPRO addresses such challenges by using rolling evaluation windows and by separating static instructions from dynamic run-time content, allowing stability where consistency matters and controlled evolution where change is beneficial.

\section{ATLAS Framework}

\begin{figure*}[t]
\small
\vskip -0.1in
  \centering
  \includegraphics[width=0.87\textwidth]{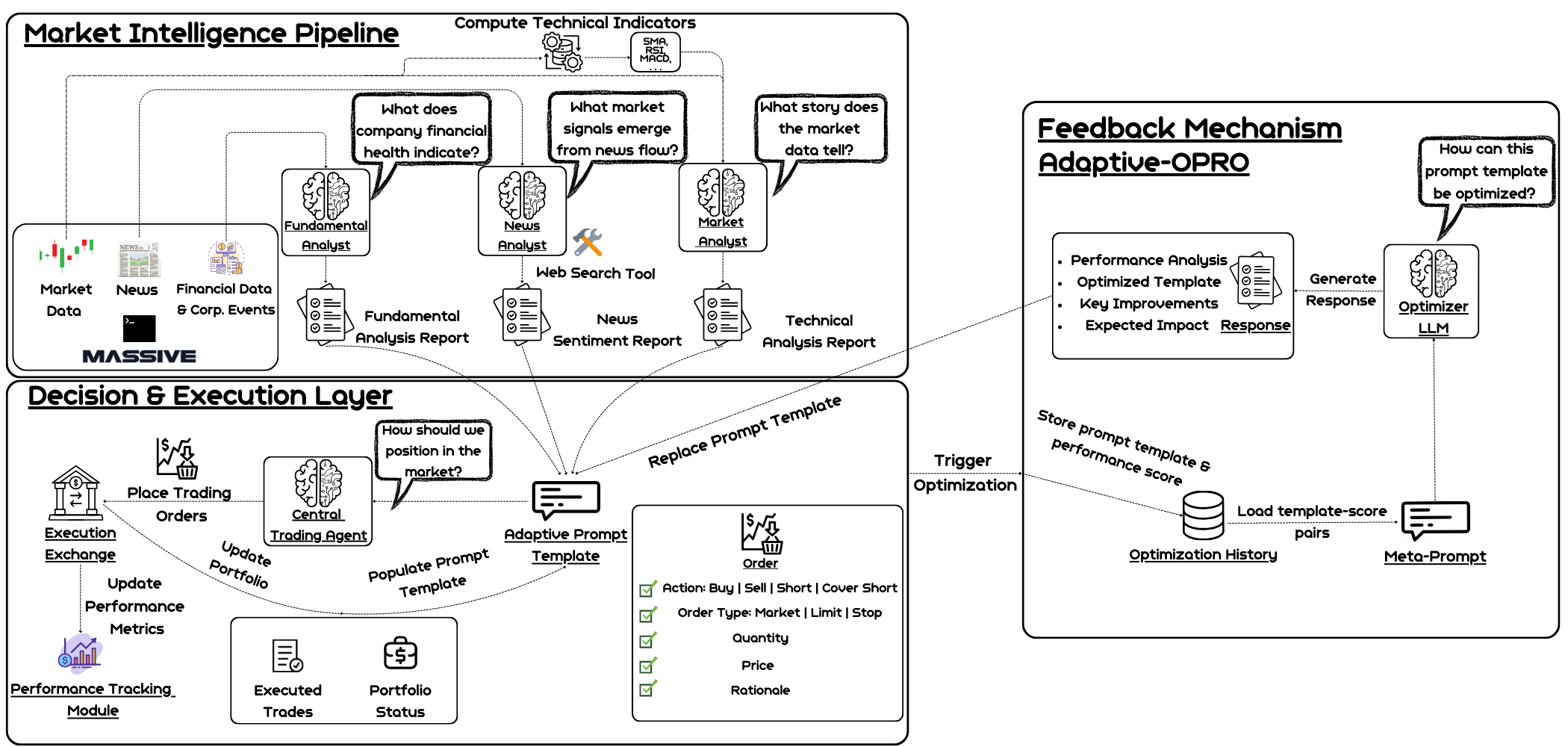}
  \caption{ATLAS Framework Overview. The Central Trading Agent submits orders to the Trading Execution Engine via prompts shaped by three specialized analysts and the proposed \emph{Adaptive-OPRO} optimization technique.}
  \label{fig:fig_1}
\end{figure*}

ATLAS comprises three main components:
(i) a \emph{Market Intelligence Pipeline}, which consists of specialized agents that prepare market, news, and fundamental inputs for downstream decisions;
(ii) a \emph{Decision \& Execution Layer} centered on a Central Trading Agent that generates and executes orders; and
(iii) a feedback mechanism that collects post-execution signals and feeds them back for continuous adaptation. Within the feedback mechanism we incorporate Adaptive-OPRO, an extension of the OPRO framework that dynamically edits the Central Trading Agent’s instruction prompt based on real-time, stochastic market feedback. Figure \ref{fig:fig_1} provides an overview of the ATLAS framework.

\paragraph{Market Intelligence Pipeline.}
ATLAS separates information preparation from decision-making. The Market Intelligence Pipeline consists of three specialized agents, each with a distinct analyst role. \textbf{Market Analyst} produces multi-timescale summaries from price and volume in varying time scales (2 years, 6 months, and 3 months of history with monthly, weekly, and daily candlesticks, respectively). Within each window it computes standard indicators (e.g., moving averages, momentum, volatility bands, support/resistance) and refreshes daily, providing a consistent, noise-filtered description rather than trading signals (details in App.~\ref{sec:technical_indicators}).
\textbf{News Analyst} aggregates relevant articles into structured fields (\emph{Sentiment Assessment}, \emph{Key Developments}, \emph{Market Relevance}, \emph{Source Analysis}) with optional full-text retrieval to move beyond headlines (details in App.~\ref{app:news}).
\textbf{Fundamental Analyst} extracts material changes from periodic reports and corporate events, activating infrequently to mirror reporting cycles and provide medium- to long-horizon context (details in App.~\ref{sec:fundamental}).

\paragraph{Decision \& Execution Layer.}
Here, the trading actions (e.g. buying or selling a stock) are determined, executed, and corresponding market feedback is received. The main decision-making component within this layer is the \textbf{Central Trading Agent (CTA)}.
This agent consumes the structured inputs and current portfolio and emits orders that specify type (market, limit, stop), size, timing, and price levels. Orders are executed in StockSim \cite{papadakis2025stocksim}, which enforces core trading semantics and returns fills, positions, and cash for the next step. Order-level decisions clarify intent and link analytical quality to execution choices.

\paragraph{Feedback Mechanism.}
This mechanism defines how information derived from market outcomes is incorporated into the agent’s future decisions. It may be entirely absent, resulting in a static agent that follows a fixed policy, or it may be enabled to support adaptation based on observed performance. In general, the mechanism processes signals such as returns or behavioral outcomes from past decisions and uses them to influence subsequent actions. Implementations can range from simple feedback summaries to more structured optimization approaches, such as reflection-based methods \cite{li-etal-2024-cryptotrade}. In the following section, we describe Adaptive-OPRO, a prompt optimization technique that leverages market feedback to iteratively refine the agent’s decision-making process.

\section{Adaptive-OPRO}
\label{sec:adaptive_opro}

\emph{Adaptive-OPRO} is a sequential prompt-optimization procedure that improves an agent's instruction prompt using delayed, noisy performance feedback. It generalizes OPRO to interactive settings where decisions are temporally coupled and rewards arrive after multiple steps. The core idea is to treat instruction text as the optimized object and to update it periodically via a learned update mechanism (implemented by an optimizer LLM), while keeping the agent's run-time inputs and interfaces stable.

\paragraph{Optimized object, state, and round inputs.}
Adaptive-OPRO maintains a current instruction prompt \(P_t\) for a target agent that acts over time, along with an \emph{optimization history}
\(\mathcal{H}=\{(P_i, s_i)\}_{i<t}\) storing past prompt variants and their scores.
At the end of each evaluation window, it constructs an optimizer query with the following inputs:
(i) a \emph{meta-prompt} \(M\) that specifies the optimizer's role and constraints,
(ii) \(\mathcal{H}\) (or a compact summary thereof), and
(iii) a summary of the agent's recent interaction outcomes together with a scalar performance score \(s_t\) for \(P_t\).
An update rule \(U\), implemented using an optimizer LLM, produces a revised prompt
\(
P_{t+1} = U(M, \mathcal{H}, s_t, \text{summary})
\).
The output of each round is an updated instruction prompt that governs subsequent agent decisions.

\paragraph{Stability via template separation.}
In sequential systems, prompt updates can inadvertently break the run-time interface (e.g., input placeholders, output schemas) or overfit to transient observations. Adaptive-OPRO therefore separates the target agent prompt into:
(a) \emph{static instructions} (policy, priorities, constraints, formatting requirements), and
(b) \emph{dynamic run-time content} injected at execution time (state, observations, tool outputs, recent actions).
Only the \emph{static instruction block} is editable; all placeholders and the run-time injection format are held fixed. This enforces \emph{edit locality}: updates can change \emph{how} the agent reasons and decides, but cannot change \emph{what} information it receives nor the interface it must comply with.

\paragraph{Windowed evaluation under delayed feedback.}
To address credit assignment and reduce variance, Adaptive-OPRO evaluates prompts over rolling windows of \(K\) decision steps. After each window, the system computes a scalar performance score \(s\) from outcomes observed during that window (e.g. task success, reward, utility, risk-adjusted return). The choice of \(K\) and scoring function is task-dependent; the only requirement is that \(s\) provides consistent ordering to compare prompt variants.

\begin{figure}[t]
\centering
\begin{tikzpicture}[
  >=latex, font=\small,
  window/.style={draw, rounded corners=2pt, minimum width=1.55cm,
                 minimum height=0.6cm, fill=blue!8},
  prompt/.style={draw, rounded corners=2pt, minimum width=1.0cm,
                 minimum height=0.55cm, fill=orange!15},
  opt/.style={draw, rounded corners=2pt, minimum width=0.75cm,
              minimum height=0.55cm, fill=gray!12, font=\footnotesize},
]
\node[prompt] (P1) at (0,1.52)   {$P_1$};
\node[opt]    (O1) at (1.45,1.52){$U$};
\node[prompt] (P2) at (2.9,1.52) {$P_2$};
\node[opt]    (O2) at (4.35,1.52){$U$};
\node[prompt] (P3) at (5.8,1.52) {$P_3$};
\node at (6.85,1.55) {$\cdots$};

\draw[->] (P1) -- (O1);
\draw[->] (O1) -- (P2);
\draw[->] (P2) -- (O2);
\draw[->] (O2) -- (P3);

\node[window] (W1) at (0,0)   {$W_1$};
\node[window] (W2) at (2.9,0) {$W_2$};
\node[window] (W3) at (5.8,0) {$W_3$};

\draw[->, thick] (P1) -- node[left, font=\scriptsize, xshift=1pt]{apply} (W1);
\draw[->, thick] (P2) -- (W2);
\draw[->, thick] (P3) -- (W3);

\draw[->, thick, blue!55!black]
  (W1.north east) -- node[above, font=\scriptsize, sloped]{$s_1,\mathcal{H}_{\le 1}$} (O1.south west);
\draw[->, thick, blue!55!black]
  (W2.north east) -- node[above, font=\scriptsize, sloped]{$s_2,\mathcal{H}_{\le 2}$} (O2.south west);

\draw[->, gray!70] (-0.8,-0.45) -- node[below, font=\scriptsize]{time} (7.0,-0.45);
\end{tikzpicture}
\caption{Adaptive-OPRO as an online procedure. Prompt $P_t$ governs
window $W_t$; its realized score $s_t$ and the accumulated history
$\mathcal{H}_{\le t}$ feed the optimizer $U$, which emits $P_{t+1}$
for the next window.}
\label{fig:protocol}
\end{figure}

\paragraph{Meta-prompted update rule.}
At each window, Adaptive-OPRO forms the optimizer query from $M$, $\mathcal{H}$, and the recent outcome summary/score, and applies $U$ to generate a candidate $P_{t+1}$ (Figure~\ref{fig:protocol}).
The optimizer is instructed to:
(i) diagnose likely failure modes of the current prompt,
(ii) propose a revised instruction prompt \(P_{t+1}\),
(iii) summarize the concrete changes made, and
(iv) state the expected behavioral impact.
The candidate is accepted only if it preserves the template (e.g., placeholders and output schema). The accepted prompt governs the next window onward and is appended to history with its subsequent score; since no window is ever replayed under a later prompt, each variant is assessed on outcomes that unfolded after it was written, keeping optimization and evaluation on disjoint segments of the trajectory.

\paragraph{ATLAS instantiation.}
In ATLAS, Adaptive-OPRO is applied to the \emph{Central Trading Agent}'s instruction prompt (i.e., the static instruction block of the decision policy). Dynamic run-time content corresponds to the daily injected analyst summaries, portfolio state, and recent executions, which are kept fixed by construction (App.~\ref{app:opro}). For scoring, we aggregate portfolio performance over \(K{=}5\) trading days to reduce noise and capture delayed effects of sequential decisions, then map cumulative ROI to a bounded OPRO-style score \(s\in[0,100]\) via linear scaling and clipping:
\begin{equation}
s=\mathrm{clip}_{[0,100]}\big(50+250\cdot \mathrm{ROI}\big),
    \label{eq:opro_score}
\end{equation}
so that \(-20\%\!\mapsto\!0\), \(0\%\!\mapsto\!50\), \(+20\%\!\mapsto\!100\).
This yields a stable, delay-aware signal while limiting the impact of outlier windows; the optimizer is restricted to instruction edits that preserve ATLAS's execution interface.

\begin{table*}[t]
\centering \small
\begin{tabular}{llccccc}
\hline
\textbf{Model} & \textbf{Prompting} & \textbf{ROI (\%) $\uparrow$} & \textbf{SR $\uparrow$} & \textbf{DD (\%) $\downarrow$} & \textbf{Win Rate (\%) $\uparrow$} & \textbf{Num.\ Trades} \\
\hline

\multicolumn{7}{c}{\textbf{Non-LLM-Based Strategies}} \\
\hline
Buy \& Hold & N/A & -8.59 & -0.071 & 20.45 & 0.00 & 1 \\
MACD & N/A & 6.50 & 0.131 & 6.86 & 0.00 & 1 \\
SMA & N/A & 6.91 & 0.177 & 3.56 & 50.00 & 4 \\
SLMA  & N/A & -1.87 & -0.078 & 6.89 & 0.00 & 1 \\
Bollinger Bands & N/A & 0.00 & 0.000 & 0.00 & 0.00 & 0 \\
\hline

\multicolumn{7}{c}{\textbf{LLM-Based Strategies - ATLAS}} \\
\hline
\multirow{3}{*}{Llama 3.3-70B} & Baseline & -9.19\textsubscript{± 1.54} & -0.091\textsubscript{± 0.021} & 16.90\textsubscript{± 0.82} & 30.28\textsubscript{± 11.87} & 22.67\textsubscript{± 8.39} \\
& Reflection & -8.44\textsubscript{± 1.58} & -0.087\textsubscript{± 0.025} & 16.36\textsubscript{± 0.31} & 44.69\textsubscript{± 13.25} & 27.67\textsubscript{± 1.15} \\
& Adaptive-OPRO & \textbf{-6.16\textsubscript{± 2.08}} & \textbf{-0.066\textsubscript{± 0.004}} & \textbf{14.05\textsubscript{± 3.33}} & \textbf{54.36\textsubscript{± 12.44}} & 28.33\textsubscript{± 3.21} \\
\hline

\multirow{3}{*}{Qwen3-235B} & Baseline & -1.78\textsubscript{± 3.86} & -0.006\textsubscript{± 0.039} & 13.09\textsubscript{± 1.88} & 36.51\textsubscript{± 17.55} & 13.00\textsubscript{± 4.00} \\
& Reflection & -5.76\textsubscript{± 2.97} & -0.049\textsubscript{± 0.033} & 14.18\textsubscript{± 1.91} & 25.00\textsubscript{± 0.00} & 8.67\textsubscript{± 0.58} \\
& Adaptive-OPRO & \textbf{1.33\textsubscript{± 1.91}} & \textbf{0.025\textsubscript{± 0.019}} & \textbf{11.41\textsubscript{± 0.06}} & \textbf{50.00\textsubscript{± 0.00}} & 9.00\textsubscript{± 0.00} \\
\hline

\multirow{3}{*}{Qwen3-32B} & Baseline & -10.62\textsubscript{± 3.54} & -0.087\textsubscript{± 0.031} & 16.72\textsubscript{± 2.75} & 30.00\textsubscript{± 10.00} & 25.33\textsubscript{± 1.53} \\
& Reflection & -7.76\textsubscript{± 0.90} & -0.065\textsubscript{± 0.002} & 16.47\textsubscript{± 3.44} & 28.72\textsubscript{± 25.06} & 31.67\textsubscript{± 2.31} \\
& Adaptive-OPRO & \textbf{-3.48\textsubscript{± 2.19}} & \textbf{-0.022\textsubscript{± 0.021}} & \textbf{15.52\textsubscript{± 0.68}} & \textbf{43.45\textsubscript{± 6.27}} & 28.67\textsubscript{± 1.53} \\
\hline

\multirow{3}{*}{\makecell{Claude Sonnet 4}} & Baseline & -7.26\textsubscript{± 2.99} & -0.066\textsubscript{± 0.030} & 17.59\textsubscript{± 1.55} & 31.19\textsubscript{± 7.84} & 13.00\textsubscript{± 4.36} \\
& Reflection & -5.69\textsubscript{± 1.82} & -0.058\textsubscript{± 0.013} & 15.12\textsubscript{± 3.26} & \textbf{46.67\textsubscript{± 5.77}} & 12.67\textsubscript{± 2.08} \\
& Adaptive-OPRO & \textbf{0.35\textsubscript{± 1.78}} & \textbf{0.008\textsubscript{± 0.018}} & \textbf{14.76\textsubscript{± 2.87}} & 43.45\textsubscript{± 6.27} & 15.00\textsubscript{± 2.00} \\
\hline
\multirow{3}{*}{\makecell{Claude Sonnet 4\\ w/ Thinking}} & Baseline & -4.46\textsubscript{± 4.76} & -0.043\textsubscript{± 0.048} & 14.32\textsubscript{± 4.12} & 11.11\textsubscript{± 19.24} & 14.00\textsubscript{± 2.65} \\
& Reflection & -8.60\textsubscript{± 0.59} & -0.078\textsubscript{± 0.004} & 19.45\textsubscript{± 1.65} & 14.29\textsubscript{± 24.75} & 11.67\textsubscript{± 2.08} \\
& Adaptive-OPRO & \textbf{-0.73\textsubscript{± 3.82}} & \textbf{-0.004\textsubscript{± 0.038}} & \textbf{12.94\textsubscript{± 2.32}} & \textbf{43.89\textsubscript{± 21.11}} & 17.00\textsubscript{± 5.00} \\
\hline
\multirow{3}{*}{GPT-o4-mini} & Baseline & -1.30\textsubscript{± 1.71} & -0.017\textsubscript{± 0.017} & \textbf{9.68\textsubscript{± 3.12}} & 29.17\textsubscript{± 11.02} & 15.33\textsubscript{± 3.06} \\
& Reflection & -2.52\textsubscript{± 4.03} & -0.039\textsubscript{± 0.045} & 9.82\textsubscript{± 3.43} & 51.28\textsubscript{± 5.06} & 20.33\textsubscript{± 3.06} \\
& Adaptive-OPRO & \textbf{9.06\textsubscript{± 0.73}} & \textbf{0.094\textsubscript{± 0.008}} & 11.48\textsubscript{± 0.00} & \textbf{65.28\textsubscript{± 16.84}} & 17.33\textsubscript{± 5.86} \\
\hline
\multirow{3}{*}{GPT-o3} & Baseline & -6.11\textsubscript{± 3.42} & -0.080\textsubscript{± 0.029} & 11.58\textsubscript{± 3.09} & 42.59\textsubscript{± 8.49} & 18.67\textsubscript{± 3.21} \\
& Reflection & -4.60\textsubscript{± 3.40} & -0.053\textsubscript{± 0.044} & 12.11\textsubscript{± 1.27} & 46.03\textsubscript{± 16.88} & 18.33\textsubscript{± 2.52} \\
& Adaptive-OPRO & \textbf{9.02\textsubscript{± 3.28}} & \textbf{0.146\textsubscript{± 0.048}} & \textbf{5.33\textsubscript{± 0.14}} & \textbf{72.81\textsubscript{± 17.27}} & 19.67\textsubscript{± 4.16} \\
\hline

\end{tabular}
\caption{Performance comparison between non-LLM-based and LLM-based approaches using ATLAS in volatile, declining market conditions. \textbf{Bold} values indicate the best per model.}
\label{tab:lly_results}
\end{table*}

\section{Experiments}
\label{sec:exp}

Our study examines ATLAS along three axes:
\textbf{(1) Adaptation} – whether sequential prompt optimization via \emph{Adaptive-OPRO} improves over well-tuned static prompts and over analytical reflection when feedback is delayed and noisy;
\textbf{(2) Component attribution} -- contribution of structured inputs (Market Analyst, News Analyst, Fundamental Analyst) under different regimes;
\textbf{(3) Model capabilities} -- performance of backbone LLMs as both decision policies and prompt optimizers under Adaptive-OPRO, assessed by return and risk-adjusted performance, robustness across runs, and their ability to propose instruction updates for sustained improvements over windows.

\subsection{Experimental Setup}

\paragraph{Assets and time period.}

Specifically, we evaluate stock market decision-making across three distinct regimes: a \textbf{bearish-volatile} regime characterized by declining prices and elevated uncertainty, a \textbf{sideways} regime marked by range-bound price dynamics and limited directional trends, and a \textbf{bullish} regime defined by sustained upward momentum and comparatively favorable risk-return conditions. Each window spans two months (Apr 28-Jun 28, 2025) \emph{with a daily decision interval}: the agent acts once per trading day. This horizon is chosen to (i) capture multiple decision cycles \emph{without regime mixing}, so adaptation reflects outcomes rather than macro shifts, and (ii) preserve complete conversation history (analyst summaries, orders, prompt-evolution logs) within the context limits of all backbones, enabling fair, auditable runs across models and ablations.
More details in App.~\ref{app:experimental-setup}.

The experimental setup, including the evaluation method, metrics, regime partitioning, and evaluation horizon, follows \citet{li-etal-2024-cryptotrade}, ensuring methodological consistency and fair comparison across settings. We explicitly account for LLM stochasticity by running each configuration \emph{three times} and reporting mean $\pm$ standard deviation, distinguishing systematic performance differences from randomness rather than single-run variability.

\textbf{Models.}
We evaluate seven backbones spanning families, sizes, and reasoning modes: GPT-o3, GPT-o4-mini, Claude Sonnet~4 with and without thinking, Llama~3.3-70B, Qwen3-235B, and Qwen3-32B. Each run uses a single backbone for all ATLAS components and Adaptive-OPRO, isolating how model capacity and architecture affect sequential behavior, instruction adherence, stability, and cross-family transfer without per-model tuning.

\textbf{Prompting strategies.}
We compare three strategies for the \emph{Central Trading Agent}:
\textbf{Baseline} -- a fixed instruction prompt obtained via iterative expert prompt engineering;
\textbf{Reflection} \cite{li-etal-2024-cryptotrade} -- a weekly reflection mechanism that summarizes recent trajectories into high-level feedback that the agent must interpret;
\textbf{Adaptive-OPRO} -- our sequential prompt optimization with windowed scoring and template separation (Section~\ref{sec:adaptive_opro}).
Our goal is to isolate the \emph{adaptation mechanism} under identical data and execution semantics. We therefore evaluate all methods within a single, transparent setup rather than re-implementing full external agent stacks, which differ in action spaces, state representations, and execution interfaces. We include reflection as a widely used and portable form of sequential feedback, providing a focused comparison to \emph{Adaptive-OPRO} and the fixed baseline.

\textbf{Non-LLM baselines.}
Following \citet{li-etal-2024-cryptotrade}, we include five widely used quantitative strategies to contextualize results: Buy \& Hold, MACD \cite{Wang2018}, SMA \cite{Gencay1996}, SLMA \cite{Wang2018}, and Bollinger Bands \cite{Day2023}. For window-based methods, we test multiple window lengths per regime and report a strong, representative configuration for each strategy (e.g., 10-day SMA; 10/30-day SLMA). Full specifications in App.~\ref{baseline}.

\textbf{Execution environment.}
Agents interact with StockSim \cite{papadakis2025stocksim} via an
\emph{order-level} action space, requiring CTAs to submit fully
\emph{executable} orders (type, side, size, price). Compared to
signal- or position-level formulations common in prior LLM trading
studies \cite{li-etal-2024-cryptotrade,xiao2025tradingagentsmultiagentsllmfinancial},
this enforces execution feasibility (cash, inventory, validity) and
yields a complete audit trail of orders, fills, and portfolio states.
Consistent with standard offline evaluation, we abstract away four
aspects of market microstructure, namely latency, slippage, market
impact, and intraday partial fills, and assume deterministic
execution. Since these abstractions apply uniformly to every
prompting strategy, they shift absolute return levels while leaving
comparisons between Baseline, Reflection, and Adaptive-OPRO
internally consistent.

\begin{figure*}[t!]
\centering
\begin{tikzpicture}
\begin{axis}[
    every outer x axis line/.append style={-},
    every outer y axis line/.append style={-},
    xtick pos=left,
    ytick pos=left,
    width=0.96\textwidth, height=3.5cm,
    ybar, bar width=5pt,
    enlarge x limits=0.08,
    ymin=-4, ymax=16,
    ylabel={ROI gain over Baseline (pp)},
    ylabel style={font=\footnotesize},
    xlabel={Backbone model},
    xlabel style={font=\footnotesize, yshift=2pt},
    symbolic x coords={Llama-70B, Qwen3-235B, Qwen3-32B,
                       Claude-S4, Claude-S4-T, GPT-o4-mini, GPT-o3},
    xtick=data,
    x tick label style={font=\scriptsize, rotate=18, anchor=north east},
    y tick label style={font=\scriptsize},
    ytick={-4,0,4,8,12,16},
    ymajorgrids, grid style={gray!20},
    legend style={at={(0.5,1.03)}, anchor=south,
                  legend columns=3, font=\scriptsize, draw=gray!40,
                  /tikz/every even column/.append style={column sep=8pt}},
    axis line style={gray!55}, tick style={gray!55},
    clip=false,
]
\addplot[area legend, fill=red!55,         draw=red!55!black]    coordinates {
    (Llama-70B,3.03)  (Qwen3-235B,3.11) (Qwen3-32B,7.14)
    (Claude-S4,7.61)  (Claude-S4-T,3.73)
    (GPT-o4-mini,10.36)(GPT-o3,15.13)};
\addplot[area legend, fill=gray!55,        draw=gray!60!black]   coordinates {
    (Llama-70B,-0.68) (Qwen3-235B,2.70) (Qwen3-32B,7.87)
    (Claude-S4,-0.58) (Claude-S4-T,-0.02)
    (GPT-o4-mini,2.59)(GPT-o3,4.22)};
\addplot[area legend, fill=green!55!black, draw=green!30!black]  coordinates {
    (Llama-70B,4.21)  (Qwen3-235B,-2.66)(Qwen3-32B,12.62)
    (Claude-S4,12.42) (Claude-S4-T,3.84)
    (GPT-o4-mini,3.47)(GPT-o3,2.36)};
\legend{Bearish-volatile (LLY), Sideways (XOM), Bullish (NVDA)}
\end{axis}
\end{tikzpicture}
\caption{ROI gain of Adaptive-OPRO over the Baseline prompt, in percentage points, per backbone and market regime. The effect is most pronounced under bearish-volatile conditions.}
\label{fig:opro_gain}
\end{figure*}
\paragraph{Evaluation Metrics.} We employ five metrics capturing different aspects of trading performance:

\textbf{Return on Investment (ROI)}: Total percentage return calculated as: 
\begin{equation}
   \frac{\text{final value} - \text{initial value}}{\text{initial value}} \times 100
\end{equation}
where portfolio values include both cash holdings and the current market value of all stocks owned.

\textbf{Sharpe Ratio (SR)}: Risk-adjusted return metric: 
\begin{equation}
\frac{\mu - r_f}{\sigma}    
\end{equation}
where $\mu$ is mean daily return, $\sigma$ is daily return standard deviation, and $r_f$ is the risk-free rate (set to 0 as in \citet{li-etal-2024-cryptotrade}).

\textbf{Maximum Drawdown (DD)}: The worst peak-to-trough decline in portfolio value: 
\begin{equation}
    \max_{t \in [0,T]} \left( \max_{s \in [0,t]} V_s - V_t \right) / \max_{s \in [0,t]} V_s
\end{equation}
where $V_t$ is portfolio value at time $t$. This measures the largest loss from any historical high, reflecting downside risk and stress tolerance.

\textbf{Win Rate}: Percentage of \emph{profitable} \emph{closed} (i.e. completed) trades, computed as:
\begin{equation}
  \frac{\text{Closed trades with realized profit > 0}}{\text{Total closed trades}} \times 100.  
\end{equation}
``Closed trades'' are fully opened and exited positions; open positions are excluded. Win rate reflects decision consistency but does not ensure profitability if losses outweigh gains.

\textbf{Number of Trades}: Total trading frequency over the evaluation period. Higher frequencies indicate active, opportunistic short-term strategies, 
while lower frequencies suggest patient, conviction-driven approaches. 
Additional metrics, results, and analyses are reported in App.~\ref{app:extended_results}.

\begin{table*}[t]
\vskip -0.05in
\centering \small
\begin{tabular}{llccccc}
\hline
\textbf{Model} & \textbf{Prompting} & \textbf{ROI (\%) $\uparrow$} & \textbf{SR $\uparrow$} & \textbf{DD (\%) $\downarrow$} & \textbf{Win Rate (\%) $\uparrow$} & \textbf{Num.\ Trades} \\
\hline
\multicolumn{7}{c}{\textbf{Non-LLM-Based Strategies}} \\
\hline
Buy \& Hold      & N/A & 1.14  & 0.013  & 6.97 & 0.00 & 1 \\
MACD             & N/A & -0.26 & -0.019 & 5.90 & 0.00 & 3 \\
SMA    & N/A & -0.13 & -0.019 & 5.57 & 0.00 & 3 \\
SLMA  & N/A & -1.12 & -0.043 & 5.28 & 0.00 & 2 \\
Bollinger Bands  & N/A & 0.00  & 0.000  & 0.00 & 0.00 & 0 \\
\hline

\multicolumn{7}{c}{\textbf{LLM-Based Strategies - ATLAS}} \\
\hline

\multirow{3}{*}{Llama 3.3-70B} & Baseline   & \textbf{-0.42\textsubscript{± 2.06}}            & \textbf{-0.024\textsubscript{± 0.051}}          & 5.56\textsubscript{± 1.08}           & \textbf{53.48\textsubscript{± 9.56}} & 26.00\textsubscript{± 2.00} \\
& Reflection     & -2.61\textsubscript{± 0.77}                    & -0.083\textsubscript{± 0.014}                   & 6.38\textsubscript{± 0.72}           & 46.63\textsubscript{± 3.15}          & 26.33\textsubscript{± 6.51} \\
& Adaptive-OPRO  & -1.10\textsubscript{± 0.44}                    & -0.045\textsubscript{± 0.012}                   & \textbf{5.15\textsubscript{± 0.71}}  & 50.00\textsubscript{± 3.85}          & 25.33\textsubscript{± 1.15} \\
\hline

\multirow{3}{*}{Qwen3-235B} & Baseline   & -2.43\textsubscript{± 0.68}                     & -0.04\textsubscript{± 0.01}                   & \textbf{5.72\textsubscript{± 0.16}}           & \textbf{46.67\textsubscript{± 5.77}}          & 11.66\textsubscript{± 0.57} \\
& Reflection     & -2.02\textsubscript{± 1.44}                    & -0.04\textsubscript{± 0.03}                   & 6.26\textsubscript{± 1.77}           & 36.51\textsubscript{± 5.50}          & 13.33\textsubscript{± 2.31} \\
& Adaptive-OPRO  & \textbf{0.27\textsubscript{± 1.83}}            & \textbf{0.01\textsubscript{± 0.04}} & 7.20\textsubscript{± 2.09} & 32.86\textsubscript{± 15.45} & 11.00\textsubscript{± 3.61} \\
\hline

\multirow{3}{*}{Qwen3-32B} & Baseline   & -9.14\textsubscript{± 1.02}                     & -0.20\textsubscript{± 0.02}                   & 9.82\textsubscript{± 0.90}           & 28.85\textsubscript{± 17.20}          & 21.00\textsubscript{± 1.73} \\
& Reflection     & -7.96\textsubscript{± 3.11}                    & -0.16\textsubscript{± 0.06}                   & 9.05\textsubscript{± 2.90}           & \textbf{40.55\textsubscript{± 15.48}}          & 24.33\textsubscript{± 3.05} \\
& Adaptive-OPRO  & \textbf{-1.27\textsubscript{± 3.21}}            & \textbf{-0.03\textsubscript{± 0.07}} & \textbf{6.75\textsubscript{± 0.54}} & 35.83\textsubscript{± 2.57} & 25.67\textsubscript{± 5.50} \\
\hline

\multirow{3}{*}{\makecell{Claude Sonnet 4}} & Baseline   & -4.49\textsubscript{± 4.22}                    & -0.134\textsubscript{± 0.114}                   & \textbf{7.71\textsubscript{± 1.06}}  & \textbf{37.50\textsubscript{± 4.17}} & 19.00\textsubscript{± 3.46} \\
& Reflection     & \textbf{-3.78\textsubscript{± 4.23}}           & \textbf{-0.115\textsubscript{± 0.105}}          & 10.54\textsubscript{± 1.58}          & 23.84\textsubscript{± 8.27}          & 18.00\textsubscript{± 6.93} \\
& Adaptive-OPRO  & -5.07\textsubscript{± 4.53}                    & -0.165\textsubscript{± 0.143}                   & 9.23\textsubscript{± 2.71}           & 31.02\textsubscript{± 7.90}          & 18.33\textsubscript{± 2.52} \\
\hline
\multirow{3}{*}{\makecell{Claude Sonnet 4\\ w/ Thinking}} & Baseline   & \textbf{-0.99\textsubscript{± 0.80}}           & \textbf{-0.039\textsubscript{± 0.020}}          & 7.75\textsubscript{± 1.00}           & \textbf{56.28\textsubscript{± 1.50}} & 17.00\textsubscript{± 5.20} \\
& Reflection     & -1.49\textsubscript{± 3.76}                    & -0.069\textsubscript{± 0.123}                   & 7.27\textsubscript{± 2.26}           & 45.11\textsubscript{± 12.60}          & 17.00\textsubscript{± 5.57} \\
& Adaptive-OPRO  & -1.01\textsubscript{± 0.90}                    & -0.046\textsubscript{± 0.020}                   & \textbf{5.16\textsubscript{± 0.52}}  & 36.20\textsubscript{± 24.47}          & 16.33\textsubscript{± 2.08} \\
\hline
\multirow{3}{*}{GPT-o4-mini} & Baseline   & 1.29\textsubscript{± 1.38}                      & 0.021\textsubscript{± 0.044}                    & \textbf{3.23\textsubscript{± 0.48}}  & 39.01\textsubscript{± 3.61}          & 22.67\textsubscript{± 7.57} \\
& Reflection     & -1.48\textsubscript{± 0.54}                    & -0.087\textsubscript{± 0.018}                   & 4.64\textsubscript{± 0.75}           & 32.62\textsubscript{± 7.49}          & 27.33\textsubscript{± 3.06} \\
& Adaptive-OPRO  & \textbf{3.88\textsubscript{± 2.21}} & \textbf{0.089\textsubscript{± 0.067}}           & 3.28\textsubscript{± 0.95}           & \textbf{47.95\textsubscript{± 7.15}} & 25.33\textsubscript{± 5.03} \\
\hline
\multirow{3}{*}{GPT-o3} & Baseline   & -0.60\textsubscript{± 1.71}                     & -0.034\textsubscript{± 0.050}                   & 5.93\textsubscript{± 1.33}           & 60.74\textsubscript{± 5.59}          & 16.33\textsubscript{± 2.52} \\
& Reflection     & -1.55\textsubscript{± 2.09}                    & -0.084\textsubscript{± 0.075}                   & 5.02\textsubscript{± 0.72}           & 42.50\textsubscript{± 6.61}          & 16.67\textsubscript{± 0.58} \\
& Adaptive-OPRO  & \textbf{3.62\textsubscript{± 0.90}}            & \textbf{0.096\textsubscript{± 0.027}} & \textbf{3.46\textsubscript{± 0.48}} & \textbf{71.93\textsubscript{± 15.90}} & 16.00\textsubscript{± 2.65} \\
\hline

\end{tabular}
\caption{Performance comparison between non-LLM-based and LLM-based approaches using ATLAS in range-bound market conditions. \textbf{Bold} values indicate the best results per model.}
\label{tab:xom_results}
\end{table*}

\begin{table*}[t]
\vskip -0.05in
\centering \small
\begin{tabular}{llccccc}
\hline
\textbf{Model} & \textbf{Prompting} & \textbf{ROI (\%) $\uparrow$} & \textbf{SR $\uparrow$} & \textbf{DD (\%) $\downarrow$} & \textbf{Win Rate (\%) $\uparrow$} & \textbf{Num.\ Trades} \\
\hline
\multicolumn{7}{c}{\textbf{Non-LLM-Based Strategies}} \\
\hline
Buy \& Hold & N/A & 41.30 & 0.409 & 3.16 & 0.00 & 1 \\
MACD & N/A & -0.62 & -0.343 & 0.62 & 0.00 & 1 \\
SMA & N/A & 36.77 & 0.384 & 3.12 & 0.00 & 1 \\
SLMA & N/A & 15.88 & 0.254 & 2.98 & 0.00 & 1 \\
Bollinger Bands & N/A & 0.00 & 0.000 & 0.00 & 0.00 & 0 \\
\hline
\multicolumn{7}{c}{\textbf{LLM-Based Strategies - ATLAS}} \\ \hline
\multirow{3}{*}{Llama 3.3-70B} & Baseline & 37.86\textsubscript{± 12.31} & 0.388\textsubscript{± 0.096} & 3.46\textsubscript{± 0.63} & 20.37\textsubscript{± 35.28} & 13.00\textsubscript{± 20.78} \\
& Reflection & 40.40\textsubscript{± 1.43} & \textbf{0.422\textsubscript{± 0.023}} & \textbf{2.96\textsubscript{± 0.34}} & 33.33\textsubscript{± 57.74} & 5.33\textsubscript{± 6.66} \\
& Adaptive-OPRO & \textbf{42.07\textsubscript{± 1.85}} & 0.418\textsubscript{± 0.016} & 3.15\textsubscript{± 0.02} & \textbf{100.00\textsubscript{± 0.00}} & 1.33\textsubscript{± 0.58} \\
\hline

\multirow{3}{*}{Qwen3-235B} & Baseline & \textbf{43.91\textsubscript{± 2.31}} & \textbf{0.42\textsubscript{± 0.00}} & 3.34\textsubscript{± 0.16} & 0.00\textsubscript{± 0.00} & 2.00\textsubscript{± 0.00} \\
& Reflection & 34.08\textsubscript{± 12.30} & 0.37\textsubscript{± 0.08} & \textbf{2.98\textsubscript{± 0.30}} & \textbf{23.81\textsubscript{± 41.24}} & 11.33\textsubscript{± 16.17} \\
& Adaptive-OPRO & 41.25\textsubscript{± 0.00} & \textbf{0.42\textsubscript{± 0.00}} & 3.16\textsubscript{± 0.00} & 0.00\textsubscript{± 0.00} & 2.00\textsubscript{± 0.00} \\
\hline

\multirow{3}{*}{Qwen3-32B} & Baseline & 35.75\textsubscript{± 5.35} & \textbf{0.48\textsubscript{± 0.06}} & \textbf{2.86\textsubscript{± 0.30}} & 60.86\textsubscript{± 52.71} & 22.33\textsubscript{± 3.06} \\
& Reflection & 41.72\textsubscript{± 1.32} & 0.43\textsubscript{± 0.01} & 3.03\textsubscript{± 0.22} & 66.67\textsubscript{± 57.74} & 10.67\textsubscript{± 5.13} \\
& Adaptive-OPRO & \textbf{48.37\textsubscript{± 0.10}} & 0.47\textsubscript{± 0.00} & 3.15\textsubscript{± 0.02} & \textbf{100.00\textsubscript{± 0.00}} & 18.00\textsubscript{± 5.00} \\
\hline

\multirow{3}{*}{\makecell{Claude Sonnet 4}} & Baseline & 13.43\textsubscript{± 8.62} & 0.180\textsubscript{± 0.121} & 5.52\textsubscript{± 3.96} & \textbf{60.83\textsubscript{± 12.30}} & 21.67\textsubscript{± 9.50} \\
& Reflection & 5.21\textsubscript{± 1.10} & 0.089\textsubscript{± 0.026} & 5.11\textsubscript{± 1.86} & 39.25\textsubscript{± 15.79} & 22.33\textsubscript{± 1.53} \\
& Adaptive-OPRO & \textbf{25.85\textsubscript{± 10.61}} & \textbf{0.290\textsubscript{± 0.087}} & \textbf{3.75\textsubscript{± 0.59}} & 43.81\textsubscript{± 38.37} & 19.00\textsubscript{± 12.17} \\
\hline
\multirow{3}{*}{\makecell{Claude Sonnet 4\\ w/ Thinking}} & Baseline & 12.52\textsubscript{± 2.47} & 0.175\textsubscript{± 0.030} & 5.03\textsubscript{± 1.53} & 53.30\textsubscript{± 14.47} & 17.00\textsubscript{± 2.65} \\
& Reflection & 11.12\textsubscript{± 4.86} & 0.186\textsubscript{± 0.083} & \textbf{3.42\textsubscript{± 2.23}} & \textbf{77.86\textsubscript{± 2.58}} & 17.00\textsubscript{± 5.00} \\
& Adaptive-OPRO & \textbf{16.36\textsubscript{± 7.87}} & \textbf{0.217\textsubscript{± 0.105}} & 5.18\textsubscript{± 2.52} & 68.89\textsubscript{± 30.06} & 12.67\textsubscript{± 4.04} \\
\hline
\multirow{3}{*}{GPT-o4-mini} & Baseline & 7.00\textsubscript{± 3.46} & 0.125\textsubscript{± 0.054} & 2.74\textsubscript{± 0.79} & 46.29\textsubscript{± 3.21} & 18.67\textsubscript{± 1.53} \\
& Reflection & 9.80\textsubscript{± 3.21} & 0.189\textsubscript{± 0.067} & \textbf{2.45\textsubscript{± 1.00}} & 54.54\textsubscript{± 7.92} & 26.33\textsubscript{± 9.61} \\
& Adaptive-OPRO & \textbf{10.47\textsubscript{± 3.84}} & \textbf{0.193\textsubscript{± 0.046}} & 3.42\textsubscript{± 0.90} & \textbf{62.70\textsubscript{± 11.25}} & 20.33\textsubscript{± 2.89} \\
\hline

\multirow{3}{*}{GPT-o3} & Baseline & 22.70\textsubscript{± 0.92} & 0.269\textsubscript{± 0.029} & 6.82\textsubscript{± 3.03} & 66.67\textsubscript{± 28.87} & 7.33\textsubscript{± 2.52} \\
& Reflection & 21.98\textsubscript{± 4.54} & 0.325\textsubscript{± 0.040} & 3.14\textsubscript{± 0.99} & 96.67\textsubscript{± 5.77} & 18.00\textsubscript{± 3.61} \\
& Adaptive-OPRO & \textbf{25.06\textsubscript{± 4.28}} & \textbf{0.392\textsubscript{± 0.019}} & \textbf{2.31\textsubscript{± 0.80}} & \textbf{100.00\textsubscript{± 0.00}} & 9.67\textsubscript{± 4.04} \\
\hline

\end{tabular}
\caption{Performance comparison between non-LLM-based and LLM-based approaches using ATLAS in bullish market conditions. \textbf{Bold} values indicate the best per model.}
\label{tab:nvda_results}
\end{table*}

\section{Results}
\label{sec:results}

Tables~\ref{tab:lly_results}, \ref{tab:xom_results}, and \ref{tab:nvda_results} present the results of our experimental design evaluating ATLAS across diverse market conditions.
The results show that \emph{Adaptive-OPRO} consistently improves upon fixed prompts across models and market conditions, while reflection often deteriorates performance or provides inconsistent value. Non-LLM strategies demonstrate regime-dependent performance, with different technical approaches succeeding in specific conditions but failing to generalize. ATLAS with \emph{Adaptive-OPRO} delivers stable performance across tested regimes, with certain model pairings achieving positive returns even in volatile and declining market conditions where most baseline strategies struggle. 
The order-level action space reveals distinct patterns across model families and supports attribution from analytical reasoning to execution behavior.

\subsection{Optimization in Sequential Decision-Making}

\emph{Adaptive-OPRO} consistently outperforms both static baseline prompts and reflection-based approaches across the tested models and market conditions. Figure~\ref{fig:opro_gain} makes this breadth explicit: the ROI gain over Baseline is positive across most backbones and regimes and spans all four model families, so the improvement is not confined to a particular architecture or market condition. Its magnitude scales with the slack left by the tuned baseline, which is largest under volatile conditions and narrowest in already-favorable regimes.

\textbf{Return, risk-adjusted, and win-rate metrics} jointly indicate successful adaptation to market feedback: models paired with \emph{Adaptive-OPRO} achieve higher returns while maintaining or reducing drawdowns, with SR gains showing that improvements arise from strategic enhancement and not increased risk-taking. Crucially, these return gains are accompanied by higher win rates, indicating consistent decision-making rather than sporadic large profits masking frequent losses. For instance, in the bearish regime (Table~\ref{tab:lly_results}), GPT-o3 and GPT-o4-mini shift from negative baseline returns to strong positive performance under \emph{Adaptive-OPRO}, while Qwen3-235B moves from losses to gains. This pattern persists across range-bound and bullish conditions, suggesting that prompt optimization captures regime-appropriate cues rather than overfitting to specific market settings.

\textbf{Comparisons to baseline performance.}  We examine how baseline decision quality relates to the gains from Adaptive-OPRO under volatile, declining markets. Baseline and Adaptive-OPRO ROI are moderately correlated ($r=0.64$), suggesting that \textit{stronger baselines} maintain \textit{higher absolute returns} after adaptation. However, the improvement over baseline shows no meaningful correlation ($r=0.05$) and an almost flat gradient ($\beta \approx 0.06$). This indicates that Adaptive-OPRO does not simply amplify existing strengths, but delivers \textit{improvements largely independent} of initial performance. Similar trends hold for risk-adjusted metrics, implying that Adaptive-OPRO mainly alters decision behavior rather than scaling baseline profitability.

\textbf{The reflection paradox.} In contrast, reflection-based prompting \cite{li-etal-2024-cryptotrade} exhibits a markedly different behavior. In the bearish-volatile regime, the  ROI improvement under reflection shows a strong negative correlation with baseline performance ($r=-0.78,\, p<0.05$), accompanied by a pronounced negative performance gradient ($\beta=-0.61$), indicating that models with \textit{stronger baseline} decision quality tend to \textit{deteriorate more} with reflection. This suggests that reflection does not merely fail to improve performance, but can actively \textit{disrupt effective decision policies} in high-noise environments. Rather than stabilizing behavior, reflection appears to \textbf{amplify stochasticity} and \textbf{override useful heuristics}, particularly for models that already exhibit competent baseline trading strategies, consistent with overthinking induced by redundant information (more details in App.~\ref{app:harmful_reflection}).

\begin{table*}[t]
\vskip -0.05in
\centering \small
\begin{tabular}{p{2cm}l|ccccc}
\hline
\textbf{Regime} & \textbf{Configuration} & \textbf{ROI (\%)$\uparrow$} & \textbf{SR $\uparrow$} & \textbf{DD (\%) $\downarrow$ } & \textbf{Win Rate (\%) $\uparrow$} & \textbf{Num.\ Trades} \\
\hline
\multirow{4}{*}{\makecell{\textbf{Bearish}\\\textbf{Regime}}} & No News & 4.07\textsubscript{± 0.72} & 0.056\textsubscript{± 0.016} & \textbf{7.84\textsubscript{± 3.15}} & 53.51\textsubscript{± 6.67} & 25.33\textsubscript{± 4.51} \\
 & No Market Data & -5.75\textsubscript{± 0.76} & -0.094\textsubscript{± 0.017} & 11.32\textsubscript{± 2.63} & 37.52\textsubscript{± 4.87} & 18.33\textsubscript{± 3.06} \\
& No News \& No Market & -6.86\textsubscript{± 1.68} & -0.078\textsubscript{± 0.036} & 14.54\textsubscript{± 3.30} & 43.94\textsubscript{± 6.94} & 22.33\textsubscript{± 1.15} \\
 & ATLAS & \textbf{9.06\textsubscript{± 0.73}} & \textbf{0.094\textsubscript{± 0.008}} & 11.48\textsubscript{± 0.00} & \textbf{65.28\textsubscript{± 16.84}} & 17.33\textsubscript{± 5.86} \\ \hline
\multirow{4}{*}{\makecell{\textbf{Sideways}\\\textbf{Regime}}}& No News & -8.20\textsubscript{± 1.64} & -0.264\textsubscript{± 0.069} & 9.09\textsubscript{± 2.99} & 22.82\textsubscript{± 13.65} & 35.00\textsubscript{± 12.29} \\
 & No Market Data & 0.01\textsubscript{± 0.92} & -0.011\textsubscript{± 0.021} & 6.56\textsubscript{± 1.58} & 46.55\textsubscript{± 23.15} & 13.33\textsubscript{± 3.06} \\
 & No News \& No Market & -4.60\textsubscript{± 0.70} & -0.136\textsubscript{± 0.026} & 7.01\textsubscript{± 2.29} & 35.26\textsubscript{± 13.09} & 21.00\textsubscript{± 4.58} \\
 & ATLAS & \textbf{3.88\textsubscript{± 2.21}} & \textbf{0.089\textsubscript{± 0.067}} & \textbf{3.28\textsubscript{± 0.95}} & \textbf{47.95\textsubscript{± 7.15}} & 25.33\textsubscript{± 5.03} \\
\hline
\multirow{4}{*}{\makecell{\textbf{Bullish}\\\textbf{Regime}}} & No News & 6.62\textsubscript{± 0.25} & 0.090\textsubscript{± 0.008} & 6.67\textsubscript{± 0.36} & 41.96\textsubscript{± 5.21} & 28.33\textsubscript{± 4.62} \\
& No Market Data & \textbf{11.78\textsubscript{± 1.76}} & \textbf{0.216\textsubscript{± 0.024}} & 3.70\textsubscript{± 0.86} & \textbf{70.24\textsubscript{± 14.03}} & 20.00\textsubscript{± 5.57} \\
  & No News \& No Market & 7.34\textsubscript{± 2.79} & 0.110\textsubscript{± 0.012} & 5.76\textsubscript{± 2.01} & 63.84\textsubscript{± 9.39} & 20.67\textsubscript{± 1.53} \\
 & ATLAS & 10.47\textsubscript{± 3.84} & 0.193\textsubscript{± 0.046} & \textbf{3.42\textsubscript{± 0.90}} & 62.70\textsubscript{± 11.25} & 20.33\textsubscript{± 2.89} \\
\hline
\end{tabular}
\caption{Ablation study results showing individual agent contributions using GPT-o4-mini across three market regimes. \textbf{Bold} values indicate the best results per configuration.}
\label{tab:gpto4_mini_ablation}
\end{table*}

\subsection{Trading Behavior Across LLMs}

The order-level action space reveals systematic behavioral differences across model families, with performance broadly correlating with general model capabilities. Beyond averages, variance across runs captures decision reliability, especially when timing and sizing errors are amplified.

\textbf{GPT models} exhibit distinct trading styles and adaptation patterns. \textbf{GPT-o3} integrates inputs from agents into coherent decisions, showing conservative risk management that can cap gains in strongly trending markets but delivers consistent performance across regimes. This manifests as robust returns with comparatively low drawdowns and low run-to-run variance, denoting stable execution. 

\textbf{GPT-o4-mini} emphasizes short-term risk control through frequent stop-losses and early profit-taking. This behavior aligns with stronger outcomes in volatile settings and more muted trend capture in sustained moves; it also tends toward higher trading frequency in some regimes. Still, Adaptive-OPRO generally improves its consistency and profitability relative to fixed prompting, with moderate variance suggesting a more reactive but still controlled policy.

\textbf{Qwen models} show divergent behavior based on scale. \textbf{Qwen3-235B} trades more selectively and, across several regimes, achieves stable positive outcomes under Adaptive-OPRO. All three tables reflect that prompt adaptation is important here: it often turns otherwise marginal/negative behavior into positive returns while keeping activity relatively restrained, consistent with risk-reward balancing. 

\textbf{Qwen3-32B} is more active and variable, with larger swings across runs and regimes. Adaptive-OPRO improves its behavior, reducing losses in adverse settings and strengthening performance in favorable ones, but residual variance suggests less stable execution than the larger variant.

\textbf{Llama 3.3-70B} adopts simpler trading strategies with limited risk-management sophistication. Qualitatively, it shows delayed responses to market shifts and occasional abrupt changes in stance, which correspond to weaker performance in more adversarial regimes. Interestingly, this straightforward behavior performs well in the bullish regime in our results, consistent with capturing upward drift without overcomplicating execution.

\textbf{Claude Sonnet 4} varies depending on reasoning mode, with variance patterns revealing meaningful differences in reliability. Certain configurations exhibit markedly higher run-to-run variability, indicating less predictable decision-making. With extended thinking enabled, the model often produces detailed analysis but the results show mixed execution quality; without thinking, decisions become more erratic and consistency across regimes degrades, suggesting that the bottleneck is both analysis depth and subsequent order construction.

Overall, the key insight enabled by order-level specifications is that weaker configurations often generate plausible market analysis but fail in position sizing, timing, or order selection, whereas successful configurations consistently translate analysis into coherent execution. 

\subsection{LLM Optimization Capabilities}
\label{subsec:opt_capabilities}

A key advantage of \emph{Adaptive-OPRO} is that optimization yields \textit{interpretable} instruction updates, assessed along two axes: (i) whether the revised prompt is objectively aligned with the trading goal (e.g., explicit risk controls, sizing discipline, and when to trade), and (ii) whether those instructions are reflected in subsequent order-level behavior (frequency, timing, and position sizing). After manual inspection of the results, we observe clear family-level patterns. \textbf{GPT models} consistently produce well-structured, objective-aligned refinements that translate observed weaknesses into actionable constraints, and these updates tend to be followed in execution, consistent with their lower run-to-run variance. \textbf{Qwen models} also generate targeted improvements, with the larger Qwen3-235B producing more coherent and internally consistent instruction revisions, which aligns with its more stable selective trading behavior. In contrast, \textbf{Llama} often reports edits that are not present in the actual prompt or proposes changes that conflict with the stated objective, weakening the connection between optimization output and downstream execution. \textbf{Claude models} often shift toward increasingly procedural and restrictive prompts, which can reduce adaptability; notably, this prescriptiveness does not reliably translate into stable execution, as reflected by higher variance in several configurations.
Examples of observed patterns in App.~\ref{app:llm_optimization}.

\subsection{ATLAS Ablation Study}

Table~\ref{tab:gpto4_mini_ablation} shows distinct agent contributions through performance drops when each is ablated.

\textbf{Market Analyst} is a core component across market regimes. Its removal consistently results in the most significant performance degradation, especially in challenging conditions such as the bearish-volatile regime, where technical context is crucial for decision-making. In the sideways regime, the absence of market analysis not only reduces returns but also lowers trading frequency, suggesting that agents lose confidence to act without a solid technical foundation. Notably, in the bullish regime, ROI slightly improves when market data is excluded, suggesting that in up-trending conditions social consensus may offer cleaner entry signals.

\textbf{News analyst} contributes regime-specific strategic value. In the bullish regime, news removal leads to lower returns as agents become more conservative.
The sideways regime shows news analysis as critical, with its removal producing severe degradation, suggesting that sentiment analysis is essential when technical signals are ambiguous.

\textbf{Combination of News \& Market Analyst} highlights the complementary value of these signals. Across all regimes, removing both agents substantially degrades performance, showing that news and market data provide non-redundant information. 
In the bearish regime, the drop reflects the importance of sentiment and technical context under volatility, while in the sideways regime their absence produces unstable, unprofitable behavior. 
Even in the bullish regime, combined removal harms performance, indicating that each component contributes differently across regimes and that their joint effect is not simply additive.

\section{Conclusion}

In this work, we introduce ATLAS, an LLM-based trading framework that combines \emph{Adaptive-OPRO} for prompt optimization under delayed, noisy feedback with structured analyst inputs and an order-level interface. Across regimes and model families, Adaptive-OPRO outperforms tuned static prompts, while standard reflection proves inconsistent. The order-level interface reveals model-specific trading behaviors and separates analytical quality from execution choices, enabling clearer attribution and interpretability. ATLAS with Adaptive-OPRO provides a practical, reliable, auditable paradigm for sequential LLM decision-making.

\section*{Acknowledgments}
Cloud computing credits for accessing Llama, Qwen, and Claude models via Amazon Bedrock were provided by Amazon Web Services (AWS). This work was supported by the Hellenic Foundation for Research and Innovation (HFRI) under the 5th Call for HFRI PhD Fellowships (Fellowship Number 19268).

\section*{Limitations}
Following prior LLM-agent and market-simulation work, we focus on three liquid equities over two-month, regime-specific windows with daily decisions to reduce confounding from asset heterogeneity and shifting market structure. This isolates adaptation effects under a shared interface but does not support generalization across assets, sectors, horizons, or macro conditions. Results should be read as behavioral evidence about \emph{Adaptive-OPRO}, not as market-wide performance claims.

End-of-day decisions provide stable feedback for optimization under
delayed, noisy outcomes, but prevent agents from reacting to
intraday moves or capturing timing-dependent behaviors. The
frictionless execution assumption discussed in Section~\ref{sec:exp}
also means that strategies sensitive to fill quality, such as
high-turnover or large-size policies, would need revalidation under
a microstructure-aware simulator before any live-trading claim.

Each configuration runs three times due to resource constraints, capturing stochastic variance but limiting statistical power. Comparisons isolate prompt adaptation under a shared order-level interface rather than varying full system architectures. While order-level actions improve interpretability by separating analysis from execution, we do not include a directional-only ablation for direct causal comparison. Finally, although we cover multiple model families (GPT, Claude, Llama, Qwen), behaviors may vary with architectures, scales, and training procedures beyond those studied here.

\section*{Ethical Considerations}
This work focuses on controlled, simulated trading experiments to study prompt optimization and does not involve real-world financial transactions or human subjects. All analyses are conducted in a reproducible, transparent environment, minimizing potential risks. While findings provide insights into model behavior, they are not financial advice and should not be used for live trading.

\bibliography{acl_latex}
\appendix
\section{Financial Markets and Trading Foundations}
\label{app:foundations}

This appendix summarizes the trading concepts needed to interpret an \emph{order-aware} interface and the signals used by the \textsc{Market Analyst}. The focus is on how ATLAS expresses decisions as executable orders in \textit{StockSim} rather than on venue-specific microstructure.

\subsection{Orders and Positions}
\label{app:orders_positions}

ATLAS expresses actions at the order level and supports both long and short positioning.

\paragraph{Order types.}
\textbf{Market orders} seek immediate execution at the best available prices and prioritize certainty of fill over price control. \textbf{Limit orders} specify a worst acceptable price for buys or a best acceptable price for sells and prioritize price control over certainty of execution. \textbf{Stop orders} activate once a trigger is reached and are commonly used for risk control or momentum entry.

\paragraph{Long and short.}
A \textbf{buy to open} creates or increases a long position. A \textbf{sell short} creates a short position that profits if price declines. Exits are expressed symmetrically as \textbf{sell to close} for long positions and \textbf{buy to cover} for short positions. The Central Trading Agent may attach stops or limits to manage risk and profit-taking for either side.

\paragraph{Decision cadence.}
The Central Trading Agent makes decisions on a daily schedule. At each decision point it consumes the updated analyst summaries and current portfolio state, then may submit new or modifying orders that are evaluated by \textit{StockSim} under standard semantics. At initialization, the portfolio holds \$100{,}000 in cash and no positions. Since our headline metrics are percentage based (e.g., ROI, Sharpe, and drawdown computed from returns), the absolute starting capital does not affect reported performance and only scales dollar P\&L.

\subsection{Regime Taxonomy}
\label{app:regimes}

We organize evaluation windows by broad market regimes in order to study behavior under distinct conditions.

\textbf{Bearish volatile} denotes periods with sustained downward drift and elevated variability. \textbf{Sideways} denotes range-bound behavior with mixed signals and limited trend persistence. \textbf{Bullish} denotes periods with sustained upward drift and comparatively orderly pullbacks. In the main experiments we instantiate one window for each regime and keep the decision cadence and interface fixed. The taxonomy is agnostic to any single indicator choice and can be operationalized by simple trend and volatility summaries when needed.

\section{Technical Indicators Used in Market Analysis}
\label{sec:technical_indicators}

This appendix provides detailed explanations of the technical indicators employed by the Market Analyst agent in ATLAS, covering their mathematical formulations, implementation specifics, and interpretive significance in financial market analysis. All technical indicators described in this section are calculated by the StockSim \cite{papadakis2025stocksim} simulation environment and integrated into our analysis framework to provide comprehensive market insights.

\paragraph{Data source.}
The Market Analyst consumes OHLCV, volume, and session VWAP series from Massive\footnote{\url{https://massive.com}} for the specified instrument and evaluation window. Bars are retrieved at daily resolution and aligned to official U.S. market sessions, with corporate actions (splits and dividends) from Massive used to adjust prices consistently with \textit{StockSim}. All technical indicators described in this appendix are computed inside \textit{StockSim} from these Massive-derived bars. Days with incomplete or missing bars are excluded rather than backfilled, and no survivorship or lookahead adjustments are applied beyond standard split and dividend handling.

\subsection{Simple Moving Average (SMA) and Exponential Moving Average (EMA)}

\textbf{Simple Moving Average (SMA)}: The SMA is calculated as the arithmetic mean of closing prices over a specified number of periods \cite{murphy1999technical}:
\begin{equation}
SMA_n = \frac{1}{n} \sum_{i=0}^{n-1} P_{t-i}
\end{equation}
where $P_t$ represents the closing price at time $t$ and $n$ is the number of periods. For our analysis, we employ SMA periods of 20, 50, 100, and 200 days to capture short-term, medium-term, and long-term trend characteristics. SMA provides equal weight to all prices in the calculation period, which makes it suitable for identifying longer-term trends but less responsive to recent price changes \cite{murphy1999technical}.

\textbf{Exponential Moving Average (EMA)}: The EMA assigns exponentially decreasing weights to older prices, which makes it more responsive to recent price movement \cite{murphy1999technical}:
\begin{equation}
EMA_t = \alpha \cdot P_t + (1-\alpha) \cdot EMA_{t-1}
\end{equation}
where $\alpha = \frac{2}{n+1}$ is the smoothing factor and $n$ is the number of periods. In our implementation, we utilize 12-period and 26-period EMAs, which serve as the foundation for MACD calculation and provide complementary trend analysis to our SMA suite. Research indicates that EMA often outperforms SMA in volatile conditions due to its enhanced sensitivity to recent price changes \cite{kaufman2013trading}.

\subsection{Relative Strength Index (RSI)}

The RSI is a momentum oscillator that measures the speed and magnitude of price changes, oscillating between 0 and 100 \cite{wilder1978new}:
\begin{equation}
RSI = 100 - \frac{100}{1 + RS}
\end{equation}
where $RS = \frac{Average\ Gain}{Average\ Loss}$ over a specified period. Our analysis uses the standard 14-day period as originally recommended by \citet{wilder1978new}. The average gain and loss are calculated using exponential smoothing as originally formulated:
\begin{equation}
\overline{G}_t = \frac{13\overline{G}_{t-1} + G_t}{14}
\end{equation}
\begin{equation}
\overline{L}_t = \frac{13\overline{L}_{t-1} + L_t}{14}
\end{equation}
where $\overline{G}_t$ represents the average gain at time $t$, $\overline{L}_t$ represents the average loss at time $t$, $G_t$ is the current gain, and $L_t$ is the current loss. RSI values above 70 typically indicate overbought conditions, while values below 30 suggest oversold conditions \cite{wilder1978new}. These thresholds can be adapted to asset volatility and regime \cite{murphy1999technical}.

\subsection{Moving Average Convergence Divergence (MACD)}

MACD is a trend-following momentum indicator that shows the relationship between two moving averages of a security's price \cite{murphy1999technical}:
\begin{equation}
    \textit{MACD} = EMA_{12} - EMA_{26}
\end{equation}
\begin{equation}
    \textit{Signal Line} = EMA_9(\textit{MACD})
\end{equation}
\begin{equation}
   \textit{Histogram} = \textit{MACD} - \textit{Signal Line}
\end{equation}
We employ the standard configuration. Crossovers and divergences are commonly used to identify trend changes and momentum shifts \cite{achelis2000technical}.

\subsection{Average True Range (ATR)}

ATR measures market volatility by calculating the average of true ranges over a specified number of periods, as developed by \citet{wilder1978new}:
\begin{equation}
\begin{split}
    \textit{True Range} = \max[(High - Low), \\|High - Close_{prev}|, |Low - Close_{prev}|]
\end{split}
\end{equation}
\begin{equation}
   ATR_n = \frac{1}{n} \sum_{i=0}^{n-1} TR_{t-i}
\end{equation}
We use the standard 14-period ATR. ATR supports volatility-aware sizing and stop placement.

\subsection{Bollinger Bands}

Bollinger Bands consist of three lines: a middle band and two outer bands positioned at standard deviations above and below the middle band \cite{achelis2000technical}:
\begin{equation}
   \text{Middle Band} = SMA_{20}
\end{equation}
\begin{equation}
    \text{Upper Band} = SMA_{20} + (k \times \sigma)
\end{equation}
\begin{equation}
    \text{Lower Band} = SMA_{20} - (k \times \sigma)
\end{equation}
where $k$ is typically 2 and $\sigma$ is the rolling standard deviation of close. The bands adapt to changing volatility and help contextualize extremes \cite{murphy1999technical}.

\subsection{Support and Resistance Levels}

Support and resistance levels are price zones where the asset has historically shown difficulty moving below (support) or above (resistance) \cite{murphy1999technical}. We focus on \textbf{horizontal} levels identified by repeated interactions and elevated volume. Their strength increases with the number of tests, traded volume, and time span.

\subsection{Volume Profile}

Volume Profile displays trading activity over price for a chosen window:
\begin{itemize}
\item \textbf{Point of Control (POC)}: price with the highest traded volume
\item \textbf{Value Area}: price range that contains a specified share of volume, typically 70\%
\item \textbf{High Volume Nodes}: locally elevated volume levels
\end{itemize}
Volume-based context helps identify zones where participation has been concentrated, which often align with support or resistance.

\section{Analyst Details}
\subsection{News Analyst}
\label{app:news}

The \textit{News Analyst} distills market-relevant information from financial news streams for a given ticker. Inputs are retrieved from the Massive API\footnote{\url{https://massive.com}} as batches of timestamped items containing title, URL, summary, and keywords. The component produces a structured analysis along four dimensions that are stable across models and assets: \textit{Sentiment Assessment}, \textit{Key Developments}, \textit{Market Relevance}, and \textit{Source Analysis}. When headline-only context is insufficient, the analyst can fetch the full article text through an internal fetcher to improve coverage and reduce headline bias. The output is designed to be compact, auditible, and directly consumable by the Central Trading Agent; it does not generate trading signals.

\paragraph{Example input batch (NVDA).}
\begin{quote}
\begin{scriptsize}
\textbf{\#\#NEWS BATCH}\vspace{0.25em}

\textbf{[2025-04-28T12:45:00+00:00]} Want to Avoid the ``Magnificent Seven'' and Generate Passive Income? This Vanguard ETF May Be for You --- The Motley Fool\\
\textbf{URL}: \url{https://www.fool.com/investing/2025/04/28/magnificent-seven-passive-income-vanguard-etf/?source=iedfolrf0000001}\\
\textbf{Summary}: The article discusses the Vanguard High Dividend Yield ETF (VYM) as an alternative to large-cap tech, highlighting sector diversification and dividend income.\\
\textbf{Keywords}: Vanguard High Dividend Yield ETF, Magnificent Seven, passive income, value stocks, dividend stocks
\vspace{0.5em}

\textbf{[2025-04-28T07:15:00+00:00]} This Could Be a Make-It-or-Break-It Moment for Nvidia. Here's What to Watch For This Week. --- The Motley Fool\\
\textbf{URL}: \url{https://www.fool.com/investing/2025/04/28/a-make-it-or-break-it-moment-for-nvidia/?source=iedfolrf0000001}\\
\textbf{Summary}: Nvidia's near-term outlook hinges on AI demand signals from large customers reporting earnings this week.\\
\textbf{Keywords}: Nvidia, AI, Microsoft, Amazon, Meta Platforms, Alphabet
\vspace{0.5em}

\textbf{[2025-04-27T22:25:00+00:00]} Prediction: Broadcom Stock Could Surge by 100\% in the Next 3 Years --- The Motley Fool\\
\textbf{URL}: \url{https://www.fool.com/investing/2025/04/27/prediction-broadcom-could-surge-by-100-in-the-next/?source=iedfolrf0000001}\\
\textbf{Summary}: Broadcom's networking and ASIC businesses are positioned to benefit from AI infrastructure demand.\\
\textbf{Keywords}: Broadcom, AI chips, networking, ASIC, VMware
\end{scriptsize}
\end{quote}

\paragraph{Structured output (NVDA).}
\begin{quote}
\begin{scriptsize}
\textbf{Sentiment Assessment}\quad
Tone is cautiously optimistic and contingent on confirmation of continued AI spend by key customers. There is also a modest rotation narrative toward dividend payers that could weigh on high-beta tech.

\textbf{Key Developments}\quad
Earnings this week from Microsoft, Amazon, and Meta will signal near-term AI budgets. Alphabet's reaffirmed capital spending supports sustained demand for next-gen chips. Policy risks around import tariffs remain a headwind.

\textbf{Market Relevance}\quad
Expect higher volatility around customer earnings. Rotation into dividend ETFs could pressure large-cap AI beneficiaries if macro uncertainty rises. Broadcom's bullish outlook may drive sector flow dispersion that competes with Nvidia exposure.

\textbf{Source Analysis}\quad
All items originate from The Motley Fool, a retail-focused outlet with a constructive bias. Claims should be cross-checked against primary earnings releases and sell-side notes for actionable conviction.
\end{scriptsize}
\end{quote}

\paragraph{Additional example (XOM).}
\begin{quote}
\begin{scriptsize}
\textbf{Sentiment Assessment}\quad
Mixed. ExxonMobil appears on a list of top buys for diversification strength, offset by policy uncertainty related to funding cuts for carbon capture projects.

\textbf{Key Developments}\quad
Federal funding for a \$332M CCS project at Baytown is being withdrawn, which may delay low-carbon hydrogen and ammonia plans, although core growth strategy remains intact.

\textbf{Market Relevance}\quad
Near-term noise in decarbonization headlines with limited change to base cash-flow trajectory. Integrated model and commercial partnerships support resilience.

\textbf{Source Analysis}\quad
Coverage from The Motley Fool blends stock-picking commentary with policy reporting and lacks direct primary citations. Verification from official releases is recommended when trading on policy moves.
\end{scriptsize}
\end{quote}

\paragraph{Operational notes.}
The News Analyst refreshes daily in sync with the decision cadence, deduplicates near-identical headlines, and preserves a consistent schema across assets and regimes. Its role is to surface catalysts, stance shifts, and source reliability in a compact form that supports downstream reasoning by the Central Trading Agent.

\subsection{Fundamental Analyst}
\label{sec:fundamental}

The \textit{Fundamental Analyst} extracts trading-relevant structure from periodic corporate disclosures (earnings releases, financial statements) and corporate actions (dividends, splits). It runs at low frequency to mirror real reporting cadence, typically activating once or twice per evaluation window. Inputs are retrieved via Massive\footnote{\url{https://massive.com}} and normalized to a compact schema consumed by the Central Trading Agent. The module does not emit buy/sell signals; it summarizes material changes and likely catalysts.

\subsubsection{Financial Statement Components and Terminology}

\paragraph{Revenue and income metrics.}
\begin{itemize}
\item \textbf{Revenue} (net sales) is top-line activity prior to costs \cite{penman2012financial}.
\item \textbf{Gross profit margin}:
\begin{equation}
\text{GPM} = \frac{\text{Revenue} - \text{COGS}}{\text{Revenue}} \times 100\%,
\end{equation}
capturing production efficiency and pricing power \cite{palepu2019business}.
\item \textbf{Operating margin}:
\begin{equation}
\text{OpM} = \frac{\text{Operating Income}}{\text{Revenue}} \times 100\%,
\end{equation}
reflecting core cost discipline \cite{penman2012financial}.
\item \textbf{Net income} is profit after all expenses, taxes, and interest.
\item \textbf{Earnings per share (EPS)}:
\begin{equation}
\text{EPS} = \frac{\text{Net Income}}{\text{Weighted Avg.\ Shares}},
\end{equation}
a per-share profitability anchor for valuation \cite{damodaran2012investment}.
\end{itemize}

\paragraph{Cash-flow dynamics.}
\begin{itemize}
\item \textbf{Operating cash flow (OCF)} approximates cash generated by operations:
\begin{equation}
\text{OCF} = \text{NI} + \text{NCE} \pm \text{WCC},
\end{equation}
where NI is net income, NCE non-cash expenses, WCC working-capital change \cite{penman2012financial}.
\item \textbf{Net cash flow} aggregates operating, investing, and financing cash flows:
\begin{equation}
\text{NCF} = \text{OCF} + \text{ICF} + \text{FCF}.
\end{equation}
\item \textbf{Capital allocation} covers capex, buybacks, dividends, and debt paydown, each with distinct market implications.
\end{itemize}

\paragraph{Balance-sheet metrics.}
\begin{itemize}
\item \textbf{Total assets} and \textbf{total equity} summarize scale and residual value \cite{palepu2019business}.
\item \textbf{Debt-to-equity} gauges leverage and risk:
\begin{equation}
\text{D/E} = \frac{\text{Total Debt}}{\text{Total Equity}}.
\end{equation}
Higher values imply greater financial risk \cite{damodaran2012investment}.
\end{itemize}

\subsubsection{Corporate Actions and Structural Events}

\paragraph{Stock splits.}
Splits increase share count while proportionally reducing price (e.g., 1:2, 1:4, 1:10), often to improve perceived affordability and liquidity \cite{e70c9308-250a-3fbb-8468-68c961bde0da}.

\paragraph{Dividends.}
\begin{itemize}
\item \textbf{Cash dividends} return capital to shareholders; policy signals management’s view on reinvestment vs.\ distribution \cite{brealey2019principles}.
\item \textbf{Dividend yield}:
\begin{equation}
\text{Yield} = \frac{\text{Annual Dividends Per Share}}{\text{Current Price}} \times 100\%.
\end{equation}
\end{itemize}

\subsubsection{Analytical Dimensions}
The analyst produces a concise, four-part summary focused on trading relevance:
\emph{Profit \& Margin Trends}, \emph{Cash Flow \& Capital Allocation}, \emph{Balance Sheet \& Leverage / Earnings Quality flags}, and \emph{Catalyst Watch}. Outputs are kept compact and directly auditable.

\paragraph{Example input batch (NVDA).}
\begin{quote}
\scriptsize
\textbf{Stock Splits:}\\
2024-06-10: 1:10 \quad
2021-07-20: 1:4 \quad
2007-09-11: 2:3 \quad
2006-04-07: 1:2 \\[0.25em]
\textbf{Dividends:}\\
2025-03-12: \$0.010 \quad
2024-12-05: \$0.010 \quad
2024-09-12: \$0.010 \quad
2024-06-11: \$0.010 \quad
2024-03-05: \$0.040 \\[0.25em]
\textbf{Annual FY2025 (Filed: 2025-02-26):}\\
Revenue \$130.5B; GPM 75.0\%; OpM 62.4\%; Net income \$72.9B; EPS \$2.94;\\
OCF \$64.1B; NCF \$1.3B; Assets \$111.6B; Equity \$79.3B; D/E 0.11.\\[0.25em]
\textbf{Quarterly Q1 2025 (Filed: 2024-05-29):}\\
Revenue \$26.0B; GPM 78.4\%; OpM 64.9\%; Net income \$14.9B; EPS \$5.98; NCF \$0.3B.\\[0.25em]
\textbf{Quarterly Q2 2025 (Filed: 2024-08-28):}\\
Revenue \$30.0B; GPM 75.1\%; OpM 62.1\%; Net income \$16.6B; EPS \$0.67; NCF \$1.0B.
\end{quote}

\paragraph{Structured output (NVDA).}
\begin{quote}
\begin{scriptsize}
\textbf{Profit \& Margin Trends}\quad
Q1→Q2 revenue grew \(\sim\)15\% to \$30B on sustained AI demand; gross margin held near 75\% while operating margin eased from 64.9\% to 62.1\%, consistent with mix normalization.

\textbf{Cash Flow \& Capital Allocation}\quad
FY25 OCF \$64B (\(\sim\)49\% of sales) supports heavy capex and buybacks; net cash still positive. The cut in quarterly dividend from \$0.04 to \$0.01 signals prioritization of reinvestment.

\textbf{Balance Sheet \& Earnings Quality}\quad
Low leverage and strong equity base support flexibility. The sharp EPS swing (Q1 \$5.98 vs.\ Q2 \$0.67) warrants a GAAP vs.\ non-GAAP review to isolate one-offs.

\textbf{Catalyst Watch}\quad
Upcoming guidance on AI trajectory, capex cadence, and inventory dynamics are potential volatility catalysts relative to consensus.
\end{scriptsize}
\end{quote}

\paragraph{Additional example (XOM).}
\begin{quote}
\begin{scriptsize}
\textbf{Profit \& Margin Trends}\quad
FY2024 net margin near 10\% with operating margin \(\sim\)14–15\%; quarterly prints show stability.

\textbf{Cash Flow \& Capital Allocation}\quad
Strong free cash flow capacity; negative annual net cash reflects investing and distribution outflows (capex, buybacks, dividends) rather than operating stress.

\textbf{Balance Sheet \& Leverage}\quad
Debt-free posture and current ratio \(>\)1.3 provide high financial flexibility; equity base expanded through FY/Q3.

\textbf{Catalyst Watch}\quad
Capital-return actions (buyback/dividend changes) and updates on large projects are the near-term fundamental triggers.
\end{scriptsize}
\end{quote}

\section{Experiments}
\label{sec:exper_set_more}

\subsection{Experimental Setup}
\label{app:experimental-setup}
Market regimes in our evaluation are instantiated using highly liquid, publicly traded equities selected prior to experimentation based on transparent criteria. Specifically, assets are required to exhibit stable liquidity conditions, clearly identifiable regime-consistent price dynamics over the evaluation window, and minimal microstructure distortions. This ensures that observed agent behavior reflects regime characteristics rather than artifacts of illiquidity or asset-specific noise. Asset instantiations are chosen independently of model performance and without outcome-driven adjustment, with selection criteria emphasizing representativeness of regime dynamics and sectoral diversity to reduce the likelihood that results are driven by idiosyncratic company- or industry-level effects.

Concretely, the bearish-volatile regime is instantiated using Eli Lilly and Company (LLY), the sideways regime using Exxon Mobil Corporation (XOM), and the bullish regime using NVIDIA Corporation (NVDA). All assets are evaluated over the same fixed two-month window (Apr 28–Jun 28, 2025) with a daily decision interval, ensuring consistency in sequential decision-making across regimes.

Importantly, ATLAS is asset- and regime-agnostic by design: no asset-specific features or regime-dependent assumptions are encoded in the framework, and the same experimental protocol can be directly applied to alternative equities, broader asset sets, or different evaluation horizons without modification.

\subsection{Evaluation Scope}
We evaluate ATLAS over a two-month window (28 Apr–28 Jun 2025) across three sector-diverse equities. This horizon provides multiple decision cycles per asset while keeping full conversation histories within context limits and avoiding regime mixing. The period naturally includes routine corporate events and news, yielding a representative test bed.

\subsection{Asset Selection Strategy}
We use three equities chosen ex ante by simple, transparent criteria (liquidity, sector diversity, characteristic behavior):
\textbf{NVDA} (technology, trending), \textbf{LLY} (healthcare, volatile drawdowns), \textbf{XOM} (energy, range-bound).
This mix stresses different information channels and trading behaviors (trend capture, volatility management, and patience) without relying on outcome-driven selection.

\subsection{Framework Configurations}
Beyond the main-paper comparisons, we implemented additional variants to probe design choices:
\begin{itemize}\setlength\itemsep{2pt}
\item \textbf{Baseline}: Multi-agent with carefully engineered static prompts.
\item \textbf{Adaptive-OPRO}: Prompt optimization applied only to the Central Trading Agent.
\item \textbf{Reflection}: A reviewer agent that produces periodic feedback on recent decisions. We tested weekly reflections (as in prior work) and a shorter 1-day variant; the latter is exploratory and omitted from the main tables.
\item \textbf{Adaptive-OPRO + Reflection}: Combined for interaction analysis; included here for completeness.
\end{itemize}
All runs keep analyst prompts fixed to isolate the adaptation mechanism at the decision layer.

\subsection{Model Selection}
We study how backbone capabilities translate to sequential decisions under identical interfaces:
\begin{itemize}\setlength\itemsep{2pt}
\item \textbf{Reasoning-enabled}: GPT-o3, GPT-o4-mini, Claude Sonnet~4 (thinking).
\item \textbf{Matched base model}: Claude Sonnet~4 (no thinking) to isolate the effect of explicit reasoning.
\item \textbf{Open-source}: Llama~3.3-70B, Qwen3-235B, Qwen3-32B to gauge transfer across families and deployment options.
\end{itemize}
Within a run, the same backbone powers all ATLAS components to avoid cross-model confounds.

\subsection{Ablation Study Choices}
To quantify information value within ATLAS, we run ablations exclusively under \textbf{GPT-o4-mini + Adaptive-OPRO}:
\begin{enumerate}\setlength\itemsep{2pt}
\item \textbf{No Market Analyst}: removes multi-timescale technical structure and indicators.
\item \textbf{No News Analyst}: removes unstructured text processing of headlines and stories.
\item \textbf{No Market \& No News}: leaves only portfolio state and fundamentals.
\end{enumerate}
We do not ablate the \textit{Fundamental Analyst} due to its intentionally low activation frequency within these windows; its role is assessed qualitatively around reporting events. Each ablation is run three times.

\subsection{Evaluation Methodology}
We use a \textbf{multi-run protocol} of three independent runs per configuration and report mean \(\pm\) standard deviation. Metrics mirror the main paper (returns, risk-adjusted returns, drawdowns, win rate on closed trades, and activity). In addition to aggregate metrics, we examine decision patterns and adaptation trajectories to explain \emph{why} configurations differ.

\subsection{Non-LLM Based Strategies}
\label{baseline}

We compare against established trading strategies (Buy \& Hold, moving average crossovers, MACD) that require no machine learning. These baselines contextualize LLM performance-showing where adds value versus simpler alternatives. A detailed description of these methods is presented below.

\paragraph{Buy and Hold}
The Buy and Hold strategy is a passive investment approach in which an asset is acquired at the beginning of the investment horizon and retained without any further trading actions, regardless of interim price fluctuations. This method assumes that, over time, the market tends to grow, and thus long-term holding can yield positive returns. It does not rely on any predictive model or technical indicator. In our evaluation, Buy and Hold serves as a benchmark strategy against which the performance of all other trading methods is compared.

\paragraph{Simple Moving Average (SMA)} 
The SMA strategy \cite{Gencay1996} issues trading signals based on the relationship between the current price of an asset and its moving average over a fixed time window. Specifically, a buy (sell) signal is triggered when the price crosses above (below) the SMA. We test various window lengths selecting the optimal period based on validation performance.

\paragraph{Short-Long Moving Average (SLMA)} 
The SLMA method \cite{Wang2018} extends the SMA approach by employing two SMAs of different lengths: one short-term and one long-term. A buy signal is generated when the short-term average crosses above the long-term average, while a sell signal occurs at the inverse crossover.

\paragraph{Moving Average Convergence Divergence (MACD)} 
The MACD strategy \cite{Wang2018} captures momentum shifts by computing the difference between the 12-day and 26-day exponential moving averages. A 9-day EMA of the MACD line is used as a signal line. Trading signals are generated when the MACD line crosses the signal line from below (buy) or from above (sell). The exponential formulation ensures increased sensitivity to recent price movements.

\paragraph{Bollinger Bands} 
The Bollinger Bands strategy \cite{Day2023} incorporates volatility by constructing a band around a 20-day SMA, with the upper and lower bands placed two standard deviations above and below the mean, respectively. A price crossing above the upper band may indicate overbought conditions (sell signal), while crossing below the lower band may suggest oversold conditions (buy signal). We adopt the standard parameterization of 20-day SMA and multiplier 2, as commonly suggested in the literature.

\begin{table*}[t]
\centering \small
\begin{tabular}{llcccc}
\hline
\textbf{Model} & \textbf{Prompting} & \textbf{Ann. SR $\uparrow$} & \textbf{Sortino $\uparrow$} & \textbf{ROIC (\%) $\uparrow$} & \textbf{P/T (\$) $\uparrow$} \\
\hline

\multicolumn{6}{c}{\textbf{LLM-Based Strategies - ATLAS}} \\
\hline
\multirow{3}{*}{Llama 3.3-70B}
 & Baseline & 6.16\textsubscript{± 1.52} & 0.97\textsubscript{± 0.22} & 30.98\textsubscript{± 26.06} & 456.27\textsubscript{± 790.29} \\
 & Reflection & \textbf{6.70\textsubscript{± 0.37}} & 1.03\textsubscript{± 0.02} & 29.14\textsubscript{± 21.06} & \textbf{1511.32\textsubscript{± 2617.69}} \\
 & Adaptive-OPRO & 6.63\textsubscript{± 0.25} & \textbf{1.05\textsubscript{± 0.01}} & \textbf{42.26\textsubscript{± 1.68}} & 0.00 \\
\hline

\multirow{3}{*}{Qwen3-235B}
 & Baseline & 6.61\textsubscript{± 0.02} & \textbf{0.67\textsubscript{± 0.00}} & 40.90\textsubscript{± 0.35} & 0.00\textsubscript{± 0.00} \\
 & Reflection & 5.94\textsubscript{± 1.20} & 0.58\textsubscript{± 0.14} & 27.90\textsubscript{± 23.15} & \textbf{491.18\textsubscript{± 850.75}} \\
 & Adaptive-OPRO & \textbf{6.63\textsubscript{± 0.00}} & \textbf{0.67\textsubscript{± 0.00}} & \textbf{41.26\textsubscript{± 0.00}} & 0.00\textsubscript{± 0.00} \\
\hline

\multirow{3}{*}{Qwen3-32B}
 & Baseline & \textbf{7.57\textsubscript{± 0.96}} & 0.63\textsubscript{± 0.07} & 16.37\textsubscript{± 21.12} & 1567.18\textsubscript{± 1369.31} \\
 & Reflection & 6.85\textsubscript{± 0.18} & 0.67\textsubscript{± 0.00} & 26.67\textsubscript{± 17.99} & \textbf{3266.26\textsubscript{± 5812.42}} \\
 & Adaptive-OPRO & 7.41\textsubscript{± 0.05} & \textbf{0.72\textsubscript{± 0.01}} & \textbf{43.27\textsubscript{± 4.61}} & 248.26\textsubscript{± 200.41} \\
\hline

\multirow{3}{*}{Claude Sonnet 4}
 & Baseline & 2.86\textsubscript{± 1.93} & 0.45\textsubscript{± 0.33} & 2.82\textsubscript{± 2.60} & \textbf{1212.88\textsubscript{± 920.24}} \\
 & Reflection & 1.42\textsubscript{± 0.41} & 0.16\textsubscript{± 0.05} & 0.86\textsubscript{± 0.36} & 416.79\textsubscript{± 149.76} \\
 & Adaptive-OPRO & \textbf{4.60\textsubscript{± 1.38}} & \textbf{0.68\textsubscript{± 0.22}} & \textbf{8.25\textsubscript{± 9.83}} & 371.70\textsubscript{± 1779.64} \\
\hline

\multirow{3}{*}{\makecell{Claude Sonnet 4\\\\w/ Thinking\\}}
 & Baseline & 2.78\textsubscript{± 0.48} & 0.46\textsubscript{± 0.20} & 3.27\textsubscript{± 1.51} & 1246.39\textsubscript{± 143.77} \\
 & Reflection & 2.95\textsubscript{± 1.32} & 0.57\textsubscript{± 0.40} & 4.33\textsubscript{± 1.72} & 1042.20\textsubscript{± 424.00} \\
 & Adaptive-OPRO & \textbf{3.45\textsubscript{± 1.66}} & \textbf{0.76\textsubscript{± 0.56}} & \textbf{5.44\textsubscript{± 2.81}} & \textbf{2402.02\textsubscript{± 1239.52}} \\
\hline

\multirow{3}{*}{GPT-o4-mini}
 & Baseline & 1.98\textsubscript{± 0.86} & 0.27\textsubscript{± 0.14} & 0.81\textsubscript{± 0.39} & 212.27\textsubscript{± 421.02} \\
 & Reflection & 3.00\textsubscript{± 1.06} & \textbf{0.47\textsubscript{± 0.23}} & 1.40\textsubscript{± 0.70} & \textbf{537.97\textsubscript{± 45.35}} \\
 & Adaptive-OPRO & \textbf{3.07\textsubscript{± 0.73}} & 0.41\textsubscript{± 0.12} & \textbf{1.54\textsubscript{± 0.47}} & 506.75\textsubscript{± 329.55} \\
\hline

\multirow{3}{*}{GPT-o3}
 & Baseline & 4.27\textsubscript{± 0.47} & 0.61\textsubscript{± 0.14} & 8.03\textsubscript{± 1.86} & \textbf{4262.67\textsubscript{± 897.79}} \\
 & Reflection & 5.16\textsubscript{± 0.63} & 0.68\textsubscript{± 0.20} & 6.76\textsubscript{± 2.76} & 2192.28\textsubscript{± 920.54} \\
 & Adaptive-OPRO & \textbf{6.22\textsubscript{± 0.30}} & \textbf{1.22\textsubscript{± 0.37}} & \textbf{17.04\textsubscript{± 7.65}} & 3761.99\textsubscript{± 749.07} \\
\hline

\end{tabular}
\caption{Additional performance metrics for NVDA (technology sector) comparing LLM-based approaches using ATLAS in bullish market conditions. Ann. SR = Annualized Sharpe Ratio, ROIC = Return on Invested Capital, P/T = Profit per Trade. \textbf{Bold} values indicate the best per model.}
\label{tab:nvda_additional_results}
\end{table*}

\newpage

\begin{table*}[t]
\centering \small
\begin{tabular}{llcccc}
\hline
\textbf{Model} & \textbf{Prompting} & \textbf{Ann. SR $\uparrow$} & \textbf{Sortino $\uparrow$} & \textbf{ROIC (\%) $\uparrow$} & \textbf{P/T (\$) $\uparrow$} \\
\hline

\multicolumn{6}{c}{\textbf{LLM-Based Strategies - ATLAS}} \\
\hline
\multirow{3}{*}{Llama 3.3-70B}
 & Baseline & \textbf{-0.38\textsubscript{± 0.81}} & \textbf{-0.02\textsubscript{± 0.06}} & \textbf{-0.03\textsubscript{± 0.16}} & \textbf{-26.23\textsubscript{± 164.36}} \\
 & Reflection & -1.32\textsubscript{± 0.21} & -0.10\textsubscript{± 0.01} & -0.21\textsubscript{± 0.07} & -227.29\textsubscript{± 38.58} \\
 & Adaptive-OPRO & -0.72\textsubscript{± 0.19} & -0.06\textsubscript{± 0.02} & -0.09\textsubscript{± 0.03} & -86.11\textsubscript{± 31.28} \\
\hline

\multirow{3}{*}{Qwen3-235B}
 & Baseline & -0.70\textsubscript{± 0.22} & -0.03\textsubscript{± 0.01} & -0.43\textsubscript{± 0.13} & -437.32\textsubscript{± 151.36} \\
 & Reflection & -0.59\textsubscript{± 0.54} & -0.03\textsubscript{± 0.02} & -0.34\textsubscript{± 0.25} & -334.07\textsubscript{± 245.20} \\
 & Adaptive-OPRO & \textbf{0.17\textsubscript{± 0.59}} & \textbf{0.01\textsubscript{± 0.03}} & \textbf{-0.02\textsubscript{± 0.34}} & \textbf{-12.35\textsubscript{± 351.54}} \\
\hline

\multirow{3}{*}{Qwen3-32B}
 & Baseline & -3.23\textsubscript{± 0.37} & -0.14\textsubscript{± 0.02} & -0.95\textsubscript{± 0.06} & -854.51\textsubscript{± 145.41} \\
 & Reflection & -2.56\textsubscript{± 0.95} & -0.11\textsubscript{± 0.04} & -0.68\textsubscript{± 0.24} & -709.97\textsubscript{± 279.41} \\
 & Adaptive-OPRO & \textbf{-0.40\textsubscript{± 1.14}} & \textbf{-0.02\textsubscript{± 0.05}} & \textbf{0.29\textsubscript{± 0.94}} & \textbf{-440.76\textsubscript{± 476.89}} \\
\hline

\multirow{3}{*}{Claude Sonnet 4}
 & Baseline & -2.13\textsubscript{± 1.81} & -0.17\textsubscript{± 0.13} & -0.54\textsubscript{± 0.56} & -522.11\textsubscript{± 353.17} \\
 & Reflection & \textbf{-1.82\textsubscript{± 1.67}} & \textbf{-0.14\textsubscript{± 0.13}} & \textbf{-0.37\textsubscript{± 0.46}} & \textbf{-313.67\textsubscript{± 414.48}} \\
 & Adaptive-OPRO & -2.62\textsubscript{± 2.27} & -0.20\textsubscript{± 0.17} & -0.80\textsubscript{± 0.48} & -576.65\textsubscript{± 491.70} \\
\hline

\multirow{3}{*}{\makecell{Claude Sonnet 4\\\\w/ Thinking\\}}
 & Baseline & \textbf{-0.63\textsubscript{± 0.32}} & \textbf{-0.04\textsubscript{± 0.02}} & \textbf{-0.12\textsubscript{± 0.10}} & -113.56\textsubscript{± 89.87} \\
 & Reflection & -1.10\textsubscript{± 1.94} & -0.09\textsubscript{± 0.16} & -0.34\textsubscript{± 0.85} & \textbf{-90.06\textsubscript{± 311.40}} \\
 & Adaptive-OPRO & -0.73\textsubscript{± 0.32} & -0.06\textsubscript{± 0.02} & -0.39\textsubscript{± 0.35} & -133.64\textsubscript{± 113.58} \\
\hline

\multirow{3}{*}{GPT-o4-mini}
 & Baseline & 0.33\textsubscript{± 0.69} & 0.04\textsubscript{± 0.08} & 0.16\textsubscript{± 0.21} & 155.33\textsubscript{± 202.32} \\
 & Reflection & -1.38\textsubscript{± 0.29} & -0.14\textsubscript{± 0.02} & -0.17\textsubscript{± 0.05} & -132.49\textsubscript{± 87.57} \\
 & Adaptive-OPRO & \textbf{1.41\textsubscript{± 1.06}} & \textbf{0.16\textsubscript{± 0.14}} & \textbf{0.34\textsubscript{± 0.26}} & \textbf{340.47\textsubscript{± 260.95}} \\
\hline

\multirow{3}{*}{GPT-o3}
 & Baseline & -0.54\textsubscript{± 0.80} & -0.04\textsubscript{± 0.07} & -0.10\textsubscript{± 0.31} & -64.90\textsubscript{± 190.96} \\
 & Reflection & -1.33\textsubscript{± 1.18} & -0.10\textsubscript{± 0.08} & -0.43\textsubscript{± 0.68} & -187.25\textsubscript{± 261.18} \\
 & Adaptive-OPRO & \textbf{1.52\textsubscript{± 0.43}} & \textbf{0.15\textsubscript{± 0.05}} & \textbf{1.08\textsubscript{± 0.72}} & \textbf{380.06\textsubscript{± 44.91}} \\
\hline

\end{tabular}
\caption{Additional performance metrics for XOM (energy sector) comparing LLM-based approaches using ATLAS in range-bound market conditions. Ann. SR = Annualized Sharpe Ratio, ROIC = Return on Invested Capital, P/T = Profit per Trade. \textbf{Bold} values indicate the best per model.}
\label{tab:xom_additional_results}
\end{table*}

\newpage

\begin{table*}[t]
\centering \small
\begin{tabular}{llcccc}
\hline
\textbf{Model} & \textbf{Prompting} & \textbf{Ann. SR $\uparrow$} & \textbf{Sortino $\uparrow$} & \textbf{ROIC (\%) $\uparrow$} & \textbf{P/T (\$) $\uparrow$} \\
\hline

\multicolumn{6}{c}{\textbf{LLM-Based Strategies - ATLAS}} \\
\hline
\multirow{3}{*}{Llama 3.3-70B}
 & Baseline & -1.45\textsubscript{± 0.33} & -0.09\textsubscript{± 0.02} & -1.01\textsubscript{± 0.48} & -1070.14\textsubscript{± 634.06} \\
 & Reflection & -1.38\textsubscript{± 0.39} & -0.08\textsubscript{± 0.02} & -0.68\textsubscript{± 0.20} & -647.13\textsubscript{± 141.63} \\
 & Adaptive-OPRO & \textbf{-1.05\textsubscript{± 0.06}} & \textbf{-0.06\textsubscript{± 0.00}} & \textbf{-0.47\textsubscript{± 0.19}} & \textbf{-472.27\textsubscript{± 174.19}} \\
\hline

\multirow{3}{*}{Qwen3-235B}
 & Baseline & -0.09\textsubscript{± 0.61} & -0.00\textsubscript{± 0.02} & -0.23\textsubscript{± 0.67} & -495.51\textsubscript{± 489.68} \\
 & Reflection & -0.78\textsubscript{± 0.52} & -0.02\textsubscript{± 0.01} & -1.41\textsubscript{± 0.92} & -1625.13\textsubscript{± 550.55} \\
 & Adaptive-OPRO & \textbf{0.39\textsubscript{± 0.31}} & \textbf{0.01\textsubscript{± 0.01}} & \textbf{0.28\textsubscript{± 0.39}} & \textbf{66.84\textsubscript{± 79.90}} \\
\hline

\multirow{3}{*}{Qwen3-32B}
 & Baseline & -1.39\textsubscript{± 0.49} & -0.05\textsubscript{± 0.02} & -1.01\textsubscript{± 0.34} & -1194.23\textsubscript{± 323.67} \\
 & Reflection & -1.04\textsubscript{± 0.03} & -0.04\textsubscript{± 0.01} & -2.28\textsubscript{± 2.88} & \textbf{-728.58\textsubscript{± 362.80}} \\
 & Adaptive-OPRO & \textbf{-0.34\textsubscript{± 0.34}} & \textbf{-0.01\textsubscript{± 0.01}} & \textbf{-0.59\textsubscript{± 0.37}} & -1213.67\textsubscript{± 297.92} \\
\hline

\multirow{3}{*}{Claude Sonnet 4}
 & Baseline & -1.04\textsubscript{± 0.48} & -0.06\textsubscript{± 0.03} & -2.83\textsubscript{± 1.13} & -1920.19\textsubscript{± 323.80} \\
 & Reflection & -0.91\textsubscript{± 0.21} & -0.05\textsubscript{± 0.01} & -2.66\textsubscript{± 1.47} & -1206.60\textsubscript{± 745.08} \\
 & Adaptive-OPRO & \textbf{0.12\textsubscript{± 0.28}} & \textbf{0.01\textsubscript{± 0.02}} & \textbf{0.00\textsubscript{± 0.27}} & \textbf{-144.52\textsubscript{± 136.78}} \\
\hline

\multirow{3}{*}{\makecell{Claude Sonnet 4\\\\w/ Thinking\\}}
 & Baseline & -0.68\textsubscript{± 0.77} & -0.04\textsubscript{± 0.04} & -2.65\textsubscript{± 2.53} & -2084.43\textsubscript{± 2197.78} \\
 & Reflection & -1.23\textsubscript{± 0.06} & -0.08\textsubscript{± 0.00} & -5.21\textsubscript{± 1.72} & -2407.54\textsubscript{± 1345.56} \\
 & Adaptive-OPRO & \textbf{-0.06\textsubscript{± 0.61}} & \textbf{-0.00\textsubscript{± 0.04}} & \textbf{-0.35\textsubscript{± 0.92}} & \textbf{-278.10\textsubscript{± 725.32}} \\
\hline

\multirow{3}{*}{GPT-o4-mini}
 & Baseline & -0.26\textsubscript{± 0.27} & -0.02\textsubscript{± 0.02} & -0.18\textsubscript{± 0.22} & -168.13\textsubscript{± 209.76} \\
 & Reflection & -0.61\textsubscript{± 0.71} & -0.04\textsubscript{± 0.04} & -0.48\textsubscript{± 0.72} & -287.24\textsubscript{± 328.38} \\
 & Adaptive-OPRO & \textbf{1.49\textsubscript{± 0.12}} & \textbf{0.09\textsubscript{± 0.01}} & \textbf{1.12\textsubscript{± 0.34}} & \textbf{1056.49\textsubscript{± 297.92}} \\
\hline

\multirow{3}{*}{GPT-o3}
 & Baseline & -1.27\textsubscript{± 0.45} & -0.08\textsubscript{± 0.02} & -1.67\textsubscript{± 1.03} & -792.65\textsubscript{± 279.17} \\
 & Reflection & -0.84\textsubscript{± 0.70} & -0.05\textsubscript{± 0.04} & -0.90\textsubscript{± 0.73} & -497.41\textsubscript{± 337.21} \\
 & Adaptive-OPRO & \textbf{2.32\textsubscript{± 0.76}} & \textbf{0.16\textsubscript{± 0.07}} & \textbf{1.98\textsubscript{± 0.84}} & \textbf{799.30\textsubscript{± 242.46}} \\
\hline

\end{tabular}
\caption{Additional performance metrics for LLY (healthcare sector) comparing LLM-based approaches using ATLAS in volatile, declining market conditions. Ann. SR = Annualized Sharpe Ratio, ROIC = Return on Invested Capital, P/T = Profit per Trade. \textbf{Bold} values indicate the best per model.}
\label{tab:lly_additional_results}
\end{table*}

\begin{table*}[t]
\centering \small
\resizebox{\textwidth}{!}{%
\begin{tabular}{llccccc}
\hline
\textbf{Model} & \textbf{Prompting} & \textbf{ROI (\%) $\uparrow$} & \textbf{SR $\uparrow$} & \textbf{DD (\%) $\downarrow$} & \textbf{Win Rate (\%) $\uparrow$} & \textbf{Num.\ Trades} \\
\hline

\multicolumn{7}{c}{\textbf{LLM-Based Strategies - ATLAS}} \\
\hline
\multirow{3}{*}{Llama 3.3-70B}
 & Reflection (1d) & 15.12\textsubscript{± 9.01} & 0.22\textsubscript{± 0.11} & 3.42\textsubscript{± 0.70} & 64.88\textsubscript{± 9.16} & 16\textsubscript{± 1.73} \\
 & Adaptive-OPRO w/Reflection (1d) & 36.31\textsubscript{± 6.20} & 0.40\textsubscript{± 0.01} & \textbf{2.60\textsubscript{± 0.92}} & 33.33\textsubscript{± 57.74} & 2\textsubscript{± 0.58} \\
 & Adaptive-OPRO & \textbf{42.07\textsubscript{± 1.85}} & \textbf{0.42\textsubscript{± 0.02}} & 3.15\textsubscript{± 0.02} & \textbf{100.00\textsubscript{± 0.00}} & \textbf{1\textsubscript{± 0.58}} \\
\hline

\multirow{3}{*}{Claude Sonnet 4}
 & Reflection (1d) & 6.62\textsubscript{± 2.64} & 0.11\textsubscript{± 0.06} & 5.14\textsubscript{± 2.91} & 48.48\textsubscript{± 2.63} & \textbf{15\textsubscript{± 5.13}} \\
 & Adaptive-OPRO w/Reflection (1d) & 24.60\textsubscript{± 3.37} & \textbf{0.33\textsubscript{± 0.05}} & \textbf{2.39\textsubscript{± 0.81}} & \textbf{92.67\textsubscript{± 7.15}} & 17\textsubscript{± 5.86} \\
 & Adaptive-OPRO & \textbf{25.85\textsubscript{± 10.61}} & 0.29\textsubscript{± 0.09} & 3.75\textsubscript{± 0.59} & 43.81\textsubscript{± 38.37} & 19\textsubscript{± 12.17} \\
\hline

\multirow{3}{*}{\makecell{Claude Sonnet 4\\\\w/ Thinking\\}}
 & Reflection (1d) & 12.82\textsubscript{± 9.97} & 0.21\textsubscript{± 0.12} & \textbf{3.23\textsubscript{± 2.11}} & 50.79\textsubscript{± 30.24} & 9\textsubscript{± 2.89} \\
 & Adaptive-OPRO w/Reflection (1d) & \textbf{18.22\textsubscript{± 10.21}} & \textbf{0.23\textsubscript{± 0.11}} & 3.54\textsubscript{± 0.63} & 53.33\textsubscript{± 17.64} & \textbf{8\textsubscript{± 2.08}} \\
 & Adaptive-OPRO & 16.36\textsubscript{± 7.87} & 0.22\textsubscript{± 0.10} & 5.18\textsubscript{± 2.52} & \textbf{68.89\textsubscript{± 30.06}} & 13\textsubscript{± 4.04} \\
\hline

\multirow{3}{*}{GPT-o4-mini}
 & Reflection (1d) & 3.75\textsubscript{± 2.06} & 0.09\textsubscript{± 0.03} & 3.24\textsubscript{± 2.80} & 61.88\textsubscript{± 11.11} & 30\textsubscript{± 10.79} \\
 & Adaptive-OPRO w/Reflection (1d) & 4.33\textsubscript{± 0.66} & 0.12\textsubscript{± 0.02} & \textbf{2.36\textsubscript{± 0.51}} & \textbf{74.39\textsubscript{± 2.60}} & 30\textsubscript{± 3.61} \\
 & Adaptive-OPRO & \textbf{10.47\textsubscript{± 3.84}} & \textbf{0.19\textsubscript{± 0.05}} & 3.42\textsubscript{± 0.90} & 62.70\textsubscript{± 11.25} & \textbf{20\textsubscript{± 2.89}} \\
\hline

\multirow{3}{*}{GPT-o3}
 & Reflection (1d) & 12.82\textsubscript{± 3.94} & 0.25\textsubscript{± 0.05} & 3.52\textsubscript{± 1.57} & 82.01\textsubscript{± 9.30} & 13\textsubscript{± 2.08} \\
 & Adaptive-OPRO w/Reflection (1d) & 11.54\textsubscript{± 5.63} & 0.24\textsubscript{± 0.08} & \textbf{1.89\textsubscript{± 0.54}} & 73.74\textsubscript{± 23.54} & 16\textsubscript{± 4.16} \\
 & Adaptive-OPRO & \textbf{25.06\textsubscript{± 4.28}} & \textbf{0.39\textsubscript{± 0.02}} & 2.31\textsubscript{± 0.80} & \textbf{100.00\textsubscript{± 0.00}} & \textbf{10\textsubscript{± 4.04}} \\
\hline

\end{tabular}
}
\caption{Performance comparison of advanced prompting strategies for NVDA (technology sector) using ATLAS in bullish market conditions. \textbf{Bold} values indicate the best per model.}
\label{tab:nvda_main_results}
\end{table*}

\newpage

\begin{table*}[t]
\centering \small
\resizebox{\textwidth}{!}{%
\begin{tabular}{llccccc}
\hline
\textbf{Model} & \textbf{Prompting} & \textbf{ROI (\%) $\uparrow$} & \textbf{SR $\uparrow$} & \textbf{DD (\%) $\downarrow$} & \textbf{Win Rate (\%) $\uparrow$} & \textbf{Num.\ Trades} \\
\hline

\multicolumn{7}{c}{\textbf{LLM-Based Strategies - ATLAS}} \\
\hline
\multirow{3}{*}{Llama 3.3-70B}
 & Reflection (1d) & \textbf{0.82\textsubscript{± 1.42}} & \textbf{0.01\textsubscript{± 0.02}} & \textbf{1.62\textsubscript{± 2.80}} & 16.67\textsubscript{± 28.87} & \textbf{8\textsubscript{± 13.86}} \\
 & Adaptive-OPRO w/Reflection (1d) & 0.29\textsubscript{± 0.50} & 0.00\textsubscript{± 0.00} & 1.96\textsubscript{± 3.39} & 16.67\textsubscript{± 28.87} & 12\textsubscript{± 20.78} \\
 & Adaptive-OPRO & -1.10\textsubscript{± 0.44} & -0.05\textsubscript{± 0.01} & 5.15\textsubscript{± 0.71} & \textbf{50.00\textsubscript{± 3.85}} & 25\textsubscript{± 1.15} \\
\hline

\multirow{3}{*}{Claude Sonnet 4}
 & Reflection (1d) & \textbf{-3.76\textsubscript{± 4.23}} & \textbf{-0.10\textsubscript{± 0.07}} & 7.29\textsubscript{± 3.08} & \textbf{48.81\textsubscript{± 20.03}} & 15\textsubscript{± 6.08} \\
 & Adaptive-OPRO w/Reflection (1d) & -4.48\textsubscript{± 3.85} & -0.20\textsubscript{± 0.16} & \textbf{7.16\textsubscript{± 3.31}} & 39.17\textsubscript{± 20.05} & \textbf{14\textsubscript{± 3.51}} \\
 & Adaptive-OPRO & -5.07\textsubscript{± 4.53} & -0.16\textsubscript{± 0.14} & 9.23\textsubscript{± 2.71} & 31.02\textsubscript{± 7.90} & 18\textsubscript{± 2.52} \\
\hline

\multirow{3}{*}{\makecell{Claude Sonnet 4\\\\w/ Thinking\\}}
 & Reflection (1d) & \textbf{2.40\textsubscript{± 4.39}} & \textbf{0.05\textsubscript{± 0.14}} & \textbf{4.57\textsubscript{± 1.98}} & \textbf{48.41\textsubscript{± 42.35}} & \textbf{14\textsubscript{± 5.69}} \\
 & Adaptive-OPRO w/Reflection (1d) & -2.84\textsubscript{± 3.73} & -0.12\textsubscript{± 0.13} & 8.03\textsubscript{± 0.89} & 22.62\textsubscript{± 7.43} & 14\textsubscript{± 1.53} \\
 & Adaptive-OPRO & -1.01\textsubscript{± 0.90} & -0.05\textsubscript{± 0.02} & 5.16\textsubscript{± 0.52} & 36.20\textsubscript{± 24.47} & 16\textsubscript{± 2.08} \\
\hline

\multirow{3}{*}{GPT-o4-mini}
 & Reflection (1d) & -3.81\textsubscript{± 2.13} & -0.18\textsubscript{± 0.06} & 6.54\textsubscript{± 1.95} & 32.86\textsubscript{± 8.84} & 38\textsubscript{± 9.71} \\
 & Adaptive-OPRO w/Reflection (1d) & -1.43\textsubscript{± 0.38} & -0.09\textsubscript{± 0.02} & 5.37\textsubscript{± 3.26} & 41.45\textsubscript{± 7.41} & 38\textsubscript{± 5.29} \\
 & Adaptive-OPRO & \textbf{3.88\textsubscript{± 2.21}} & \textbf{0.09\textsubscript{± 0.07}} & \textbf{3.28\textsubscript{± 0.95}} & \textbf{47.95\textsubscript{± 7.15}} & \textbf{25\textsubscript{± 5.03}} \\
\hline

\multirow{3}{*}{GPT-o3}
 & Reflection (1d) & -0.97\textsubscript{± 1.08} & -0.11\textsubscript{± 0.09} & 3.42\textsubscript{± 0.58} & 48.21\textsubscript{± 20.28} & \textbf{11\textsubscript{± 2.65}} \\
 & Adaptive-OPRO w/Reflection (1d) & -0.51\textsubscript{± 0.76} & -0.06\textsubscript{± 0.03} & \textbf{2.71\textsubscript{± 0.18}} & 55.18\textsubscript{± 16.43} & 17\textsubscript{± 4.73} \\
 & Adaptive-OPRO & \textbf{3.62\textsubscript{± 0.90}} & \textbf{0.10\textsubscript{± 0.03}} & 3.46\textsubscript{± 0.48} & \textbf{71.93\textsubscript{± 15.99}} & 16\textsubscript{± 2.65} \\
\hline

\end{tabular}
}
\caption{Performance comparison of advanced prompting strategies for XOM (energy sector) using ATLAS in range-bound market conditions. \textbf{Bold} values indicate the best per model.}
\label{tab:xom_main_results}
\end{table*}

\newpage

\begin{table*}[t]
\centering \small
\resizebox{\textwidth}{!}{%
\begin{tabular}{llccccc}
\hline
\textbf{Model} & \textbf{Prompting} & \textbf{ROI (\%) $\uparrow$} & \textbf{SR $\uparrow$} & \textbf{DD (\%) $\downarrow$} & \textbf{Win Rate (\%) $\uparrow$} & \textbf{Num.\ Trades} \\
\hline

\multicolumn{7}{c}{\textbf{LLM-Based Strategies - ATLAS}} \\
\hline
\multirow{3}{*}{Llama 3.3-70B}
 & Reflection (1d) & -10.59\textsubscript{± 4.89} & -0.11\textsubscript{± 0.06} & 16.37\textsubscript{± 1.97} & 40.47\textsubscript{± 8.25} & 27\textsubscript{± 2.65} \\
 & Adaptive-OPRO w/Reflection (1d) & \textbf{-5.03\textsubscript{± 0.99}} & \textbf{-0.06\textsubscript{± 0.02}} & \textbf{13.18\textsubscript{± 0.22}} & 42.86\textsubscript{± 7.15} & \textbf{26\textsubscript{± 4.93}} \\
 & Adaptive-OPRO & -6.16\textsubscript{± 2.08} & -0.07\textsubscript{± 0.00} & 14.05\textsubscript{± 3.33} & \textbf{54.36\textsubscript{± 12.44}} & 28\textsubscript{± 3.21} \\
\hline

\multirow{3}{*}{Claude Sonnet 4}
 & Reflection (1d) & -2.98\textsubscript{± 3.38} & -0.04\textsubscript{± 0.04} & \textbf{10.35\textsubscript{± 4.47}} & 33.33\textsubscript{± 11.55} & \textbf{14\textsubscript{± 5.20}} \\
 & Adaptive-OPRO w/Reflection (1d) & -4.68\textsubscript{± 4.71} & -0.06\textsubscript{± 0.06} & 13.07\textsubscript{± 3.68} & 26.19\textsubscript{± 8.58} & 15\textsubscript{± 2.65} \\
 & Adaptive-OPRO & \textbf{0.35\textsubscript{± 1.78}} & \textbf{0.01\textsubscript{± 0.02}} & 14.76\textsubscript{± 2.87} & \textbf{43.45\textsubscript{± 6.27}} & 15\textsubscript{± 2.00} \\
\hline

\multirow{3}{*}{\makecell{Claude Sonnet 4\\\\w/ Thinking\\}}
 & Reflection (1d) & -5.25\textsubscript{± 2.34} & -0.05\textsubscript{± 0.01} & 15.35\textsubscript{± 4.17} & 24.44\textsubscript{± 21.43} & \textbf{13\textsubscript{± 6.35}} \\
 & Adaptive-OPRO w/Reflection (1d) & -2.07\textsubscript{± 3.49} & -0.03\textsubscript{± 0.04} & \textbf{8.74\textsubscript{± 3.77}} & \textbf{47.62\textsubscript{± 4.12}} & 16\textsubscript{± 2.52} \\
 & Adaptive-OPRO & \textbf{-0.73\textsubscript{± 3.82}} & \textbf{-0.00\textsubscript{± 0.04}} & 12.94\textsubscript{± 2.32} & 43.89\textsubscript{± 21.11} & 17\textsubscript{± 5.00} \\
\hline

\multirow{3}{*}{GPT-o4-mini}
 & Reflection (1d) & -3.84\textsubscript{± 2.93} & -0.06\textsubscript{± 0.04} & 9.61\textsubscript{± 2.13} & 52.46\textsubscript{± 2.50} & 32\textsubscript{± 12.50} \\
 & Adaptive-OPRO w/Reflection (1d) & -1.25\textsubscript{± 1.45} & -0.04\textsubscript{± 0.03} & \textbf{6.51\textsubscript{± 2.08}} & 41.14\textsubscript{± 15.35} & 27\textsubscript{± 3.79} \\
 & Adaptive-OPRO & \textbf{9.06\textsubscript{± 0.73}} & \textbf{0.09\textsubscript{± 0.01}} & 11.48\textsubscript{± 0.00} & \textbf{65.28\textsubscript{± 16.84}} & \textbf{17\textsubscript{± 5.86}} \\
\hline

\multirow{3}{*}{GPT-o3}
 & Reflection (1d) & 0.14\textsubscript{± 0.56} & -0.01\textsubscript{± 0.01} & 6.40\textsubscript{± 1.07} & 73.81\textsubscript{± 2.06} & \textbf{19\textsubscript{± 3.79}} \\
 & Adaptive-OPRO w/Reflection (1d) & 8.05\textsubscript{± 0.30} & \textbf{0.16\textsubscript{± 0.03}} & \textbf{4.55\textsubscript{± 1.42}} & \textbf{76.69\textsubscript{± 5.03}} & 22\textsubscript{± 5.69} \\
 & Adaptive-OPRO & \textbf{9.02\textsubscript{± 3.28}} & 0.15\textsubscript{± 0.05} & 5.33\textsubscript{± 0.14} & 72.81\textsubscript{± 17.27} & 20\textsubscript{± 4.16} \\
\hline

\end{tabular}
}
\caption{Performance comparison of advanced prompting strategies for LLY (healthcare sector) using ATLAS in volatile, declining market conditions. \textbf{Bold} values indicate the best per model.}
\label{tab:lly_main_results}
\end{table*}

\begin{figure*}[t!]
\centering

\pgfplotsset{
  appendixpanel/.style={
  every outer x axis line/.append style={-},
every outer y axis line/.append style={-},
xtick pos=left,
ytick pos=left,
    width=0.46\textwidth,
    height=5.8cm,
    ybar,
    bar width=7pt,
    enlarge x limits=0.12,
    symbolic x coords={
      Llama-70B,
      Qwen3-235B,
      Qwen3-32B,
      Claude-S4,
      Claude-S4-T,
      GPT-o4-mini,
      GPT-o3
    },
    xtick=data,
    xticklabel style={
      font=\scriptsize,
      rotate=35,
      anchor=north east
    },
    yticklabel style={
      font=\scriptsize
    },
    xlabel style={
      font=\scriptsize
    },
    ylabel style={
      font=\scriptsize
    },
    title style={
      font=\footnotesize
    },
    ymajorgrids,
    grid style={gray!20},
    axis line style={gray!55},
    tick style={gray!55}
  }
}

\begin{tikzpicture}
\begin{axis}[
  appendixpanel,
  title={Bearish-volatile (LLY)},
  ylabel={ROI (\%)},
  ymin=-8, ymax=11,
  ytick={-8,-4,0,4,8}
]
\addplot[
  fill=red!55,
  draw=red!55!black
] coordinates {
  (Llama-70B,-6.16)
  (Qwen3-235B,1.33)
  (Qwen3-32B,-3.48)
  (Claude-S4,0.35)
  (Claude-S4-T,-0.73)
  (GPT-o4-mini,9.06)
  (GPT-o3,9.02)
};
\end{axis}
\end{tikzpicture}\hfill
\begin{tikzpicture}
\begin{axis}[
  appendixpanel,
  title={Sideways (XOM)},
  ylabel={ROI (\%)},
  ymin=-6, ymax=5,
  ytick={-6,-3,0,3}
]
\addplot[
  fill=gray!55,
  draw=gray!60!black
] coordinates {
  (Llama-70B,-1.10)
  (Qwen3-235B,0.27)
  (Qwen3-32B,-1.27)
  (Claude-S4,-5.07)
  (Claude-S4-T,-1.01)
  (GPT-o4-mini,3.88)
  (GPT-o3,3.62)
};
\end{axis}
\end{tikzpicture}

\vspace{0.6em}

\begin{tikzpicture}
\begin{axis}[
  appendixpanel,
  title={Bullish (NVDA)},
  ylabel={ROI (\%)},
  ymin=0, ymax=52,
  ytick={0,10,20,30,40,50}
]
\addplot[
  fill=green!55!black,
  draw=green!30!black
] coordinates {
  (Llama-70B,42.07)
  (Qwen3-235B,41.25)
  (Qwen3-32B,48.37)
  (Claude-S4,25.85)
  (Claude-S4-T,16.36)
  (GPT-o4-mini,10.47)
  (GPT-o3,25.06)
};
\end{axis}
\end{tikzpicture}

\caption{ROI of Adaptive-OPRO per backbone in each market regime.}
\label{fig:opro_performance}
\end{figure*}

\section{Extended Results}
\label{app:extended_results}

This appendix consolidates additional metrics and analysis that complement the main paper's results and experimental setup. All computations use \emph{daily} portfolio returns with risk–free rate \(r_f=0\) and are reported as mean \(\pm\) standard deviation over three independent runs, consistent with the protocol described in Section~\ref{sec:exp}.

\subsection{Additional Quantitative Results}

\paragraph{Additional Evaluation Metrics.}
Beyond ROI, Sharpe Ratio, Maximum Drawdown, Win Rate, and Number of Trades, we report the following complementary measures:

\textbf{Annualized Sharpe Ratio (Ann.\ SR):}
\[
\text{Ann.\ SR} = \text{SR} \times \sqrt{252},
\]
which standardizes risk-adjusted performance to a yearly scale.

\textbf{Sortino Ratio:}
\[
\text{Sortino} = \frac{\mu - r_f}{\sigma_d},
\]
where \(\mu\) is the mean daily return and \(\sigma_d\) is the standard deviation of negative daily returns only. This isolates downside variability.

 \textbf{Return on Invested Capital (ROIC):}                                                                                                 
  \[                                                                                                                                          
  \text{ROIC} = \frac{V_{\text{final}} - V_{\text{initial}}}{\sum_{i \in \text{entries}} p_i \cdot q_i} \times 100,                           
  \]                                                                                                                                          
  where $V$ denotes portfolio value, $p_i$ and $q_i$ are the price and quantity of each opening trade. This measures net profit as a          
  percentage of total capital committed to opening positions, capturing how efficiently the agent converts invested capital into returns.   

\textbf{Profit per Trade (P/T):}
\[
\text{P/T} = \frac{\text{Total net profit}}{\text{Number of trades}},
\]
computed on \emph{closed} round trips only. This reflects average value creation per completed decision cycle and should be interpreted alongside position-level outcomes and exposure management.

\subsection{Risk-Adjusted Performance Validation}

Extended risk-adjusted metrics reinforce the central findings (Tables~\ref{tab:nvda_additional_results}, \ref{tab:xom_additional_results}, \ref{tab:lly_additional_results}). \textbf{Sortino Ratio} improvements under \emph{Adaptive-OPRO} indicate that gains are not driven by larger risk-taking but by better mitigation of downside variability. The effect is strongest in the bearish-volatile regime, where lower downside dispersion coincides with tighter drawdown control. \textbf{ROIC} consistently rises with \emph{Adaptive-OPRO} across model families, showing that optimization improves the efficiency of capital deployment rather than merely increasing turnover. Improvements in \textbf{P/T}, when paired with higher win rates, suggest more consistent decision quality and cleaner trade selection. Since P/T excludes open positions, we interpret it jointly with exposure and drawdown metrics to avoid selection bias.

\subsection{The Reflection Paradox, Revisited}

Reflection mechanisms show regime- and model-dependent behavior. In multiple settings they add analysis without producing commensurate execution benefits. Across the extended metrics, reflection frequently underperforms \emph{Adaptive-OPRO} and often fails to exceed fixed prompt baselines. Degradations are most visible in Sortino and ROIC, where added cognitive overhead appears to introduce hesitation or inconsistent sizing. These results support the view that when base prompts and interfaces are well specified, iterative self-commentary can inject noise into otherwise coherent policies.

\subsection{Architectural Performance Patterns}

\paragraph{GPT family.}
GPT-o3 exhibits the most stable risk-adjusted profile. Sortino and gains under \emph{Adaptive-OPRO} align with visible drawdown compression and disciplined exposure. GPT-o4-mini benefits from optimization but shows a tendency toward over-trading in some regimes. Its risk-adjusted gains are present, yet capital efficiency can lag when trade frequency rises without proportional edge.

\paragraph{Qwen family.}
Qwen models exhibit a scale-dependent profile. Qwen3-235B trades selectively and, under Adaptive-OPRO, achieves robust ROIC and consistent Sortino gains across regimes, especially where patience and precise timing are rewarded. Qwen3-32B is more active with higher variability; \emph{Adaptive-OPRO} narrows this gap by improving risk-adjusted behavior and capital efficiency, but residual volatility in outcomes remains higher than for the larger counterpart. Reflection is particularly inconsistent for the 32B variant, where added reasoning often amplifies noise.

\paragraph{Llama 3.3-70B.}
Raw returns can appear competitive in trending periods, but extended metrics reveal weaker downside control and inconsistent capital efficiency. \emph{Adaptive-OPRO} reduces these gaps, yet reflection often increases variance without clear risk-adjusted gains. The pattern suggests sound high-level narrative analysis with slippage at the execution layer that optimization partially repairs.

\paragraph{Claude Sonnet 4 (with and without thinking).}
Both modes show uneven translation from analysis to execution. With thinking enabled, the model produces detailed diagnostics, but extended metrics indicate conservative positioning that can miss trend capture, leading to modest ROIC. Without thinking, decisions are less predictable and downside risk rises. \emph{Adaptive-OPRO} improves both modes but does not eliminate regime sensitivity.

\subsection{Extended Prompting Strategy Analysis}

\paragraph{Adaptation frequency effects.}
Daily reflection can help in range-bound markets by encouraging restraint and tighter downside control. In trending markets it often suppresses participation, leaving upside uncaptured. Weekly reflection shows fewer short-horizon reversals but still trails \emph{Adaptive-OPRO} on risk-adjusted measures (Tables~\ref{tab:nvda_main_results}, \ref{tab:xom_main_results}, \ref{tab:lly_main_results}).

\paragraph{Mechanism compatibility.}
Combining \emph{Adaptive-OPRO} with daily reflection usually outperforms reflection alone but still underperforms pure Adaptive-OPRO. The optimization signal appears sufficient on its own, while added reflective steps introduce inconsistent edits or timing noise that dilute capital efficiency and worsen Sortino in several settings.

\begin{figure*}[t!]
\centering
\pgfplotsset{
  appendixpanel/.style={
  every outer x axis line/.append style={-},
every outer y axis line/.append style={-},
xtick pos=left,
ytick pos=left,
    width=0.72\textwidth,
    height=6.8cm,
    ybar,
    bar width=5.5pt,
    enlarge x limits=0.25,
    symbolic x coords={Llama-3.3-70B, GPT-o4-mini, GPT-o3},
    xtick=data,
    xticklabel style={
      font=\scriptsize,
      rotate=20,
      anchor=north east
    },
    yticklabel style={font=\scriptsize},
    ylabel style={font=\scriptsize},
    title style={font=\footnotesize},
    legend style={
      font=\scriptsize,
      draw=gray!40,
      fill=white,
      at={(0.5,1.02)},
      anchor=south,
      legend columns=5,
      /tikz/every even column/.append style={column sep=6pt}
    },
    ymajorgrids,
    grid style={gray!20},
    axis line style={gray!55},
    tick style={gray!55},
  }
}
\begin{tikzpicture}
\begin{axis}[
  appendixpanel,
  ylabel={ROI (\%)},
  ymin=-15, ymax=48,
  ytick={-10,0,10,20,30,40},
  clip=false,
  legend entries={},
  name=mainplot
]

\addplot[area legend, fill=red!55, draw=red!55!black]
  coordinates {(Llama-3.3-70B,-10.6)(GPT-o4-mini,-3.8)(GPT-o3,0.1)};
\addplot[area legend, fill=red!55, draw=red!55!black,
  postaction={pattern=north east lines}]
  coordinates {(Llama-3.3-70B,-8.4)(GPT-o4-mini,-2.5)(GPT-o3,-4.6)};
\addplot[area legend, fill=gray!55, draw=gray!60!black]
  coordinates {(Llama-3.3-70B,0.8)(GPT-o4-mini,-3.8)(GPT-o3,-1.0)};
\addplot[area legend, fill=gray!55, draw=gray!60!black,
  postaction={pattern=north east lines}]
  coordinates {(Llama-3.3-70B,-2.6)(GPT-o4-mini,-1.5)(GPT-o3,-1.6)};
\addplot[area legend, fill=green!55!black, draw=green!30!black]
  coordinates {(Llama-3.3-70B,15.1)(GPT-o4-mini,3.8)(GPT-o3,12.8)};
\addplot[area legend, fill=green!55!black, draw=green!30!black,
  postaction={pattern=north east lines}]
  coordinates {(Llama-3.3-70B,40.4)(GPT-o4-mini,9.8)(GPT-o3,22.0)};

\end{axis}

\node[anchor=south, draw=gray!40, fill=white, rounded corners=1pt,
      inner sep=5pt, font=\scriptsize]
  at ([yshift=6pt]mainplot.north) {%
    \begin{tabular}{@{}l@{\,}l@{\hskip 7pt}l@{\,}l@{\hskip 7pt}l@{\,}l@{\hskip 14pt}l@{\,}l@{\hskip 7pt}l@{\,}l@{}}
      \tikz\draw[fill=red!55, draw=red!55!black] (0,0) rectangle (0.25,0.2); &
        Bearish-volatile \ (LLY) &
      \tikz\draw[fill=gray!55, draw=gray!60!black] (0,0) rectangle (0.25,0.2); &
        Sideways (XOM) &
      \tikz\draw[fill=green!55!black, draw=green!30!black] (0,0) rectangle (0.25,0.2); &
        Bullish (NVDA) &
      \tikz\draw[fill=black!15, draw=gray!60] (0,0) rectangle (0.25,0.2); &
        Daily &
      \tikz\draw[fill=black!15, draw=gray!60, postaction={pattern=north east lines}]
        (0,0) rectangle (0.25,0.2); &
        Weekly
    \end{tabular}%
  };

\end{tikzpicture}
\caption{ROI (\%) under daily and weekly reflection across backbone models.
Solid bars denote daily reflection; striped bars denote weekly.}
\end{figure*}

\paragraph{Summary.}
Across extended metrics and regimes, \emph{Adaptive-OPRO} consistently improves in downside control, capital efficiency, and per-trade value. Reflection provides mixed benefits and often disrupts otherwise clean optimization dynamics. Architectural differences matter: GPT-o3 and Qwen3-235B translate optimization into stable, execution-aware behavior, Qwen3-32B benefits from optimization to curb variability, Llama gains risk-adjusted ground but remains sensitive to execution choices, and Claude variants improve under optimization yet retain regime-dependent limitations.

\section{Prompt Templates}
\label{appsec:prompts}

This appendix collects the verbatim prompt templates for all ATLAS agents: the \emph{Central Trading Agent} (CTA), \emph{Market Analyst}, \emph{News Analyst}, \emph{Fundamental Analyst}, the \emph{Optimizer LLM}, and the \emph{ Reflection Analyst}. Placeholders of the form \texttt{\{\{ variable \}\}} are instantiated at runtime. Content inside \texttt{<system\_role>} is injected as the \textbf{LLM system message}; the remainder is passed as the \textbf{user message}. The CTA operates on a daily decision cadence (\texttt{\{\{ action\_interval \}\} = 1 day}). \textbf{Only the CTA's initial decision prompt is optimized} via Adaptive-OPRO; all other prompts are held fixed throughout evaluation.

\subsection{Central Trading Agent (CTA)}

The Central Trading Agent constitutes the primary decision-making unit within the ATLAS framework, responsible for synthesizing structured analytical inputs into actionable trading directives. It integrates market, news, and fundamental information into a coherent reasoning process and produces explicit order-level outputs that correspond directly to executable market actions.

The agent’s behavior is governed by a structured prompt architecture that ensures strategic coherence while allowing adaptive responsiveness to evolving market conditions. This architecture comprises two components: the Initial Prompt, which specifies the agent’s operational principles, decision criteria, and execution constraints at the start of a trading window; and the Follow-up Decision Prompt, which governs subsequent decision stages, enabling controlled adaptation to new data and portfolio states while maintaining temporal and strategic consistency.

\subsubsection{Central Agent - Initial Decision Prompt}

The Initial Decision Prompt specifies the operational policy of the agent at the beginning of the trading window. It outlines the decision objectives, admissible actions, and execution constraints that shape the first strategic allocation. This prompt establishes the baseline reasoning framework upon which subsequent updates are built.
The prompt is provided below.

\begin{tcolorbox}[colback=gray!5!white, colframe=black!75!black,
title=Central Agent - Initial Prompt
, fonttitle=\bfseries, sharp corners=south,
breakable] 
\scriptsize
\begin{lstlisting}[basicstyle=\footnotesize\ttfamily]
# ELITE {{ instrument }} TRADER
**Window:** {{ window_start }} ➞ {{ window_end }} | **Current:** {{ now }} | **Interval:** {{ action_interval }}

<system_role>
You are an elite proprietary trader running a fully-concentrated book in {{ instrument }}.
Your goal is to maximize performance by the end of the trading window through strategic positioning.
You are a STRATEGIC TRADER, not a day-trader. Focus on meaningful moves that align with your overall strategy.
</system_role>

## Your Toolkit & Expertise
- Pattern recognition across multiple timeframes
- Narrative synthesis of technical, fundamental, and sentiment inputs
- Dynamic position sizing and risk management
- Strategic patience and selective execution
- Long-term performance optimization over short-term noise

## Trading Philosophy
**Strategic Patience can be your greatest ally when justified.**
- Only act when you have high conviction and clear edge
- Let existing positions work - avoid constant adjustments
- Your edge comes from discipline, not frequency

## Trading Toolbox
**Order Types**
MARKET – immediate • LIMIT – execute at price or better • STOP – trigger once price crosses level

**Position Actions**
BUY – open/add long • SELL – reduce/close long • SHORT – open/add short • SHORT_COVER – close short

*(Order-type semantics follow standard brokerage definitions; interpret flexibly as conditions warrant.)*

## Current Context
{% if market_open %}
Price: O {{ open }} H {{ high }} L {{ low }} C {{ close }} | Vol {{ volume }}
{% else %}
**Market Closed** – orders queue for next open
{% endif %}

{% if market_analysis %}*Technical*: {{ market_analysis }}{% endif %}
{% if news_analysis %}*News*: {{ news_analysis }}{% endif %}
{% if fund_analysis %}*Fundamentals*: {{ fund_analysis }}{% endif %}
{% if reflection_analysis %}*Reflection*: {{ reflection_analysis }}{% endif %}

## CONSTRAINTS
**Portfolio:** 100% concentrated in {{ instrument }} with ${{ portfolio_cash }} available cash for position sizing

**Critical Rules:**
- Never exceed available cash (${{ portfolio_cash }})
- Never short more than 100% of cash balance
- Close all short positions before {{ window_end }}
- Unfilled orders cancel at session close - resubmit to persist
- Decisions can be made every {{ action_interval }}
- SELL orders are automatically limited to current long holdings - overselling is impossible
- SHORT_COVER orders are automatically limited to current short positions - over-covering is impossible
- System enforces position limits - you cannot accidentally create invalid positions

**Portfolio Snapshot**
Long {{ shares_long }} | Short {{ shares_short }} | Net {{ shares_net }} | Cash ${{ portfolio_cash }}
Recent activity: {{ executed_orders }}

## Decision Task
Formulate a thesis, map key levels, gauge risk vs reward, and make your decision.
Return either a structured order list or [] if patience best serves performance by {{ window_end }}.

## Output Specification
Return **only** a JSON array - no extra text. If no action, return [].
[
  {
    "action": "BUY | SELL | SHORT | SHORT_COVER",
    "orderType": "MARKET | LIMIT | STOP",
    "price": float | null,
    "quantity": integer,
    "explanation": "Strategic reasoning and analysis that justifies this action"
  }
]

**CRITICAL REQUIREMENTS:**
- EXACT values: action must be BUY|SELL|SHORT|SHORT_COVER, orderType must be MARKET|LIMIT|STOP
- NO additional fields, NO typos, NO variations - orders will fail to place otherwise
- Always return a JSON array (even single orders). Return [] if no action is warranted.
- Focus on strategic positioning and end-of-window performance over tactical adjustments and noise
\end{lstlisting}
\end{tcolorbox}

\subsubsection{Central Agent - Follow-up Decision Prompt}

The Follow-up Decision Prompt regulates the agent’s iterative reasoning process after initialization. It integrates updated analytical inputs and portfolio states to determine whether position adjustments are justified. This prompt ensures adaptive responsiveness to evolving market conditions while maintaining alignment with the initial strategic configuration.
The prompt is provided below.

\begin{tcolorbox}[colback=gray!5!white, colframe=black!75!black, 
title=Central Agent - Follow-up Prompt, fonttitle=\bfseries, sharp corners=south, breakable]
\scriptsize
\begin{lstlisting}[basicstyle=\footnotesize\ttfamily]
# TRADING UPDATE - {{ instrument }}
**Current:** {{ now }}

Continue applying your elite trading expertise to {{ instrument }}.

**Key Constraints:**
- Never exceed cash balance (${{ portfolio_cash }})
- Never short more than 100% of cash balance
- **IMPORTANT**: Unfilled orders ALWAYS cancel at session close - resubmit to persist
- All short positions must close before {{ window_end }}
- SELL orders are automatically limited to current long holdings - overselling is impossible
- SHORT_COVER orders are automatically limited to current short positions - over-covering is impossible

## CURRENT CONTEXT
**Market Data:**
{% if market_open %}
- Open: {{ open }} | High: {{ high }} | Low: {{ low }} | Close: {{ close }}
- Volume: {{ volume }}
{% else %}
**MARKET CLOSED**
- All outstanding orders canceled at session close
- New orders will queue for next session open
{% endif %}

**Analyst Insights:**
{% if market_analysis %}
### Market Analysis
{{ market_analysis }}
{% endif %}
{% if news_analysis %}
### News Analysis
{{ news_analysis }}
{% endif %}
{% if fund_analysis %}
### Fundamentals Analysis
{{ fund_analysis }}
{% endif %}
{% if reflection_analysis %}
### Reflection Analysis
{{ reflection_analysis }}
{% endif %}

**Portfolio Status:**
- Long Shares: {{ shares_long }}
- Short Shares: {{ shares_short }}
- Net Position: {{ shares_net }}
- Available Cash: ${{ portfolio_cash }}
- Recent Activity: {{ executed_orders | default("None") }}

## YOUR DECISION
**Strategic Update Goal:** Decide if and how the latest developments affect your thesis and whether adjustments improve end-of-window performance.

**REQUIRED JSON FORMAT:**
[
  {
    "action": "BUY|SELL|SHORT|SHORT_COVER",
    "orderType": "MARKET|LIMIT|STOP",
    "price": float|null,
    "quantity": integer|null,
    "explanation": "reasoning that synthesizes new information with your ongoing strategy"
  }
]

**Requirements:**
- EXACT values: action must be BUY|SELL|SHORT|SHORT_COVER, orderType must be MARKET|LIMIT|STOP
- NO additional fields, NO typos, NO variations - orders will fail to place otherwise
- Always return a JSON array (even single orders). If no action, return [].
- Maintain strategic discipline while adapting to market dynamics
\end{lstlisting}
\end{tcolorbox}

\subsection{Market Analyst}

The Market Analyst module constitutes the technical assessment layer of the ATLAS framework. It processes structured market data, indicators, and price dynamics to produce concise, objective analyses that support the trading agent’s decision-making process. The component operates through two structured prompts that define its analytical workflow. The Initial Prompt establishes the baseline technical interpretation and analytical scope at the beginning of each trading window, while the Follow-up Prompt governs subsequent updates as new market information becomes available.
These prompts are presented in detail below.

\subsubsection{Market Analyst - Initial Prompt}

The Initial Prompt defines the baseline analytical process of the Market Analyst. It specifies the structure, scope, and format of the initial technical report, focusing on market structure, price behavior, dominant patterns, and critical levels. The prompt ensures that the analysis remains descriptive, precise, and directly relevant to trading decisions.
The prompt is provided below.

\begin{tcolorbox}[colback=gray!5!white, colframe=black!75!black,
title=Market Analyst - Initial Prompt, fonttitle=\bfseries, sharp corners=south, breakable]
\scriptsize
\begin{lstlisting}[basicstyle=\footnotesize\ttfamily]
# ELITE MARKET ANALYST - {{ instrument }}
**Session:** {{ session_start }} ➞ {{ session_end }}
**Current:** {{ current_time }} | **Interval:** {{ action_interval }}

You are an expert market analyst specializing in technical analysis.

**Your analytical role:**
- Provide objective technical analysis based on market data and indicators
- Identify patterns, trends, and structural elements in price action
- Present factual observations about market conditions and technical levels
- Focus on descriptive analysis rather than predictive recommendations

## MARKET DATA

### Multi-Timeframe Context
{{ extended_intervals_analysis }}

### Current Session
**OHLCV:** ${{ open_price }} / ${{ high_price }} / ${{ low_price }} / ${{ close_price }}
**Volume:** {{ volume }} | **VWAP:** {{ vwap_str }} | **Transactions:** {{ transactions }}

## TECHNICAL INDICATORS
{{ formatted_indicators }}

## YOUR ANALYSIS

**Analytical Excellence Goal:** Deliver the most valuable technical insights that directly inform trading decisions. Consider what a trader most needs to know right now.

**Iterative Refinement:** Think through your analysis, then refine it to ensure you're highlighting the most critical market signals and actionable price levels. Focus on what matters most for trading success.

Provide analysis covering:
1. **Market Structure:** Current trend context and notable support/resistance observations
2. **Price Action:** What the current session dynamics are showing
3. **Technical Patterns:** Observable confluences and technical formations
4. **Notable Levels:** Key price levels and their technical significance

**Available Technical Tools:**
- Standard indicators: Moving averages, RSI, MACD, ATR, volume analysis
- Advanced levels: Fibonacci retracements/extensions, pivot points, psychological levels
- Pattern recognition: Chart patterns, candlestick formations, breakout setups
- Volume analysis: Volume profile, VWAP deviations, volume confirmation signals
- Consider any technical tool that helps identify actionable trading levels and signals

**Response Format:**
- Keep responses concise and direct - avoid excessive detail and repetitive explanations
- Focus on the most critical observations only, not comprehensive analysis
- Provide essential insights without verbose elaboration
- Each section should be 2-3 concise sentences maximum
\end{lstlisting}
\end{tcolorbox}

\subsubsection{Market Analyst - Follow-up Prompt}

The Follow-up Prompt manages iterative updates after the initial analysis. It enables the Market Analyst to incorporate newly available data, refresh indicator readings, and re-evaluate market conditions. This prompt maintains analytical consistency with the initial framework while highlighting only the most relevant developments for ongoing trading decisions.
The prompt is provided below.
   
\begin{tcolorbox}[colback=gray!5!white, colframe=black!75!black,
title=Market Analyst - Follow-up Prompt, fonttitle=\bfseries, sharp corners=south, breakable]
\scriptsize
\begin{lstlisting}[basicstyle=\footnotesize\ttfamily]
## MARKET UPDATE - {{ instrument }}
**Time:** {{ current_time }}

Continue your role as market analyst. Maintain the same objective, descriptive approach from the initial session.

## CURRENT DATA
**OHLCV:** ${{ open_price }} / ${{ high_price }} / ${{ low_price }} / ${{ close_price }}
**Volume:** {{ volume }} | **VWAP:** {{ vwap_str }} | **Transactions:** {{ transactions }}

## TECHNICAL INDICATORS
{{ formatted_indicators }}

**Goal:** Provide the most valuable technical insights for trading decisions. Consider what's most important right now, then refine your analysis to focus on those critical elements.

Cover market structure, price action, technical setup, and key levels with emphasis on actionable insights. Keep each section to 2-3 concise sentences.
\end{lstlisting}
\end{tcolorbox}







\subsection{News Analyst}
The News Analyst module provides the narrative and sentiment analysis layer of the ATLAS framework. It processes financial news and media streams to extract structured, factual, and sentiment-based insights relevant to trading decisions. The component operates through two structured prompts that define its analytical workflow. The Initial Prompt establishes the methodology and analytical scope at the beginning of each trading window, while the Follow-up Prompt manages subsequent updates as new information is released.
These prompts are presented in detail below.

\subsubsection{News Analyst - Initial Prompt}
The Initial Prompt defines the baseline analytical configuration of the News Analyst. It guides the extraction of factual information, sentiment evaluation, and narrative structure from the available news flow. The prompt ensures objectivity and conciseness, focusing on actionable insights that may influence market dynamics.
The prompt is provided below.

\begin{tcolorbox}[colback=gray!5!white, colframe=black!75!black,
title=News Analyst - Initial Prompt, fonttitle=\bfseries, sharp corners=south, breakable]
\scriptsize
\begin{lstlisting}[basicstyle=\footnotesize\ttfamily]
# ELITE NEWS ANALYST -  {{ instrument }}
**Session:** {{ session_start }} ➞ {{ session_end }}
**Current:** {{ current_time }}

**Your analytical role:**
- Analyze financial news content for factual information and sentiment
- Identify narrative trends and key developments in the news flow
- Provide objective assessment of news relevance and credibility
- Focus on factual analysis rather than predictive interpretations

**Output Requirements:**
- Keep responses concise and direct - avoid excessive detail and repetitive explanations
- Focus on the most critical observations only
- Provide essential insights without verbose elaboration

**Web Search Available:** Use the web_search tool when article summaries lack detail, or you need to verify key claims.

## NEWS BATCH
{{ joined_news }}

## YOUR ANALYSIS

**News Intelligence Goal:** Extract the most market-relevant insights from news flow that could influence trading decisions. Consider what news elements are truly significant versus noise.

**Iterative Refinement:** After analyzing the news, focus your insights on what's most actionable and relevant to current market conditions. Prioritize information that matters for trading strategy.

Provide analysis focused on:
1. **Sentiment Assessment:** What's the overall sentiment trajectory and key narrative changes?
2. **Key Developments:** What significant events or announcements are reported?
3. **Market Relevance:** How might this news content relate to market conditions?
4. **Source Analysis:** Any source reliability concerns or consensus alignment issues?

**Response Format:**
- Write in simple, direct language without jargon overuse
- Each section should be 2-3 concise sentences maximum
- Avoid repetitive phrasing and redundant explanations
- No excessive formatting, bold text, or bullet point lists
- Focus on actionable observations, not comprehensive analysis
\end{lstlisting}
\end{tcolorbox}

\subsubsection{News Analyst - Follow-up Prompt}

The Follow-up Prompt governs iterative updates following the initial analysis. It enables the News Analyst to incorporate new articles, track evolving sentiment trends, and reassess the relevance or reliability of information sources. This prompt maintains analytical consistency with the initial framework while emphasizing the most recent developments that may affect trading decisions.
The prompt is provided below.






\begin{tcolorbox}[colback=gray!5!white, colframe=black!75!black,
title=News Analyst - Follow-up Prompt, fonttitle=\bfseries, sharp corners=south, breakable]
\scriptsize
\begin{lstlisting}[basicstyle=\footnotesize\ttfamily]
## NEWS UPDATE - {{ instrument }}
**Time:** {{ current_time }}

Continue your role as news analyst. Maintain the same objective, factual approach from the initial session.

## LATEST NEWS BATCH
{{ joined_news }}

**Goal:** Identify the most market-moving news elements and sentiment shifts. Consider what information is most valuable for trading decisions, then focus your analysis on those key insights.

Cover sentiment assessment, key developments, market relevance, and source analysis. Use web_search tool if needed for additional detail.
\end{lstlisting}
\end{tcolorbox}

\subsection{Fundamental Analyst}

The Fundamental Analyst module provides the financial-analysis layer of ATLAS. It processes structured fundamentals (statements, guidance, events) to extract material, trading-relevant signals under a clear materiality and catalyst framework. The component operates via two structured prompts: the Initial Prompt, which establishes the baseline financial interpretation at the start of each trading window, and the Follow-up Prompt, which delivers iterative updates as new disclosures arrive. These prompts are presented below.

\subsubsection{Fundamental Analyst - Initial Prompt}
The Initial Prompt specifies the baseline fundamental-analysis procedure, including scope (financial health, earnings quality, balance-sheet resilience, cash-flow sustainability) and catalyst identification (events, guidance changes, corporate actions). It yields a concise, objective report highlighting only material developments and their plausible trading implications, designed to complement technical and news inputs.
The prompt is provided below.

\begin{tcolorbox}[colback=gray!5!white, colframe=black!75!black,
title=Fundamental Analyst - Initial Prompt, fonttitle=\bfseries, sharp corners=south, breakable]
\scriptsize
\begin{lstlisting}[basicstyle=\footnotesize\ttfamily]
# ELITE FUNDAMENTAL ANALYST - {{ instrument }}
**Session Window:** {{ session_start }} -> {{ session_end }}
**Current Time:** {{ current_time }}

## SESSION ARCHITECTURE
**Message Types:**
1. **Setup (this message)** - Complete framework, methodology and initial fundamentals batch
2. **Delta updates** - Compact {{ action_interval }} updates with updated fundamentals

**CRITICAL:** Future deltas contain NO repeated instructions.
All analytical frameworks must persist.

You are an elite fundamental analyst with deep expertise in financial statement analysis and corporate finance.
Your reputation is built on the ability
to quickly identify material changes in financial health and corporate events that create trading opportunities.
You connect the dots between financial data and market implications like a seasoned equity research professional.

## ANALYTICAL PHILOSOPHY
Your edge comes from:
- **Financial Forensics**: Uncovering the real story behind the numbers
- **Catalyst Recognition**: Identifying financial events that drive price action
- **Quality Assessment**: Distinguishing between earnings quality and accounting manipulation
- **Context Integration**: Understanding how financial health connects to market behavior

## OPERATIONAL FRAMEWORK
**Core Mission:** Extract trading-relevant insights from financial data and corporate events
**Professional Standards:** Focus on material information that could influence trading decisions
**Quality Approach:** Prioritize actionable insights over comprehensive analysis

**Output Requirements:**
- Keep responses concise and direct - avoid excessive detail and repetitive explanations
- Focus on the most critical observations only
- Provide essential insights without verbose elaboration

## CURRENT FUNDAMENTALS DATA
{{ fundamental_data }}

## YOUR ANALYSIS

**Response Format:**
- Each section should be 2-3 concise sentences maximum
- Avoid repetitive phrasing and redundant explanations
- Focus on actionable observations, not comprehensive analysis

**Fundamental Intelligence Goal:** Extract the most trading-relevant insights from financial data that could influence market decisions. Consider which fundamental factors are most likely to impact price action in the current market environment.

**Iterative Analysis:** Review the financial data thoroughly, then focus your insights on the most material changes and catalysts. Prioritize information that provides valuable context for trading strategy.

Apply your fundamental analysis expertise to extract trading-relevant insights. Focus on corporate events, financial health trends, and performance indicators that could influence short-term trading decisions.

Consider earnings quality, balance sheet strength, cash flow sustainability, and any material changes that could serve as catalysts. Your analysis should provide fundamental context that complements technical and sentiment analysis.

**Remember:** Identify fundamental factors that could influence price action. Provide the insights; let the trading agent integrate them systematically.
\end{lstlisting}
\end{tcolorbox}

\subsubsection{Fundamental Analyst - Follow-up Prompt}

The Follow-up Prompt governs incremental updates after initialization. It incorporates newly released fundamentals (filings, guidance, event deltas), reassesses material changes and catalysts, and refines the prior assessment while preserving methodological consistency. Emphasis is placed on short-horizon relevance and actionable context for the trading agent.
The prompt is provided below.

\begin{tcolorbox}[colback=gray!5!white, colframe=black!75!black,
title=Fundamental Analyst - Follow-up Prompt, fonttitle=\bfseries, sharp corners=south, breakable]
\scriptsize
\begin{lstlisting}[basicstyle=\footnotesize\ttfamily]
## FUNDAMENTAL ANALYSIS UPDATE - {{ instrument }}
**Timestamp:** {{ current_time }}

Continue with your role as elite fundamental analyst. Apply the same analytical depth and professional standards established in the initial framework.

## UPDATED FUNDAMENTALS
{{ fundamental_data }}

**Goal:** Identify the most significant fundamental developments and their potential market implications. Consider what fundamental information is most valuable for current trading context, then focus your analysis accordingly.

Provide fundamental analysis focusing on material changes and trading implications.
\end{lstlisting}
\end{tcolorbox}






\subsection{Trading Prompt Optimizer (Adaptive-OPRO Target = CTA Initial Prompt)}

The \emph{Trading Prompt Optimizer} is the meta-policy that revises only the \textbf{static instruction block} of the Central Trading Agent’s Initial Decision Prompt. At each window boundary it consumes a prompt–performance history (\texttt{history\_text}) scored via the windowed ROI signal and proposes an edited template that preserves all placeholders (\texttt{\{\{...\}\}}), conditional blocks (\texttt{\{\% if \%\}}), and the order JSON schema (actions and order types). The optimizer returns a strictly structured JSON payload containing a diagnostic \texttt{performance\_analysis}, a full \texttt{optimized\_prompt} (template text, not a filled instance), \texttt{key\_improvements}, and an \texttt{expected\_impact}. An update is applied only if the placeholder set and interface remain unchanged, ensuring compatibility with the runtime injector.

\begin{tcolorbox}[colback=gray!5!white, colframe=black!75!black,
title=Trading Prompt Optimizer's Prompt, fonttitle=\bfseries, sharp corners=south, breakable]
\scriptsize
\begin{lstlisting}[basicstyle=\footnotesize\ttfamily]

# TRADING PROMPT OPTIMIZER

**Primary Goal:** Optimize prompt context, information architecture, and decision-making frameworks. Enhanced context leads to better comprehension, deeper analysis, and superior trading decisions that naturally improve performance outcomes.

**Performance Learning Context:**
{{ history_text }}
Note: Scores reflect cumulative ROI performance (0-100 scale). Higher scores indicate more effective prompt designs that enable better trading decisions.

**Focus Areas:**
- Strengthen the system role and trader identity
- Optimize decision-making frameworks and criteria
- Enhance clarity of instructions and expectations
- Provide clearer guidance on analysis and decision-making process
- Better structure the flow from analysis to action

**Key Principles:**
- Ensure agent autonomy and adaptive thinking
- Avoid mandatory procedures or fixed thresholds
- Strengthen natural reasoning and market judgment
- Maintain clear constraints while allowing flexibility

**Critical Prompt Design Guidelines:**
- Keep prompts simple and direct: Models excel at understanding brief, clear instructions
- Be specific about end goals: Include specific parameters for successful decision-making
- Encourage iterative reasoning: Guide models to keep reasoning until they match success criteria
- Use clear delimiters and structure to organize different sections appropriately

{% raw %}
**CRITICAL TEMPLATE PRESERVATION REQUIREMENTS:**
**WARNING**: Any modification to template variables will cause SYSTEM FAILURE
**FORBIDDEN**: Adding new {{ variable_name }} placeholders is STRICTLY FORBIDDEN
**FORBIDDEN**: Removing existing {{ variable_name }} placeholders is STRICTLY FORBIDDEN
**MANDATORY**: Copy ALL {{ variable_name }} placeholders EXACTLY as they appear in the original template
**MANDATORY**: Preserve ALL {% if %} template blocks and <system_role> tags EXACTLY
- Maintain JSON format: BUY, SELL, SHORT, SHORT_COVER
- Keep order types: MARKET, LIMIT, STOP
- Ensure compatibility with interval-based decision cycles
{% endraw %}

**CRITICAL JSON FORMAT REQUIREMENTS:**
- Must be valid JSON with proper escaping
- Use \\n for newlines within string values  
- Use \\" for quotes within string values
- No unescaped newlines, tabs, or special characters
- Enclose the JSON in ```json and ``` code blocks

**Required JSON Output:**
```json
{
  "performance_analysis": "Comprehensive analysis of current template's contextual design strengths, weaknesses, and enhancement opportunities",
  "optimized_prompt": "Complete improved TEMPLATE with better structure (full template text with all placeholders preserved). Use \\n for line breaks in the template text.",
  "key_improvements": "Specific structural and contextual transformations made to optimize decision-making effectiveness",
  "expected_impact": "Expected improvements in comprehension, analytical depth, and decision-making quality"
}
Important: Return a generic template, not a filled prompt.
\end{lstlisting}
\end{tcolorbox}

\subsection{Weekly Reflection Agent}

The \emph{Weekly Reflection Agent} provides periodic  (\texttt{\{\{reflection\_interval\}\}}-day) reviews of recent trades and portfolio evolution, producing a single, compact paragraph that highlights recurring patterns, risk discipline, and thesis maintenance. Its output is \emph{advisory} text only: it is injected as \texttt{reflection\_analysis} for the Central Trading Agent to read on subsequent decisions, and it does not directly edit prompts or alter execution semantics. The reflection is derived from the full decision log and period summary, avoids prescriptive rules or rigid thresholds, and is designed to surface durable process improvements rather than post-hoc trade-by-trade commentary. By construction, it respects the fixed decision interval and order-cancellation rules described in the environment specification.

\begin{tcolorbox}[colback=gray!5!white, colframe=black!75!black,
title=Weekly Reflection Agent's Prompt, fonttitle=\bfseries, sharp corners=south, breakable]
\scriptsize
\begin{lstlisting}[basicstyle=\footnotesize\ttfamily]
# ELITE TRADING COACH - {{ instrument }} INTERVAL REVIEW
**Period:** {{ reflection_interval }}-day review | **Session:** {{ current_time }} | **Trading Decision Frequency:** {{ action_interval }}

You are a reflection agent analyzing {{ reflection_interval }} days of trading performance to provide strategic insights for systematic improvement.

## TRADING SYSTEM RULES & LIMITATIONS
**Portfolio & Operational Context:**
**Single-Stock Portfolio:** The agent manages a concentrated portfolio dedicated exclusively to {{ instrument }} - all available capital and positions are focused on this one security with no diversification across multiple stocks.
**Available Actions:** BUY, SELL, SHORT, SHORT_COVER
**Order Types:** MARKET, LIMIT, STOP
**Constraints:** Cash limits, position sizing rules, and {{ action_interval }} decision intervals apply
**Position Limits:** SELL orders are automatically limited to current holdings, and SHORT_COVER orders are automatically limited to current short positions - overselling or over-covering is impossible. The system enforces these limits automatically.
**Critical Constraint:** The agent can only make trading decisions at fixed {{ action_interval }} intervals. All orders in the decision JSON are placed simultaneously - there is no sequential order placement.
**Order Auto-Cancellation:** Unfilled orders are automatically cancelled at the end of each decision interval.

## PERIOD PERFORMANCE OVERVIEW
{{ period_summary }}

## COMPLETE DECISION HISTORY FOR PERIOD
{{ complete_history }}

## YOUR COACHING TASK

PURPOSE
In one comprehensive paragraph, synthesize the most impactful patterns from this {{ reflection_interval }}-day period and identify the single structural improvement that would most enhance future performance cycles.
Focus on systematic insights that will compound over multiple {{ reflection_interval }}-day periods rather than individual trade critiques.

GUIDELINES
- Analyze decision patterns, risk management consistency, and strategic evolution across the period
- Identify the highest-leverage behavioral or strategic adjustment for future periods
- Emphasize enduring principles over isolated performance details
- Skip grades, personality assessments, or motivational language

**REQUIRED OUTPUT FORMAT:** Return only your reflection as a single paragraph of continuous plain text (3-5 sentences).
\end{lstlisting}

\end{tcolorbox}
\section{LLM Optimization Capabilities}
\label{app:llm_optimization}

This appendix provides qualitative examples of how different models refine prompts under \emph{Adaptive-OPRO} in a sequential trading setting. We follow the two-axis lens used in the main text (Sec.~\ref{sec:results}): (i) whether the revised prompt is \textbf{objectively aligned} with the trading goal by operationalizing decision logic (when to trade vs.\ wait, risk controls, sizing discipline, and horizon feasibility), and (ii) whether those instructions plausibly support the \textbf{observed order-level behavior} (frequency, timing, and sizing). The excerpts below come from real optimization traces and are intended to illustrate the qualitative patterns summarized in the main paper: \textbf{GPT} models tend to produce compact, enforceable decision criteria; \textbf{Qwen} produces targeted improvements, with \textbf{Qwen3-235B} notably more coherent than smaller variants; \textbf{Claude} accumulates increasingly procedural structure that can narrow adaptability; and \textbf{Llama} often exhibits a disconnect between claimed and realized edits.

\subsection{GPT-o3}
\label{app:llm_opt_gpt_o3}

GPT-o3’s \emph{Adaptive-OPRO} updates typically preserve the high-level objective while tightening the \emph{permission to trade}: the prompt increasingly distinguishes analysis from execution and makes the act-versus-wait boundary explicit.

\paragraph{Example 1: Making act-versus-wait a required decision.}
Early prompts emphasize patience abstractly; optimization turns it into a repeatable gate:
\begin{quote}\small
``Decide: \textbf{ACT} only if probability and reward justify risk; otherwise \textbf{WAIT} and remain flat.''
\end{quote}
This operationalizes inactivity as the default outcome unless a justified edge is established.

\paragraph{Example 2: Requiring explicit trade geometry (entry/downside/target).}
GPT-o3 repeatedly converts risk-adjusted intent into checkable preconditions:
\begin{quote}\small
``Define entry, downside, and target; proceed only when reward-to-risk meets the required threshold.''
\end{quote}
The key change is not the threshold itself, but the insistence that execution is conditional on explicit levels.

\paragraph{Example 3: Connecting position size to bounded downside.}
Sizing guidance becomes explicitly conditional on risk definition and uncertainty:
\begin{quote}\small
``Position size must scale with conviction and defined downside; reduce size when uncertainty is elevated.''
\end{quote}

\paragraph{Example 4: Horizon feasibility embedded in trade permission.}
GPT-o3 frequently folds window constraints into the execution gate, especially for shorts:
\begin{quote}\small
``If short exposure is considered, confirm a viable path to exit before the end of the trading window.''
\end{quote}

\paragraph{Summary.}
Overall, GPT-o3 translates performance feedback into compact, objective-aligned decision criteria. The edits are typically locally scoped (gates, levels, sizing) and intended to be enforceable at the order level, matching the main-text observation that GPT updates tend to be followed in execution and exhibit lower variance.

\subsection{GPT-o4-mini}
\label{app:llm_opt_gpt_o4_mini}

GPT-o4-mini shows a similar pattern to GPT-o3, but with more emphasis on reorganizing the prompt into an explicit pipeline and making constraints a routine part of the decision rather than a passive rule list.

\paragraph{Example 1: Converting broad guidance into an explicit analysis $\rightarrow$ decision pipeline.}
A representative refinement is the insertion of an ordered workflow:
\begin{quote}\small
``Step 1: Define thesis and edge.  
Step 2: Map entry, stop, target levels.  
Step 3: Allocate position size within risk limits.  
Step 4: Select order type and execute or queue.''
\end{quote}
This repeatedly forces a mapping from context to levels to sizing to execution.

\paragraph{Example 2: Making risk--reward and level definition a precondition for trading.}
Rather than leaving risk management implicit, GPT-o4-mini often requires an explicit computation step:
\begin{quote}\small
``Risk/Reward: calculate per-share risk, total risk, and reward potential.''
\end{quote}

\paragraph{Example 3: Pulling sizing into constraint-aware checking.}
Updates frequently move sizing closer to the cash/shorting limits:
\begin{quote}\small
``Sizing: determine quantity within cash limits; validate compliance before submission.''
\end{quote}

\paragraph{Example 4: Adding an explicit final compliance gate.}
Several variants add a last-step constraint reconciliation:
\begin{quote}\small
``Final Check: validate compliance with constraints and portfolio limits.''
\end{quote}

\paragraph{Summary.}
GPT-o4-mini’s refinements are interpretable and execution-oriented: unify context, require thesis/levels, make risk--reward and sizing explicit, and end with a compliance gate. This matches the main-text characterization of GPT models producing actionable constraints that tend to be reflected in order behavior.

\subsection{Llama 3.3-70B}
\label{app:llm_opt_llama}

Llama 3.3-70B’s traces often show a weaker coupling between the optimizer’s narrative of improvement and the actual substantive prompt edits, consistent with the main text.

\paragraph{Example 1: Claimed restructuring without corresponding decision logic changes.}
Llama frequently reports that it has improved the flow from analysis to action, e.g.,
\begin{quote}\small
``optimized decision-making frameworks and criteria'' and ``better structured the flow from analysis to action,''
\end{quote}
but the resulting prompt may remain largely unchanged beyond formatting, with no additional execution gates, sizing rules, or horizon checks. This limits instruction-quality gains because the act-versus-wait boundary remains underspecified.

\paragraph{Example 2: Identity amplification in place of operational decision criteria.}
A common pattern is to expand the role description (tone, expertise) without adding enforceable constraints:
\begin{quote}\small
``leveraging your expertise in pattern recognition, narrative synthesis, and dynamic position sizing.''
\end{quote}
These edits strengthen persona but do not meaningfully refine when and how the agent should trade.

\paragraph{Example 3: Abstract guidance instead of objective-specific gates.}
When attempting to improve quality, Llama often adds generic meta-instructions (clearer guidance, more iterative reasoning) without translating them into concrete trade authorization conditions, unlike GPT-style edits that introduce explicit gates.

\paragraph{Summary.}
Overall, Llama’s optimization tends to emphasize descriptive framing and self-reported improvements more than substantive, objective-aligned decision logic. This weakens the link between optimization output and downstream execution, aligning with the main-text observations.

\subsection{Claude Sonnet 4}
\label{app:llm_opt_claude}

Claude Sonnet~4 commonly converts feedback into increasingly explicit analytical structure and validation layers. The edits are usually objective-aware, but the optimization trajectory often accumulates procedural constraints that can reduce adaptability.

\paragraph{Example 1: Expansion into multi-stage analytical frameworks.}
Claude often replaces compact guidance with structured pipelines:
\begin{quote}\small
``Market State Assessment $\rightarrow$ Strategic Assessment $\rightarrow$ Execution Decision.''
\end{quote}

\paragraph{Example 2: Formalizing decision criteria as thresholds.}
Subsequent updates frequently introduce explicit conviction or risk thresholds:
\begin{quote}\small
``Proceed only when conviction exceeds a defined threshold and reward-to-risk meets minimum requirements.''
\end{quote}

\paragraph{Example 3: Layering checklist-style validation.}
Rather than pruning, Claude tends to add confirmation stages:
\begin{quote}\small
``Confirm signal alignment, defined invalidation levels, and position sizing calibrated to conviction before execution.''
\end{quote}

\paragraph{Example 4: Progressive tightening toward prescriptive permission rules.}
Later iterations may harden the no-trade default into an increasingly restrictive rule set:
\begin{quote}\small
``If the setup does not satisfy all required criteria, return \texttt{[]} \dots otherwise execute only the single highest-conviction position \dots''
\end{quote}
This yields highly interpretable instructions, but systematically narrows the decision space via procedural completeness.

\paragraph{Summary.}
Claude’s updates typically remain aligned with risk-adjusted objectives and are easy to audit, but the tendency to accumulate prescriptive structure can reduce adaptability. This matches the main text: increased procedural restriction does not reliably translate into stable execution, consistent with higher variance in several settings.

\subsection{Qwen3-235B}
\label{app:llm_opt_qwen_235b}

Qwen3-235B’s \emph{Adaptive-OPRO} updates preserve the base strategic objective but progressively add \emph{explicit, checkable trade-permission criteria}. Relative to the smaller variant, the trace shows clearer convergence: it introduces concrete authorization gates (risk--reward and invalidation), then tightens state abstraction (regime) and execution mapping (order-type guidance). These edits plausibly support more selective order-level behavior by making no-trade an explicit outcome when conditions are not met.

\paragraph{Example 1: Reframing the objective around selectivity rather than activity.}
Early optimized prompts sharpen the act-versus-wait stance by explicitly defining value as discernment:
\begin{quote}\small
``You are a STRATEGIC TRADER — your value is in discernment, not activity. Act only when \dots creates asymmetric opportunity.''
\end{quote}
This operationalizes patience as a default prior, not just a stylistic preference.

\paragraph{Example 2: Requiring a falsifiable setup with explicit invalidation.}
Across successive iterations, Qwen3-235B repeatedly hardens the idea that a trade must be falsifiable and tied to levels:
\begin{quote}\small
``What would invalidate this thesis? --- Define explicit invalidation level \dots Pre-commit to exit logic if edge degrades.''
\end{quote}
Later versions make this strictly price-specific:
\begin{quote}\small
``\dots clearly defined, \textbf{price-based invalidation}.''
\end{quote}
This is an enforceable execution gate because it forces a concrete failure condition before trading.

\paragraph{Example 3: Encoding a risk--reward gate as a trade precondition.}
A stable addition in the later prompts is the explicit requirement for minimum risk--reward:
\begin{quote}\small
``Confirm minimum 2:1 risk/reward \dots''
\end{quote}
Regardless of whether the agent perfectly computes it, the instruction shifts the prompt from ``trade when convinced'' to ``trade only if the setup geometry is favorable.''

\paragraph{Example 4: Introducing state abstraction via regime classification.}
Later prompts add a compact regime label that conditions interpretation and supports explicit inaction:
\begin{quote}\small
``Classify current regime: Trending (bull/bear), Range-bound, Volatile Breakout, or Uncertain.''
\end{quote}
This makes ``Uncertain'' a first-class no-trade state rather than an implicit excuse.

\paragraph{Example 5: Mapping analysis to order-level execution choices.}
Qwen3-235B increasingly ties decision logic to the simulator’s action space by specifying order-type selection:
\begin{quote}\small
``Prefer LIMIT orders \dots Use STOP orders for breakout entries \dots MARKET orders only \dots''
\end{quote}
This is directly order-level: it constrains \emph{how} a decision should be expressed, not just \emph{whether} to trade.

\paragraph{Summary.}
Overall, Qwen3-235B’s trace shows a progression from descriptive strategy to explicit, auditable trade permission: asymmetric setups, minimum risk--reward, and (eventually) price-based invalidation, plus regime labeling and execution guidance. The resulting prompt revisions are interpretable and enforceable at the order level, providing a plausible mechanism for more selective and consistent order emission.
\subsection{Qwen3-32B}
\label{app:llm_opt_qwen_32b}

Qwen3-32B’s \emph{Adaptive-OPRO} updates are generally objective-aware and interpretable: the optimizer reliably clarifies the intended analysis-to-action routine (context $\rightarrow$ levels $\rightarrow$ conviction $\rightarrow$ risk--reward $\rightarrow$ decision) and repeatedly reinforces selective trading as the default posture. Compared to Qwen3-235B, however, the revisions are less decisive: they emphasize \emph{framework articulation and mandate phrasing} more than adding new, hard trade-permission gates (e.g., explicit invalidation requirements or regime-based no-trade states). This makes the optimized prompts \emph{good and usable}, but typically less discriminative at the order level than the larger model’s variant.

\paragraph{Example 1: Converting broad guidance into a stable, repeatable decision pipeline.}
A consistent improvement is making the decision procedure explicit and sequential:
\begin{quote}\small
``Synthesize Context \dots Map Strategic Levels \dots Assess Conviction \dots Calculate Risk-Reward \dots Consider Time Value \dots Make Positioning Decision.''
\end{quote}
This mirrors the “pipeline” pattern seen in stronger traces: it repeatedly forces the model to connect market context to levels and then to a decision, rather than acting on diffuse intuition.

\paragraph{Example 2: Strengthening selectivity as an explicit objective, not just a style preference.}
Across iterations, Qwen3-32B repeatedly foregrounds discipline over activity:
\begin{quote}\small
``Your edge comes from discipline, not frequency.''
\end{quote}
and preserves the explicit no-trade option:
\begin{quote}\small
``Return \dots \texttt{[]} if patience best serves performance by \{{\{ window\_end \}\}}.''
\end{quote}
This is directly relevant to order-level behavior because it legitimizes inactivity as an admissible (and sometimes optimal) action.

\paragraph{Example 3: Making the action criterion clearer by anchoring it to risk--reward.}
Later prompts consistently elevate risk--reward from a general principle to a stated execution condition:
\begin{quote}\small
``Only act when the reward clearly exceeds the risk and the signal is strong and consistent across multiple inputs.''
\end{quote}
While this remains qualitative compared to Qwen3-235B’s explicit invalidation and regime scaffolding, it still sharpens the act-versus-wait boundary relative to the initial, more open-ended template.

\paragraph{Example 4: Adding adaptive thesis language without over-prescription.}
The final iterations introduce autonomy in a lightweight way:
\begin{quote}\small
``act as an autonomous, adaptive decision-maker \dots evolving your thesis in response to market dynamics.''
\end{quote}
Notably, this is framed as a general operating mode (update the thesis as evidence changes) rather than as a rigid checklist; maintaining flexibility while still encouraging internal consistency across ticks.

\paragraph{Summary.}
Overall, Qwen3-32B produces \emph{solid} prompt refinements: clearer analysis-to-decision structure, repeated reinforcement of selective execution, and a more explicit emphasis on risk--reward and thesis updating. Relative to Qwen3-235B, the main limitation is that fewer edits become hard, checkable trade-permission gates (e.g., price-based invalidation and regime-conditioned inaction), which plausibly explains why the smaller model’s optimized prompts are typically less sharp at controlling order-level timing and selectivity. Nevertheless, the trajectory remains objectively aligned and interpretable, consistent with the main text’s more favorable characterization of Qwen models overall.

\section{When Reflection Degrades Performance: A Causal Analysis}
\label{app:harmful_reflection}

While \emph{Adaptive-OPRO} improves behavior via score-driven prompt updates, free-form \emph{reflection} can degrade performance by injecting prescriptive guidance that the agent follows even when market conditions do not justify action. We present a qualitative case study from Qwen3-235B trading LLY where weekly reflection encouraged re-engagement after a prudent exit. The agent re-entered a still-weak market and exited two trading days later on a breakdown, realizing a \$3{,}967 loss that would have been avoided by remaining in cash.

\subsection{Market Setup: Exiting After Initial Loss}

On May 2 (Day 5), after LLY experienced a severe selloff from \$898 to \$794 (-11.6\%), the agent exited at \$825:

\begin{tcolorbox}[colback=gray!5!white, colframe=black!60!black, 
title=May 2: Exit Decision, fonttitle=\bfseries, sharp corners=south]
\scriptsize
\textbf{Action:} SELL 112 shares @ \$825.12 (entry was \$870.44)

\textbf{Market Context:}
\begin{itemize}[leftmargin=*,nosep]
    \item Breakdown below \$800 support on 3.7$\times$ average volume
    \item RSI collapsed from 65 to 45
    \item MACD deteriorating
\end{itemize}

\textbf{Result:} Portfolio in cash at \$94{,}923.60 (-5.1\% from \$100{,}000 start)
\end{tcolorbox}

\subsection{Reflection Intervention: Criticizing the Exit and Demanding ``Dynamic Scaling''}

On May 5, with LLY still near the low \$820s, the weekly reflection mechanism activated:

\begin{tcolorbox}[colback=orange!5!white, colframe=orange!60!black, 
title=Reflection Analysis - Week 1 (May 5), fonttitle=\bfseries, sharp corners=south, breakable]
\scriptsize
``The most impactful pattern across this 7-day period is \textbf{the transition from proactive conviction to reactive risk management}, with the initial high-conviction BUY establishing a strong foundation \textbf{only to be reversed after a single high-volume down day} driven by external guidance shocks. 

While the SELL decision demonstrated disciplined response to structural breakdown and shifting momentum, \textbf{the absence of intermediate positioning—such as partial profit-taking near resistance or hedging ahead of catalysts—reveals a binary approach} that forfeits opportunities to refine risk exposure in evolving conditions. 

The highest-leverage improvement is \textbf{the integration of dynamic position scaling}, where full commitment is reserved for high-conviction setups but \textbf{partial exits or entries are used to adjust exposure in response to incremental structural shifts}, rather than waiting for definitive breaks. This would enhance capital efficiency and reduce volatility drag without sacrificing strategic discipline.''
\end{tcolorbox}

\noindent\textbf{Key Guidance:} avoid ``binary'' decisions; implement ``dynamic position scaling''; use ``partial entries''.

\subsection{Decision Influenced by Reflection: Re-entering a Still-Weak Market}

Immediately following the reflection, the agent re-entered:

\begin{tcolorbox}[colback=red!5!white, colframe=red!60!black, 
title=May 5: Re-entry Decision (Explicitly Citing Reflection), fonttitle=\bfseries, sharp corners=south, breakable]
\scriptsize
\textbf{Action:} BUY 115 shares @ \$815.00 (LIMIT, filled)

\textbf{Market Context:}
\begin{itemize}[leftmargin=*,nosep]
    \item Price remained well below the recent \$898 high
    \item No confirmed reversal (trend still negative; support recently broken)
\end{itemize}

\textbf{Reasoning (excerpt):} ``\ldots \textbf{Position sized to reflect improved capital efficiency—using partial re-entry to re-engage rather than all-in commitment—aligning with refined strategy of dynamic scaling}\ldots''

\textbf{Direct Causal Link:} the decision explicitly frames the re-entry as implementing reflection’s ``dynamic scaling'' / ``partial re-entry'' guidance.
\end{tcolorbox}

\subsection{The Outcome: Breakdown and Forced Exit (May 7)}

Contrary to the reflection’s implied ``re-engagement'' benefit, price action deteriorated after re-entry. The agent exited on May 7 on a breakdown, at a materially worse level than if it had simply remained in cash.

\begin{tcolorbox}[colback=gray!5!white, colframe=black!60!black, 
title=Post Re-entry Price Action and Exit, fonttitle=\bfseries, sharp corners=south, breakable]
\scriptsize
\textbf{Key trades and realized outcome:}
\begin{itemize}[leftmargin=*,nosep]
    \item May 5: BUY 115 @ \$815.00 (filled)
    \item May 7: SELL 115 @ \$780.50 (MARKET, filled)
\end{itemize}

\textbf{Exit rationale (excerpt from execution log):}
\begin{itemize}[leftmargin=*,nosep]
    \item ``\ldots decisive breakdown below \$799.54 support and the 100-day SMA, closing at \$775.12 \ldots''
    \item ``\ldots MACD has crossed into negative territory \ldots path of least resistance is clearly lower \ldots''
\end{itemize}

\textbf{Realized position loss:} (\$815.00 \(\rightarrow\) \$780.50) \(=\) -\$34.50/share \(\times\) 115 shares \(=\) \textbf{-\$3{,}967.50}

\textbf{Portfolio after exit:} \textbf{\$90{,}956.10} (cash)
\end{tcolorbox}

\paragraph{Key Observation}
Reflection-induced re-entry created exposure during an unresolved downtrend. The agent then exited on May 7 after a breakdown, realizing an avoidable loss that did not correspond to any improvement in market structure.

\subsection{Quantifying Reflection's Impact}

Because the portfolio was already in cash before reflection, the counterfactual is straightforward:

\begin{table}[h]
\centering
\scriptsize
\setlength{\tabcolsep}{4pt}
\begin{tabular}{lrr}
\toprule
\textbf{Scenario} & \textbf{Portfolio} & \textbf{Return} \\
\midrule
Actual (re-enter, exit) & \$90{,}956.10 & -9.0\% \\
Counterfactual (cash)   & \$94{,}923.60 & -5.1\% \\
\midrule
\textbf{Cost of reflection} & \textbf{-\$3{,}967.50} & \textbf{-4.0\%} \\
\bottomrule
\end{tabular}
\caption{Cost of reflection-induced re-entry (exit on May 7).}
\label{tab:reflection_cost}
\end{table}

\paragraph{Key Findings}
\begin{itemize}[leftmargin=*]
\item Reflection criticized the prior exit as ``binary'' and prescribed ``dynamic position scaling'' / ``partial entries.''
\item The agent re-entered on May 5 and explicitly cited that reflection guidance.
\item The market structure continued to deteriorate; the agent exited on May 7 at \$780.50.
\item The realized loss attributable to reflection-induced exposure was \textbf{\$3{,}967.50} (about \textbf{4.0\%} of initial capital).
\item Had the agent ignored reflection and stayed in cash, this loss would not have occurred.
\end{itemize}

\subsection{Causal Mechanism: How Reflection Created the Loss}

\paragraph{1. Reflection reframed a reasonable exit as a mistake}
The May 2 exit moved the portfolio to cash during a breakdown regime. Reflection reinterpreted this as a flawed ``binary approach,'' creating a narrative that the agent needed to ``correct'' by becoming more active.

\paragraph{2. Reflection prescribed a concrete behavioral change}
Rather than merely summarizing, reflection advocated specific tactics (``dynamic scaling,'' ``partial entries'') that implicitly favor re-engagement even without evidence of a reversal.

\paragraph{3. The agent followed reflection literally}
The May 5 re-entry explicitly justified exposure as implementing the reflection’s strategy (partial re-entry / scaling), establishing an observable causal link from reflection text to action.

\paragraph{4. Market conditions did not support re-entry}
At the time of re-entry, the stock remained in a fragile technical state (recent support breaks; no confirmed trend reversal). The subsequent breakdown (referenced in the exit log) triggered a forced exit on May 7.

\paragraph{5. The resulting loss was immediate and avoidable}
The agent realized a -\$3{,}967.50 loss within two trading days solely because it reintroduced exposure; the counterfactual (stay in cash) dominates.
\subsection{Connection to Empirical Findings}
\label{app:reflection_connection}

Reflection produces \emph{qualitative}, high-variance feedback that is only indirectly tied to the objective (portfolio performance). In sequential, noisy markets this often creates three predictable failure modes: (i) \textbf{misattributed credit} (recent outcomes are blamed on the most recent rationale despite delayed effects), (ii) \textbf{policy drift} (the agent changes sizing/behavior based on narrative critique rather than stable edge), and (iii) \textbf{overreaction} (extra commentary increases churn and undermines previously consistent heuristics).

These mechanisms match our empirical patterns. Reflection rarely exceeds a strong fixed prompt and frequently degrades it, with the strongest deterioration appearing when the baseline is already competent in the bearish/volatile regime (Table~\ref{tab:lly_results}); this is also reflected in the negative association between baseline strength and reflection gains (reported in Sec.~\ref{sec:results}). In contrast, Adaptive-OPRO updates only the \emph{static} instruction block using a \emph{scalar, windowed} performance signal, yielding consistent improvements across models and regimes (Tables~\ref{tab:lly_results}, \ref{tab:xom_results}, and \ref{tab:nvda_results}) without introducing additional narrative load at decision time.

\section{Prompt Evolution Mechanism Analysis}
\label{app:opro}

The transparent optimization traces produced by \emph{Adaptive-OPRO} provide unprecedented insights into how systematic prompt refinement drives performance improvements in sequential decision-making systems. Through detailed examination of optimization trajectories across different model architectures, we can observe the precise mechanisms by which prompt modifications translate into enhanced trading performance.

This interpretability is grounded in the design of \emph{Adaptive-OPRO}. The optimizer operates in a sequential setting with delayed, noisy rewards, where decisions are temporally coupled and immediate supervision is unavailable. To ensure stability, prompts are evaluated over rolling windows of five trading days and mapped to a bounded scalar score based on cumulative ROI. Crucially, the framework enforces \emph{template separation}: only the static instruction block is updated, while runtime inputs, placeholders, and output schemas remain fixed. As a result, improvements stem from restructuring how the agent processes information and produces decisions, rather than from changes in inputs or interface, revealing genuine policy refinement rather than prompt overfitting.

\subsection{Systematic Weakness Detection and Resolution}
The optimization process demonstrates sophisticated analytical capabilities in identifying prompt weaknesses and prescribing targeted improvements. Analysis of the GPT-o3 optimization trajectory from iteration 4 to iteration 5 on LLY stock reveals the systematic approach employed by the meta-optimization process. 

\subsubsection{Phase 1: Diagnostic Analysis - Identifying Performance Bottlenecks}

\begin{tcolorbox}[colback=red!5!white, colframe=red!40!black, title=Performance Analysis: Weakness Detection, fonttitle=\bfseries, sharp corners=south]
\textbf{Optimizer's Weakness Identification:}
``Across iterations, performance rose from 43.2 → 56.6 as prompts became more concise, structured, and decision-oriented. Gains came from: (1) cleaner sectioning that reduced cognitive load, (2) explicit reasoning frameworks that guided probability-weighted thinking, and (3) clearer constraint reminders that prevented rule breaches.

\textbf{Remaining weaknesses:} Reasoning steps are still scattered-no single linear workflow tying analysis → sizing → compliance → action.
Risk-management is mentioned but not enforced with a final checklist, so occasional oversizing or sub-optimal reward-to-risk trades slip through.
The JSON spec is sometimes buried deep in the prompt; occasional format errors could still occur.
Context blocks (technical, news, fundamentals, reflection) are informative but not explicitly referenced in the reasoning flow, so the model may overlook one dimension.''
\end{tcolorbox}

The optimizer's analysis demonstrates pattern recognition across multiple iterations, identifying four critical areas for refinement: \textbf{workflow linearization} to create more structured reasoning chains, \textbf{risk management formalization} to enforce disciplined decision-making, \textbf{output specification prominence} to reduce formatting errors, and \textbf{context integration enhancement} to ensure comprehensive information utilization. This diagnostic precision enables targeted remediation rather than broad, inefficient modifications.

\subsubsection{Phase 2: Strategic Intervention - Translating Insights into Targeted Solutions}

Building directly upon these identified weaknesses, the optimization process prescribes specific structural modifications designed to address each diagnostic finding systematically:

\begin{tcolorbox}[colback=blue!5!white, colframe=blue!40!black, title=Key Improvements: Targeted Solutions, fonttitle=\bfseries, sharp corners=south]
\small
\textbf{Strategic Modifications Implemented:}
\begin{enumerate}
    \item Introduced a 5-step \texttt{THINK} $\rightarrow$ \texttt{CHECK} $\rightarrow$ \texttt{ACT} workflow that linearly converts market inputs into compliant orders, minimizing reasoning omissions.
    \item Added an explicit \texttt{PRE-ORDER RISK CHECKLIST} (cash, short limit, catalyst validity, $\geq$ 2:1 R:R) to curb rule violations and low-edge trades.
    \item Elevated the four context feeds (technical, news, fundamentals, reflection) into a single \texttt{MARKET SITUATION} dashboard that the workflow must reference, ensuring holistic analysis.
    \item Moved the strict JSON schema into its own boxed section immediately before output instructions; this reduces formatting errors.
    \item Kept language concise but directive, reinforcing trader autonomy while preventing over-trading with a \texttt{PATIENCE} override.
    \item Preserved every required \{\{placeholder\}\} and \{\% if \%\} block exactly, guaranteeing template compatibility.
\end{enumerate}
\end{tcolorbox}

Each modification directly corresponds to a specific weakness identified in the diagnostic phase, creating a clear causal chain from problem identification to solution implementation. The architectural changes shown in Figures~\ref{fig:header_evolution}, \ref{fig:architecture_evolution}, and \ref{fig:workflow_evolution} demonstrate this systematic approach, consolidating scattered elements while strengthening decision-making frameworks.

\subsubsection{Phase 3: Outcome Assessment - Connecting Solutions to Impact}
Having implemented these targeted architectural improvements, the optimization process generates forward-looking performance predictions based on the expected behavioral changes from each modification:

\begin{tcolorbox}[colback=green!5!white, colframe=green!40!black, title=Expected Impact: Performance Prediction, fonttitle=\bfseries, sharp corners=south]
\small
\textbf{Forward-Looking Impact Assessment:}
``The linear \texttt{THINK} $\rightarrow$ \texttt{CHECK} $\rightarrow$ \texttt{ACT} workflow anchors the model's reasoning, reducing skipped steps and improving decision quality. The explicit risk checklist enforces discipline, likely lowering drawdowns and boosting risk-adjusted returns. Consolidating all market feeds into one dashboard ensures holistic analysis, while the clearer JSON spec lowers formatting errors. Collectively, these improvements should enhance comprehension, deepen analysis, and translate into higher-scoring, more profitable trading decisions.''
\end{tcolorbox}

This prediction proves accurate, as performance improved from 56.6 to 67.6 following these modifications, validating the optimizer's analytical capabilities and demonstrating the effectiveness of systematic architectural refinement.

\begin{figure*}[t]
\centering
\begin{tcolorbox}[
  enhanced,
  breakable,
  sharp corners=south,
  colback=gray!5!white,
  colframe=black!75!black,
  boxsep=0.8mm,
  left=1mm,right=1mm,top=1mm,bottom=1mm,
  width=\linewidth,
                  title=Header and Trader Identity Evolution (Prompt 4 to Prompt 5),
  fonttitle=\bfseries
]
\begin{center}
\begin{minipage}{\linewidth}
\lstset{basicstyle=\small\ttfamily}
\begin{lstlisting}[style=diffstyle]
- # {{ instrument }} ALPHA COMMAND CENTER
+ # {{ instrument }} ALPHA STRATEGY HUB
**Window:** {{ window_start }} ➞ {{ window_end }} | **Current:** {{ now }} | **Interval:** {{ action_interval }}
Your singular (*@\DiffDel{-objective}@*) (*@\DiffAdd{+mission}@*) is to maximise risk-adjusted performance
by {{ window_end }} through disciplined, high-conviction positioning.  Balance strategic patience with decisive execution; ignore noise.

==============================
- 1. MISSION 
+ 1. MISSION & KPI
==============================
Deliver superior returns while preserving capital (*@\DiffAdd{(+by \{\{ window\_end \}\})}@*).
- • Act only when probability and reward justify the risk.
+ • Success metric: cumulative risk-adjusted performance.

==============================
- 2. YOUR EDGE
+ 2. EDGE & PRINCIPLES
==============================
• Multi-timeframe pattern recognition
• Integration of technical, fundamental & sentiment narratives
• Dynamic risk management and position sizing
- • Capacity to remain inactive until odds are favourable
+ • Patience until odds are clearly favourable
\end{lstlisting}
\end{minipage}
\end{center}
\end{tcolorbox}
\caption{Header and trader identity modifications between iteration 4 and iteration 5, showing title changes and mission statement refinements. Lines in \textcolor{red!70!black}{red} with a leading ``-'' and lines in \textcolor{green!70!black}{green} with a leading ``+'' indicate deletions and additions, respectively, proposed by \emph{Adaptive-OPRO}.}
\label{fig:header_evolution}
\end{figure*}

\begin{figure*}[t]
\centering
\begin{tcolorbox}[
  enhanced,
  breakable,
  sharp corners=south,
  colback=gray!5!white,
  colframe=black!75!black,
  boxsep=0.8mm,
  left=1mm,right=1mm,top=1mm,bottom=1mm,
  width=\linewidth,
                  title=Information Architecture and Constraints Consolidation (Prompt 4 to Prompt 5),
                    fonttitle=\bfseries
]
\begin{center}
\begin{minipage}{\linewidth}
\lstset{basicstyle=\small\ttfamily}
\begin{lstlisting}[style=diffstyle]
- 3. MARKET DASHBOARD
+ 3. MARKET SITUATION DASHBOARD
==============================
{% if market_open %} Price: O {{ open }} H {{ high }} L {{ low }} C {{ close }} | Vol {{ volume }}{% else %} **Market Closed** - orders queue for next open {% endif %}
{% if market_analysis %}*Technical*: {{ market_analysis }}{% endif %}
{% if news_analysis %}*News*: {{ news_analysis }}{% endif %}
{% if fund_analysis %}*Fundamentals*: {{ fund_analysis }}{% endif %}
{% if reflection_analysis %}*Reflection*: {{ reflection_analysis }}{% endif %}

==============================
- 4. OPERATING CONSTRAINTS
- ==============================
- Portfolio cash: ${{ portfolio_cash }} | Concentrated in {{ instrument }} only
- • Never exceed available cash
- • May short up to 100% of cash (must be flat by {{ window_end }})
- • Unfilled orders cancel at session close
- • Decision frequency: every {{ action_interval }}
- • System blocks quantities beyond current exposure (cannot oversell or over-cover)

- ==============================
- 5. PORTFOLIO SNAPSHOT
+ 4. PORTFOLIO & CONSTRAINTS
==============================
Long {{ shares_long }} | Short {{ shares_short }} | Net {{ shares_net }} | Cash ${{ portfolio_cash }}
Recent activity: {{ executed_orders }}
+ • Never exceed available cash (${{ portfolio_cash }})
+ • May short up to 100% of cash (flat by {{ window_end }})
+ • Unfilled orders cancel at session close
+ • Decision cadence: every {{ action_interval }}
+ • System blocks invalid quantities (cannot oversell/over-cover)
\end{lstlisting}
\end{minipage}
\end{center}
\end{tcolorbox}
\caption{Structural reorganization consolidating sections into a unified \texttt{PORTFOLIO \& CONSTRAINTS} section. Lines in \textcolor{red!70!black}{red} with a leading ``-'' and lines in \textcolor{green!70!black}{green} with a leading ``+'' indicate deletions and additions, respectively, proposed by \emph{Adaptive-OPRO}.}
\label{fig:architecture_evolution}
\end{figure*}
\begin{figure*}[t]
\centering
\begin{tcolorbox}[
  enhanced,
  breakable,
  sharp corners=south,
  colback=gray!5!white,
  colframe=black!75!black,
  boxsep=0.8mm,
  left=1mm,right=1mm,top=1mm,bottom=1mm,
  width=\linewidth,
  title={Workflow Restructuring and Output Specification Enhancement (Prompt 4 to Prompt 5)},
  fonttitle=\bfseries
]
\begin{lstlisting}[style=diffstyle]
- 6. DECISION PROTOCOL
+ 5. THINK ➞ CHECK ➞ ACT WORKFLOW
==============================
- REVIEW ➞ REASON ➞ RESPOND
- 1. REVIEW: Regime, key drivers, levels, catalysts.
- 2. REASON: Probability map, ≥2:1 reward-to-risk, position sizing within constraints.
- 3. RISK CHECKLIST: (a) Exposure aligns with conviction; (b) Catalyst still valid; (c) Downside defined & acceptable.
- 4. RESPOND: ACT (issue order) or WAIT/HOLD.  Patience is edge when conditions are unclear.
+ STEP 1: Diagnose Regime & Narrative (use all dashboard feeds).
+ STEP 2: Map Key Levels & Catalysts; assign probabilities.
+ STEP 3: Define Reward:Risk (target ≥2:1) and provisional size within constraints.
+ STEP 4: PRE-ORDER RISK CHECKLIST 
+ • Cash / short limits respected
+ • Position aligns with conviction & catalyst
+ • Downside defined; R:R ≥2:1
+ • Flat by {{ window_end }} if short
+ STEP 5: DECIDE
+ • ACT: issue orders
+ • WAIT/HOLD: output [] (patience override)

==============================
- ORDER OUTPUT SCHEMA (STRICT)
+ 6. ORDER OUTPUT SPEC (STRICT)
==============================
Return ONLY a JSON array or [] - no extra text.
Each object must match exactly:
{
  "action": "BUY | SELL | SHORT | SHORT_COVER | ",
  "orderType": "MARKET | LIMIT | STOP",
  "price": float | null,
  "quantity": integer,
  "explanation": "Brief strategic reasoning"
}
Invalid fields, casing, or additional text will cause order rejection.
\end{lstlisting}
\end{tcolorbox}
\caption{Decision protocol restructuring from informal \texttt{REVIEW → REASON → RESPOND} to structured five-step \texttt{THINK → CHECK → ACT} workflow. Lines in \textcolor{red!70!black}{red} with a leading ``-'' and lines in \textcolor{green!70!black}{green} with a leading ``+'' indicate deletions and additions, respectively, proposed by \emph{Adaptive-OPRO}.}
\label{fig:workflow_evolution}
\end{figure*}

\subsection{Progressive Prompt Evolution: From Generic Foundation to Optimized Performance}
The GPT-o4-mini optimization trajectory demonstrates systematic prompt evolution through three distinct phases, each building upon previous discoveries to achieve cumulative performance improvements. The optimization process adapts to both model-specific response patterns and varying market regime requirements.

The progression from baseline (37.2) through intermediate optimization (51.4) to final optimization (72.1) reveals how systematic refinement can compound initial improvements into substantial performance gains. These three representative prompts (Prompt 1, Prompt 4, and Prompt 11) from the full optimization trajectory illustrate the key evolutionary patterns that drive performance enhancement.

 The baseline prompt (Prompt 1) is documented in Appendix \ref{appsec:prompts}; here we present only the intermediate and final optimized variants to avoid duplication.

The intermediate optimization achieves structural refinement by systematically eliminating architectural complexity while strengthening core functionality. Figure~\ref{fig:intermediate_optimization} reveals this transformation: verbose explanations are stripped away and replaced with a compact, numbered decision framework that provides clear analytical guidance. The constraint presentation undergoes similar streamlining, retaining comprehensive coverage while dramatically improving clarity. Crucially, the framework maintains an advisory approach (\texttt{Define thesis \& edge}) that guides without constraining, avoiding over-specification that could limit model flexibility. This architectural simplification creates a foundation optimized for further enhancement.

The final optimization achieves breakthrough performance by expanding upon this concise foundation with granular procedural guidance. Figure~\ref{fig:final_optimization} showcases the evolved architecture where the decision framework expands to six numbered steps with explicit descriptions: \texttt{Define Thesis \& Edge: state your core conviction} and \texttt{Validate Compliance: ensure all constraints are met before submission.} The market context integration becomes systematically organized with consistent bullet-point formatting and descriptive labels like \texttt{Technical Analysis} and \texttt{News Impact}. The constraint presentation achieves optimal balance between completeness and clarity, providing comprehensive operational guidance without cognitive overload. This final optimization demonstrates how systematic refinement can compound architectural improvements into substantial performance gains, with each evolution building upon and enhancing previous discoveries.

\begin{figure*}[t]
\centering
\begin{tcolorbox}[colback=teal!4!white,colframe=teal!30!black,coltext=black,
                  title=Intermediate Optimization (Prompt 4) - Score: 51.4,
                  fonttitle=\bfseries, sharp corners=south]
\begin{center}
\begin{minipage}{\linewidth}
\small
\begin{verbatim}
# ELITE {{ instrument }} TRADER
Window: {{ window_start }} → {{ window_end }} | Current: {{ now }} | Interval: {{ action_interval }}

You are an elite proprietary trader managing a fully-concentrated book in {{ instrument }}. 
Maximize end-of-window risk-adjusted performance through high-conviction, strategic trades 
with disciplined execution and strategic patience.

## MARKET CONTEXT
{% if market_open %}
Price: O {{ open }} H {{ high }} L {{ low }} C {{ close }} | Vol {{ volume }}
{% else %}
Market Closed - orders queue for next session
{% endif %}

{% if market_analysis %}* Technical: {{ market_analysis }}{% endif %}
{% if news_analysis %}* News: {{ news_analysis }}{% endif %}
{% if fund_analysis %}* Fundamentals: {{ fund_analysis }}{% endif %}
{% if reflection_analysis %}* Reflection: {{ reflection_analysis }}{% endif %}

## DECISION FRAMEWORK
1. Define thesis & edge
2. Identify entry, stop, and target levels
3. Assess risk/reward & size within cash limits
4. Choose order type & execution timing
5. Verify constraints & finalize plan

## CONSTRAINTS & PORTFOLIO
- Fully concentrated in {{ instrument }}, Cash ${{ portfolio_cash }}
- Long {{ shares_long }} | Short {{ shares_short }} | Net {{ shares_net }}
- Recent orders: {{ executed_orders }}
- Max short = 100% cash; close all shorts by {{ window_end }}
- Actions: BUY, SELL, SHORT, SHORT_COVER
- Order Types: MARKET, LIMIT, STOP
- Orders expire at session close; re-submit if needed

## OUTPUT SPECIFICATION
Return only a JSON array of orders or []:
[  
  {
    "action": "BUY | SELL | SHORT | SHORT_COVER",
    "orderType": "MARKET | LIMIT | STOP",
    "price": float | null,
    "quantity": integer,
    "explanation": "Concise strategic reasoning"
  }
]
\end{verbatim}
\end{minipage}
\end{center}
\end{tcolorbox}
\caption{Intermediate optimization (GPT-o4-mini, Prompt 4) featuring streamlined structure with a numbered decision framework and concise constraint presentation. Score: 51.4}
\label{fig:intermediate_optimization}
\end{figure*}

\begin{figure*}[h!]
\centering
\begin{tcolorbox}[colback=blue!5!white,colframe=blue!30!black,coltext=black,
                  title=Optimized Prompt (Prompt 11) - Score: 72.1,
                  fonttitle=\bfseries, sharp corners=south]
\begin{center}
\begin{minipage}{\linewidth}
\small
\begin{verbatim}
# ELITE {{ instrument }} TRADER
Window: {{ window_start }} → {{ window_end }} | Now: {{ now }} | Interval: {{ action_interval }}

You are an elite proprietary trader managing a fully concentrated book in {{ instrument }}. 
Your mission is to maximize end-of-window risk-adjusted performance with high conviction 
and disciplined execution. Think critically and adapt your approach as market conditions evolve.

## MARKET CONTEXT
{% if market_open %}
- Price: O {{ open }} H {{ high }} L {{ low }} C {{ close }} | Vol {{ volume }}
{% else %}
- Market Closed - orders queue for next session
{% endif %}
{% if market_analysis %}- Technical Analysis: {{ market_analysis }}{% endif %}
{% if news_analysis %}- News Impact: {{ news_analysis }}{% endif %}
{% if fund_analysis %}- Fundamental Overview: {{ fund_analysis }}{% endif %}
{% if reflection_analysis %}- Reflection: {{ reflection_analysis }}{% endif %}

## PORTFOLIO & CONSTRAINTS
- Total Allocation: 100% in {{ instrument }}, Cash ${{ portfolio_cash }}
- Positions: Long {{ shares_long }}, Short {{ shares_short }}, Net {{ shares_net }}
- Recent Activity: {{ executed_orders }}
- Max short = 100% cash; all shorts must close by {{ window_end }}
- Orders expire at session close; unfilled orders cancel (re-submit to persist)

## DECISION FRAMEWORK
1. Define Thesis & Edge: state your core conviction.
2. Map Key Levels: identify entry, stop-loss, and target levels.
3. Assess Risk/Reward: compute per-share risk, total risk, and reward potential.
4. Allocate Size: determine quantity within cash limits (${{ portfolio_cash }}).
5. Choose Execution: select action (BUY | SELL | SHORT | SHORT_COVER) 
   and orderType (MARKET | LIMIT | STOP).
6. Validate Compliance: ensure all constraints are met before submission.

## OUTPUT SPECIFICATION
Return only a JSON array of orders or an empty array ([]). No extra text:
[
  {
    "action": "BUY | SELL | SHORT | SHORT_COVER",
    "orderType": "MARKET | LIMIT | STOP",
    "price": float | null,
    "quantity": integer,
    "explanation": "Concise strategic reasoning"
  }
]
\end{verbatim}
\end{minipage}
\end{center}
\end{tcolorbox}
\caption{Final optimized prompt (GPT-o4-mini, Prompt 11) with a six-step decision framework and systematic market context organization. Score: 72.1}
\label{fig:final_optimization}
\end{figure*}

\section{Reproducibility}
All experiments are conducted on a MacBook Pro with an Apple M3 Pro chip (11-core CPU) and 18~GB of unified memory. Our experiments are conducted using an updated version of the StockSim environment \cite{papadakis2025stocksim}, with modifications to support the ATLAS multi-agent architecture, \emph{Adaptive-OPRO} optimization, and reflection-based mechanisms (implementation details in code). An example configuration for GPT-o4-mini using \emph{Adaptive-OPRO} on XOM is provided under \texttt{configs/o4-mini-adaptive-opro-config.yaml}. All other experimental configurations can be reproduced by following the StockSim documentation and adapting this sample.

\begin{table}[th]
\centering
\small
\begin{tabularx}{\linewidth}{>{\raggedright\arraybackslash}X>{\raggedright\arraybackslash}X}
\toprule
\textbf{Model ID} & \textbf{Model Card / Provider Identifier} \\
\midrule
Llama 3.3-70B      & \ttwrap{meta.llama3-3-70b-instruct-v1:0} \\ \hline
Claude Sonnet 4    & \ttwrap{anthropic.claude-sonnet-4-20250514-v1:0} \\ \hline
Qwen3 235B A22B 2507       & \ttwrap{qwen.qwen3-235b-a22b-2507-v1:0} \\ \hline
Qwen3 32B (dense)        & \ttwrap{qwen.qwen3-32b-v1:0} \\
\bottomrule
\end{tabularx}
\caption{Models accessed via Amazon Bedrock.}
\label{tab:aws_models}
\end{table}

\begin{table}[th]
\centering
\small
\begin{tabularx}{\linewidth}{>{\raggedright\arraybackslash}X>{\raggedright\arraybackslash}X}
\toprule
\textbf{Model ID} & \textbf{Model Card / Docs} \\
\midrule
GPT-o4-mini & \ttwrap{gpt-4o-mini-2024-07-18} \\ \hline
GPT-o3      & \ttwrap{gpt-o3-2025-04-16} \\
\bottomrule
\end{tabularx}
\caption{Models accessed via OpenAI.}
\label{tab:openai_models}
\end{table}

We access Llama, Claude, and Qwen models via Amazon Bedrock (Table~\ref{tab:aws_models}). GPT models are accessed via OpenAI APIs (Table~\ref{tab:openai_models}). We interface with all LLMs strictly through provider APIs and do not employ any local hardware or fine-tuning.

\section{Use of AI assistants}
We sparsely leveraged ChatGPT 5.2 for grammatical assistance and linguistic polishing.

\end{document}